\documentclass[fleqn,usenatbib]{mnras}

\usepackage[T1]{fontenc}
\usepackage{ae,aecompl}

\usepackage{graphicx}	
\usepackage{amsmath}	
\usepackage{amssymb}	

\newcommand{\Nexus}{\textsc{NEXUS}}
\newcommand{\nexus}{\textsc{NEXUS+}}
\newcommand{\logFilter}{\textsc{logFilter}}
\usepackage{newtxtext,newtxmath}
\usepackage{anyfontsize}

\title[Cosmic Web Dynamics: Forces and Strains]{Cosmic Web Dynamics: Forces and Strains}

\author[R. Kugel \& R. van de Weygaert]{
Roi Kugel,$^{1,2}$
Rien van de Weygaert,$^{1}$
\\
$^{1}$Kapteyn Astronomical Institute, University of Groningen,PO Box 800, 9747 AD, Groningen, The Netherlands\\
$^{2}$Leiden Observatory, Leiden University, PO Box 9513, NL-2300 RA Leiden, the Netherlands
}

\date{Accepted XXX. Received YYY; in original form ZZZ}

\pubyear{2022}

\begin{document}
\label{firstpage}
\pagerange{\pageref{firstpage}--\pageref{lastpage}}
\maketitle

\begin{abstract}
This study concerns an inventory of the gravitational force and tidal field induced by filaments, walls, cluster nodes and voids on Megaparsec scales and how they assemble and shape the Cosmic Web. The study is based on a N$_{\rm Part}=512^3$ $\Lambda$CDM dark matter only N-body simulation in a (300$h^{-1}$~Mpc)$^3$ box at $z=0$. We invoke the density field NEXUS+ multiscale morphological procedure to assign the appropriate morphological feature to each location. We then determine the contribution by each of the cosmic web components to the local gravitational and tidal forces. We find that filaments are, by far, the dominant dynamical component in the interior of filaments, in the majority of underdense void regions and in all wall regions. The gravitational influence of cluster nodes is limited, and they are only dominant in their immediate vicinity. The force field induced by voids is marked by divergent outflowing patterns, yielding the impression of a segmented volume in which voids push matter towards their boundaries. Voids manifest themselves strongly in the tidal field as a cellular tapestry that is closely linked to the multiscale cosmic web. However, even within the interior of voids, the dynamical influence of the surrounding filaments is stronger than the outward push by voids. Therefore, the dynamics of voids cannot be understood without taking into account the influence of the environment. We conclude that filaments constitute the overpowering gravitational agent of the cosmic web, while voids are responsible for the cosmic web's spatial organisation and hence of its spatial connectivity.
\end{abstract}




\section{Introduction}
This study concerns a systematic inventory of the gravitational force and tidal field on Megaparsec scales and its role in determining the structure of the Cosmic Web. We assess the force field and tidal field induced by filaments, walls, cluster nodes and voids, and assess in how far they contribute, and dominate, the gravity and tides in the various regions of the cosmic web. It allows us to investigate the question which morphological features in the large scale universe dominate - and drive - the gravitationally driven formation and evolution of the largest structure in the universe. Also, as this will depend to a considerable extent on location, we include a systematic inventory of the identity of the regions over which voids, filaments and clusters dominate the the gravitational and tidal force. In an accompanying study, we specifically focus on the dynamical influence of cosmic voids in the large scale matter distribution, which represent the major share of the cosmic volume and who - along with filaments - dominate the dynamics of the large scale Universe. 

\begin{figure*}
  \mbox{\includegraphics[width=1.0\textwidth]{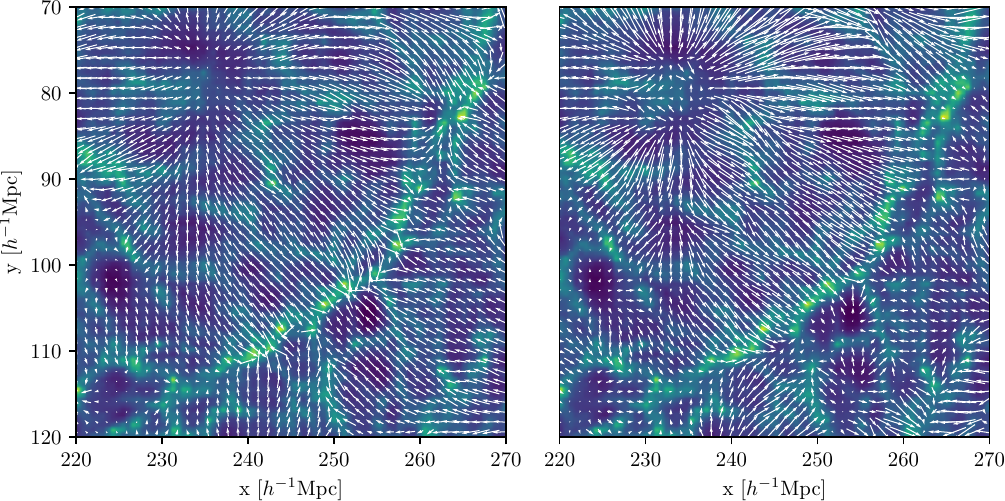}}
    \caption{Gravitational force field in and near a void. Left: the force vectors representing the local direction and amplitude of the total gravitational force, ie. the force exerted by the total cosmic mass distribution. Right: the local force vectors resulting from the combined influence of mass elements in void regions. The vector lengths are normalised with respect to the local total velocity and show the relative contribution of the field at that location.} 
    \label{fig:voidforcefld}
\end{figure*}

\subsection{The Cosmic Web: structure and detection} 
The \textit{Cosmic Web} is the intricate multiscale network defined by the matter and galaxy distribution on Megaparsec scale~\citep{zeldovich1970,einasto1977,bond1996,weybond2008,NEXUS_CW_EVO}. It represents the fundamental spatial organisation of matter on scales of a few up to a hundred Megaparsec. Galaxies, intergalactic gas and dark matter arrange themselves in a salient wispy pattern of dense compact clusters, long elongated filaments, and sheetlike tenuous walls surrounding near-empty void regions. Filaments are the most visually outstanding features of the Megaparsec Universe, in which around $50\%$ of the mass and galaxies in the Universe resides. On the other hand, almost 80\% of the cosmic volume belongs to the interior of voids~\citep[see e.g.][]{NEXUS_CW_EVO,Punya2018}. Together, they define a complex spatial pattern of intricately connected structures, displaying a rich geometry with multiple morphologies and shapes. This complexity is considerably enhanced by its intrinsic multiscale nature, including objects over a considerable range of spatial scales and densities. The connectivity of this rich pallet of features, the nature of how the various structures connect to establish the pervasive network, has only recently been recognised as an important defining - topological - aspect \citep{MMF3,codis2021,wilding2021,Feldbrugge2023}. For a recent up-to-date report on a wide range of relevant aspects of the cosmic web, we refer to the volume by~\cite{iau308}.

In the observational reality, the existence and structure of the Cosmic Web has been revealed in the most detail by maps of the nearby cosmos produced by large galaxy redshift surveys. Starting from first revelation of the web-like arrangement of galaxies by the CfA2 survey \citep[e.g.][]{Lapperent1986}, subsequent surveys such as 2dFGRS, the SDSS, the 2MASS and GAMA redshift surveys~\citep{colless2003,tegmark2004cosmological,huchra20122mass,GAMA2015} established the web-like arrangement of galaxies as a fundamental characteristic of cosmic structure. Maps of the galaxy distribution at larger cosmic depths, such as VIPERS~\citep{vipers2013}, showed its existence over a sizeable fraction of cosmic time. 

Also the intergalactic gaseous medium, closely follows the web-like structure defined by the dark matter, the principal component of the cosmic web. A range of observational probes have detected the web-like structure over which intergalactic gas, in a range of thermodynamic states \citep[see][for a review]{meiksin2009}, has diffused itself. Ly$\alpha$ absorption lines in the spectra of bright background sources such as QSOs are piercing through the web-like assembly of neutral hydrogen gas in the cosmic web at high redshifts \citep{cen1996,cen1997}. The combination of sufficiently close linear probes even allows a reconstruction of the full three-dimensional intergalactic hydrogen lanes \citep{pichon2000}.
\begin{figure*}
  \includegraphics[width=.9\textwidth]{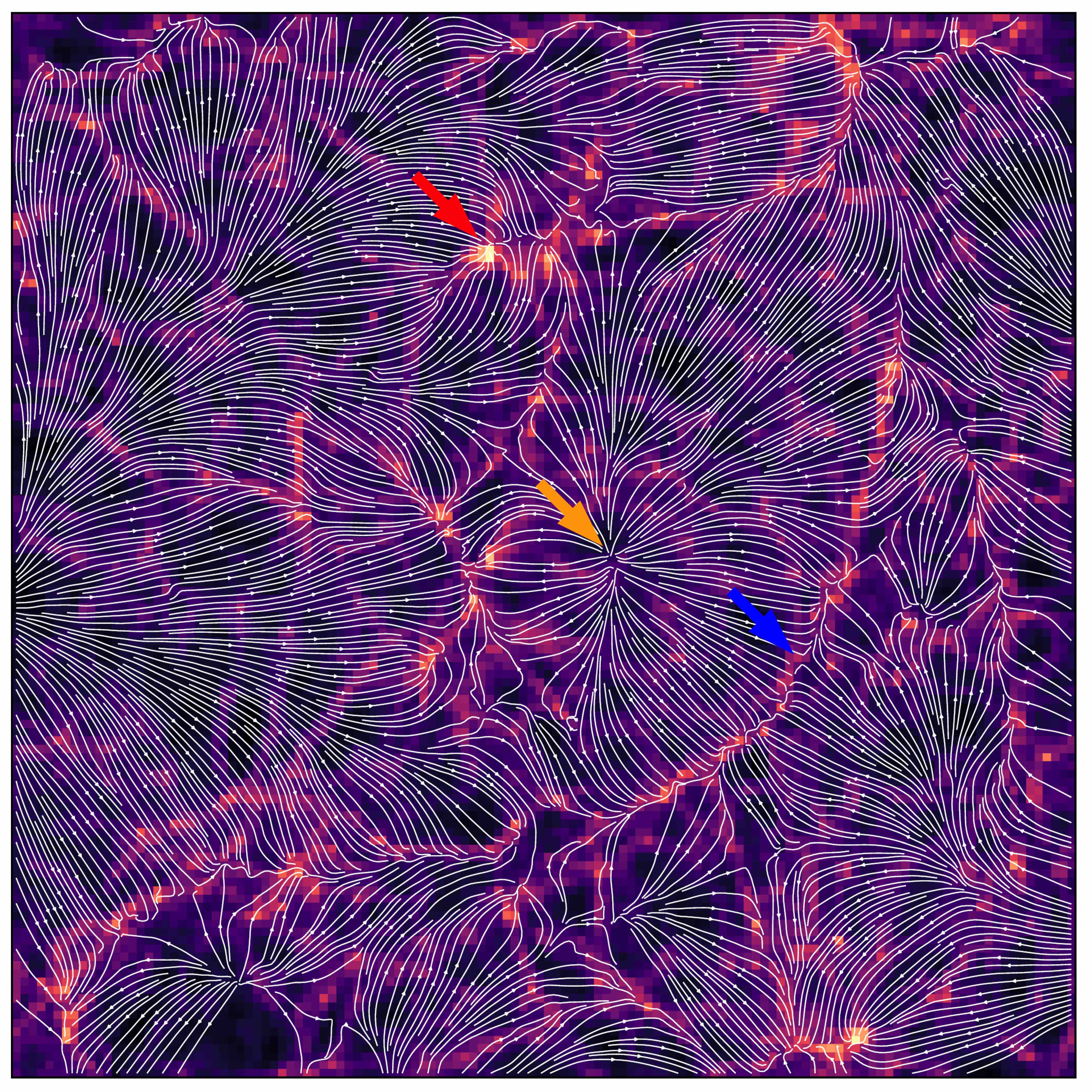}
    \caption{Flows of matter in the cosmic web. Superimposed on the (logarithmic) map of the cosmic density field in a slice of the simulation box  are the streamlines of the corresponding cosmic velocity field. Arrows indicate three different aspects of the flow field: The red arrow highlights velocity inflow into a cluster node (negative divergence), the orange arrow highlights velocity outflow from a void (positive divergence) and the blue arrow highlights shear flow along a filament.}
    \label{fig:cosmicweb_flowfield}
\end{figure*}

It has already led the Clamato survey \citep{khanlee2018} to successfully produce fascinating maps of the full three-dimensional gaseous cosmic web at high redshifts. Recent observations by the MUSE integral field unit on the very large telescope, even managed to see the Ly$\alpha$ emission from the filamentary gaseous extensions around clusters directly. At lower redshifts, most of the intergalactic gas has heated up as it settled in the deepening potential wells of the dark matter cosmic web. This warm gas, the so called WHIM, is assumed to represent the major share of baryons in the current Universe. As such is a prime target for detection and mapping \citep{IGM2018,IGMNatureMacquart2020}, although it had proven to be notoriously hard to detect. It has been more straightforward to detect the hot gas residing in the strongest filamentary features in the cosmic web at high redshifts, filling the short dense bridges between two adjacent clusters. The hot gas reveals itself through the Sunyaev-Zeldovich upscattering of CMB photons. It has even allowed the detection of a few individual high-redshift gaseous filaments, where their ubiquitous presence has been revealed by stacking numerous cluster pair Sunyaev-Zeldovich observations \citep{bonjean2018,degraaff2019}.

\begin{figure*}
  \includegraphics[width=0.8\textwidth]{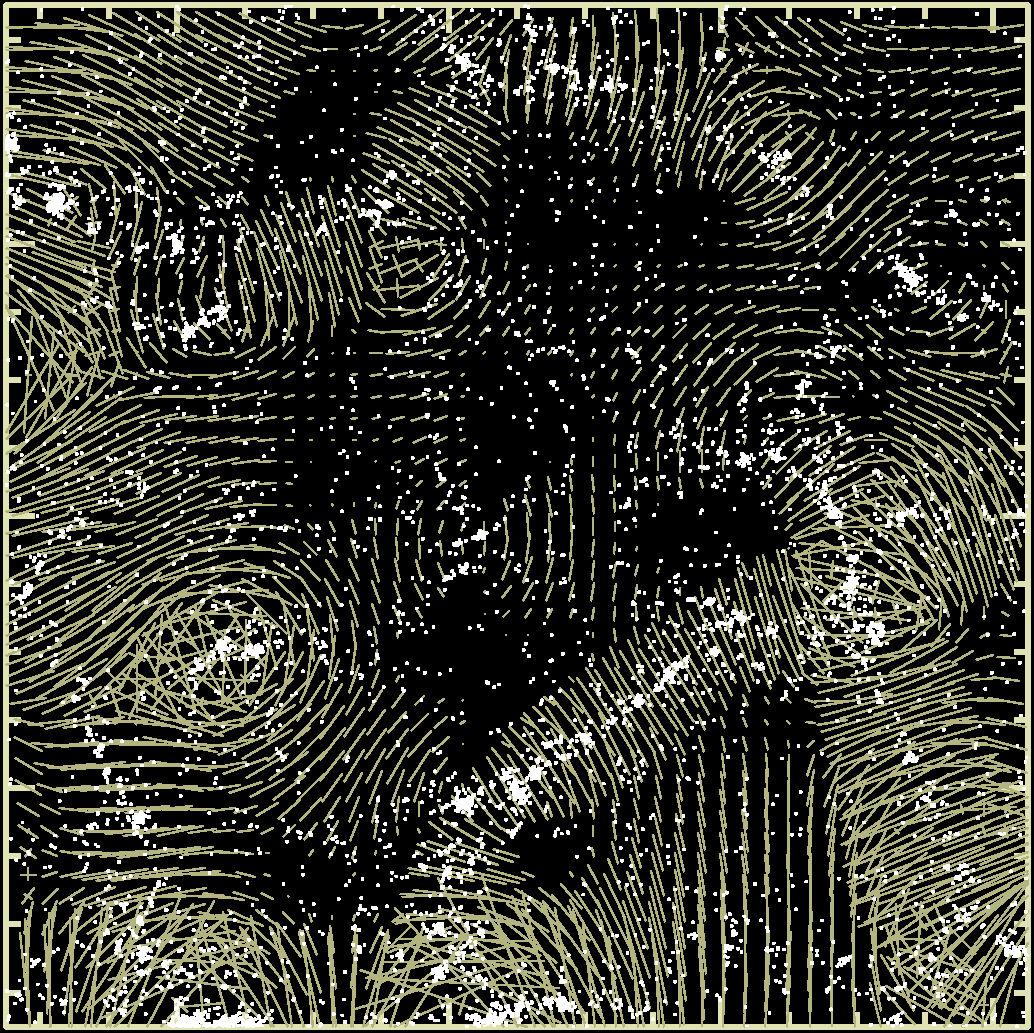}
  \caption{Cosmic web and tidal force field: the relation between the cosmic web and the spatial pattern of the
    corresponding compressional tidal field components. The image shows
    an LCDM matter distribution at the present cosmic epoch, along with the (compressional component) tidal bars in a
    a $5h^{-1}$ Mpc thin central slice. The simulation is a realization of a dark matter only CDM based scenario (in an open,
    $\Omega=0.3$ Universe. The tides are determined on a scale $R_G = 2h^{-1}$ Mpc). The matter distribution, displaying a
    pronounced weblike geometry, is clearly intimately linked with the characteristic coherent compressional tidal bar pattern.}
    \label{fig:cosmicweb_tidalbar}
\end{figure*}

\subsection{The Cosmic Web: dynamics, formation and evolution} 
The origin of the cosmic web is to be found in the tiny inhomogeneities in the primordial spatial matter distribution. These induce an inhomogeneous and anisotropic gravitational force field. It sets off the migration of matter, culminating in the continuous growth of density and velocity inhomogeneities \citep{Peebles1980}. Ultimately, as the matter fluctuations become nonlinear, it leads to the contraction and collapse of structures and expansion of underdense regions. At the transition from the initial long phase of linear growth of the primordial (Gaussian) density field and density field to the complex structured nonlinear mass distribution, we find the emergence of the intriguing and complex spatial pattern of the \textit{Cosmic Web}. It marks the cosmic web as a key phase in the dynamical buildup of structure in the Universe. 

A plethora of cosmological computer simulations \citep{defw1987,Springel2005,Illustris2018,Miratitan2020,Angulo2021}, including more elaborate cosmological hydrodynamics simulations \citep[e.g.][]{Magneticum2014,Eaglemain,BAHAMAS2017,Illustris2018,FLAMINGO2023} revealed that through gravitationally driven evolution the Gaussian initial density and velocity perturbations morph into an intricate web-like pattern that
resembles that seen in the spatial distribution of galaxies. The simulations show that the cosmic web is a fundamental aspect of structure formation in the standard $\Lambda$CDM cosmology, although this also turns out to be true for variations around the standard $\Lambda$CDM model \citep[for a review see][]{DEReview2008}. 

Within the context of the gravitational buildup of the cosmic web, the gravitational forces induced by the inhomogeneous mass distribution
are the agent for the buildup of structure. They are instrumental in directing the cosmic migration streams that transport matter from low density areas to emerging matter concentrations. The accompanying tidal force field is the key towards shaping the mass distribution into a weblike pattern. This had already been recognized and accurately described in the mildly nonlinear stage by the Zel'dovich formalism \citep{zeldovich1970,shandarin1989}. The formation and evolution of the characteristic anisotropic structures, ie. the filaments and walls, are the
product of the anisotropic tidal strains and resulting anisotropic flow field and deformations \citep{bond1996,hahn2007,weybond2008,lee2009,lee2010,hahn2010,wang2014,job2018Caustics,Feldbrugge2024}. It establishes a close relation between the weblike structures and the anisotropy of
the induced migration flows, which has been the focus of a few insightful studies \citep{kitaura2012b,hoffman2012,wang2014}. 
Recent work has also revealed the extent to which the spatial structure of the tidal force field determines the connections between the components of the cosmic web. The connectivity can even already be recognised in the primordial tidal and deformation field, showing the extent towards which the primordial anisotropic force field is steering and shaping the structure of the emerging cosmic web \citep[see eg.][]{Wilding2022,Feldbrugge2023b,Feldbrugge2024}. It even leads to the realization that embryonic outline of the cosmic web, in particular its filamentary network, can already be seen in the primordial tidal eigenvalue field \citep{Wilding2022,Feldbrugge2023b} 
(see figure~\ref{fig:cosmicweb_tidalbar}).

For insight into the nature and origin of the characteristic properties of the cosmic web and for identifying the dependence of these on cosmology and cosmological parameters, theoretical understanding of the physical mechanisms and processes behind the emergence of the cosmic web. Analytical approximation and model have been essential in interpreting the results of surveys, as well as of simulations \citep[e.g.][]{zeldovich1970,hidding2012Adhesion,wang2014,hidding2016Adhesion,job2018Caustics,Feldbrugge2024}. An important aspect of the formation process are the forces and strains that shaped cosmic structure. For example, the phase-space based \textit{Caustic Skeleton} model by \cite{job2018Caustics} demonstrated that a full understanding of the cosmic web structure is obtained through the spatial characteristics of the \textit{eigenvalue} and \textit{eigenvector} fields of the cosmic tidal force field. Motivated by these consideration, the intention of the present investigation and inventory of the dynamics of the cosmic web is to establish the role of the various morphological features of the cosmic web in its formation and dynamical evolution. It involves the assessment of in how far the various elements of the cosmic web are formed and have evolved, and how they connect up in the complex, intricate pervasive network of the cosmic web.

While many structural aspects of the cosmic web have been addressed by numerous studies, its dynamics and dynamical evolution has only been - superficially - explored within a purely theoretical or simulation context. Over recent years, we have seen that the dynamics behind the formation and evolution of the cosmic web is becoming increasingly accessible to observational investigation \citep[see eg.][]{kitaura2012}. The forces and tides that shape the complex spatial pattern induce nonlinear migration currents, marked by distinct divergent and shear-like flows, have recently been traced in new, densely probed, galaxy peculiar velocity surveys. The most notable examples of this are the Cosmicflows-3 and Cosmicflows-4 \citep{Cosmicflows3,Kourkchi2020} surveys, a view that will be considerably extended by DESI \citep{DESI}. Including the peculiar velocity information even allowed to identify the impact of some individual components of the cosmic web is also found in the local universe. This concerns in particular the gravitational influence of voids: \citet{Tully2008localvoid} concluded that the Local Void has a large contribution of no less than $\sim 240 $km/s to the peculiar velocity of the Local Group. 

Also the subtle morphing influence of the tidal forces has become susceptible to observational scrutiny. The distinct anisotropic shape of the filaments and walls in the cosmic web is the direct outcome of the gravitationally driven formation of the cosmic web by the large scale anisotropic force field \citep{weybond2008}. Even the population of voids on Megaparsec scales is noticeably shaped and aligned by the large scale tidal force field \citep{parklee2007,platen2008}.

Arguably the most prominent manifestation of the (tidal) anisotropic force field induced by the inhomogeneous matter distribution in the Universe is that of gravitational lensing. Hence, at a global level the tidal dynamics of the cosmic web is reflected in the corresponding deformation of galaxy images. In case the distortions are linear, inversion allows the study of the generating (projected) mass distribution. At present, gravitational lensing has become one of the most powerful probes of the global Universe and the cosmological parameters characterising it. With the availability of the upcoming powerful and accurate cosmological surveys, such as enabled by the Vera Rubin observatory and Euclid, the tedious lensing studies may even resolve the web-like nature of dark matter distribution. A few gravitational lensing studies have even already managed to detect and resolve filamentary dark matter bridges between nearby massive clusters \citep{dietrich2012}. Amongst the most massive representatives amongst the filament population \citep{NEXUS_CW_EVO}, they leave a rather accessible and detectable lensing imprint.

Also at galaxy scales, the large scale tidal forces induce noticeable signatures \citep[see e.g.][]{weybabul1994,paranjape2021,alam2024}. Perhaps the most outstanding manifestation is that of the alignment of the spin axis of collapsing dark matter halos with the filaments in which they are embedded, and hence that of the corresponding rotation axis of galaxies. It is clear from observations \citep[see][]{Jones2010_allign,Tempel2013,Welker2020} that the rotation axis of galaxies preferentially align with the components of the cosmic web. Largely the result of the imparted tidal torques on the collapsing halos \citep{Hoyle1949,Peebles69TTT,efstathiou1980,White1984ttt,LeePen2000ttt,porciani2002a,porciani2002b,schaeffer2008}, additional secondary effects responsible for the mass dependence of the spin orientation \citep{MMF3,hahn2007,hahn2010,codis2012,Tempel2013,Punya2018,Punya2019TTT,lopez2021,zhang2023} may be the result of the induced filamentary inflow of mass \citep{ganeshaiah2021}. The latter has become a key issue of attention, given the implications for gravitational lensing studies. 

\subsection{Morphology of the Cosmic Web}
While the cosmic web has four morphologically well defined features, there are many ways to identify the different components. Over the past decades, a range of methods and formalisms have been put forward for the detection and classification of filaments. A review and comparison of more than a dozen formalisms can be found in \cite{Libeskind2018}. They distinguish at least five classes of formalisms to classify and analyse the cosmic web. Geometric filament finders are usually based on the Hessian of the density or gravitational potential at each location. It includes the Tweb and Vweb formalisms \citep{hahn2007,foreroromero2009,hoffman2012},  which emphasize and exploit the intimate link between the
tidal foce field and induced anisotropic velocity flows and the spatial structure and connectivity of the weblike pattern that is
emanating as a result of these.
The most sophisticated ones explicitly take into account the multiscale nature of the mass distribution, of which MMF/Nexus is a particular example \citep{MMF3,NEXUS_MAIN}. Topological methods address the connections between structural singularities in the mass distribution. They are amongst the most widely used formalisms in the studies of the cosmic web, in particular, that of Disperse \citep{Sousbie2011a,Sousbie2011b}. Other representatives are Spineweb \citep{MMFb} and Felix \citep{Shivashankar2016}. 

In addition to the geometric and topological formalisms, several alternative methods have played a substantial role in the study of the cosmic web. Bisous is a well-known stochastic formalism, involving Bayesian exploration based on stochastic geometric modelling of filaments \citep{Tempel2014}. It forms the basis for the widely used filament catalogue extracted from the SDSS survey \citep{Tempel2014b}. More recent developments often incorporate machine learning codes \citep[see \textit{e.g.}][]{Awad2023}. Perhaps the oldest representatives for a systematic analysis of filamentary patterns are graph-based methods. The Minimal Spanning Tree (MST) is a prime example and is figuring prominently in the cosmic web analysis of the GAMA survey \citep{Alpaslan2014}. There is even a class of cosmic web identifiers that exploit the resemblance of the cosmic web to biological branching networks. The Monte Carlo Physarum Machine, inspired by the growth of Physarum polycephalum 'slime mold', has been successfully applied to the structural analysis of the cosmic web in both simulations and observations \citep{Elek2020,Elek2022,Wilde2023}. 

Possibly the most profound techniques for classification of the cosmic web are those emanating from the analysis of the 6D phase-space structure of the cosmic mass distribution \citep{Shandarin2011,Shandarin2012,Neyrinck2012,Abel2012}. Restricted to situations in which the initial conditions are known, they yield the identification of the matter streams constituting the migration of mass in the buildup of structure. It allows the definition of an objective physical criterion for what constitutes the various structural elements of the cosmic web. The recent study by \cite{Feldbrugge2024} on the phase-space based dynamical specification of the nature of cosmic filaments emanates from a detailed phase-space based assessment of cosmic structure formation, within the context of the Caustic Skeleton model for the formation of the cosmic web \citep{Feldbrugge2018}. 

The present study is based on the MMF/Nexus morphology classification formalism\citep{MMFa,MMFb,MMF3,NEXUS_MAIN,NEXUS_CW_EVO}, that is unique in addressing both the geometric nature as well as the multiscale nature of the cosmic matter distribution. By means of the Hessian of the density field \Nexus{} to find the geometric/morphological signature of the individual components. Of particular interest for our purpose is the \nexus{} flavour of \Nexus{}. It applies log-space filtering to the density field before identification. Like shown by \citet{NEXUS_CW_EVO} this allows \nexus{} to find even the more tenuous elements of the cosmic web.

\subsection{Cosmic Web Dynamical Inventory}
The present study extends the previous cosmic web inventory of \cite{NEXUS_CW_EVO} to that of the corresponding force and tidal field influence as a function of cosmic web environment. The distinct structure of the morphological elements of the cosmic web implies that each component will also be dynamically distinct, both in its local dynamics, but also in the influence it will have on its surroundings through the force and tidal field. \cite{bond1996} argue that the main origin of filaments in the cosmic web is due to the distribution of nodes in the cosmic web, and not much work has been done on the topic since. In this work we aim to extend our knowledge about the dynamical influence and origin of the cosmic web.

To quantify the dynamical influence, we analyse a $\Lambda$CDM (dark matter only) simulation, and asses at each location in a cosmic volume the contribution by filaments, walls, voids and nodes to the local gravitational and tidal force (see eg. fig.~\ref{fig:voidforcefld} for the force field contribution by voids). This allows a statistical inventory of the dynamical dominance of the various morphological elements of the cosmic web. To this end we apply \nexus{} to identify for each point in the simulation whether it belongs to a filaments, wall, void or node. Using direct summation over the gridcells, we are then able to construct the force and tidal field originating from each component.

The obtained fields are full vector fields, for the force field, or matrix fields, for the tidal field. This allows us to study both the morphology and the strength of the fields belonging to each component. For each field we look at both the (relative) amplitude and the direction. In addition, the spatial coherence and pattern of the induced flow field is studied on the basis of streamline maps
(see fig.~\ref{fig:cosmicweb_flowfield}). We are also able to zoom in on a few specific regions to study the interplay between the different components. In this way we are able to find where, and how the different components contribute to the total force field. By looking at the tidal field we are also able assess which component is responsible for creating the an-isotropic nature of the cosmic web (see fig.~\ref{fig:cosmicweb_tidalbar}).

\subsection{Outline}
This paper is structured as follows. In section~\ref{sec:methsim} we describe the simulation we use, how we obtain the identification of the cosmic web components and how we calculate the force and tidal fields. In section~\ref{sec:totforce} we describe the force field of the individual components for the complete $(300 \text{Mpc})^2$ slice. In section~\ref{sec:zooms} we investigate two smaller regions of the full slice, and look at the force fields around a void and a node. In section~\ref{sec:Tidalfield} we describe our results for the tidal field. We finish by summarising our conclusions in section~\ref{sec:conc}.

\section{Data and Analysis: N-body simulations}\label{sec:methsim}
For the dynamical inventory of the cosmic web in the present study, we analyze a LCDM dark matter only cosmological
N-body simulation. The force and tidal inventory in the present study largely pertains to the current cosmic epoch, redshift
$z=0$. The evolution of the force and tidal force fields will be analyzed in an accompanying upcoming publication.
In this section we describe the specific methods and formalisms used to get to the inventory of
the gravitational force and tidal field within this simulation.

The simulation is processed and analyzed with a set of instruments that allow the optimal identification of the
multiscale structure and dynamics of the cosmic web. These include the translation from the discrete set of
particle position and velocities towards continuous density and flow fields that optimally retains the anisotropic
pattern of the cosmic web as well as its multiscale structure. This is accomplished through the use of the
DTFE formalism \citep{schaap2000,weyschaap2009,Cautun2011}.

Instrumental in our study is the classification and identification of the morphological components of the cosmic web.
The morphological identification and classification of the geometric components of the cosmic web is based on the use
of the MMF/Nexus formalism, specifically that of \nexus{} version of the MMF/Nexus pipeline \citep{MMFb,NEXUS_MAIN}.

The second major aspect of our analysis concerns the calculation of the (peculiar) gravity and tidal fields, globally as well specific for those induced by the individual components of the cosmic web. The computational details are described in section~\ref{sec:force_field}. To appreciate the spatial structure of the force and tidal fields better, and to interpret the results obtained, in section~\ref{sec:visual} we include a description of the various aspects of visualisation of these fields.

The necessary details of the Nbody simulation and applied analysis tools are outlined below. 

\subsection{N-body Simulation} \label{sec:simus}
The study is based on the analysis of the N-body, dark matter only simulation described in \cite{patrickthesis}. The cosmological
context is that of a $\Lambda$CDM cosmology with WMAP 3-year parameter values:
$\Omega_{\text{m}}=0.268,\ \Omega_{\Lambda}=0.732,\ \Omega_{\text{b}Zel}=0.044,\ h=0.704,\ \sigma_8=0.776$ and $n=0.947$. The simulation contains $512^3$ dark
matter particles in a simulation box with a boxlength of 300 $h^{-1}$ Mpc. The simulations start at $z=60$, following the initial
displacement and velocity set by the Zeldovich approximation \citep{zeldovich1970}. The analysis described in the present study focuses on the 
on the inventory of the force and tidal field at the current epoch, $z=0$.

\begin{figure*}
  \includegraphics[width=0.65\textwidth]{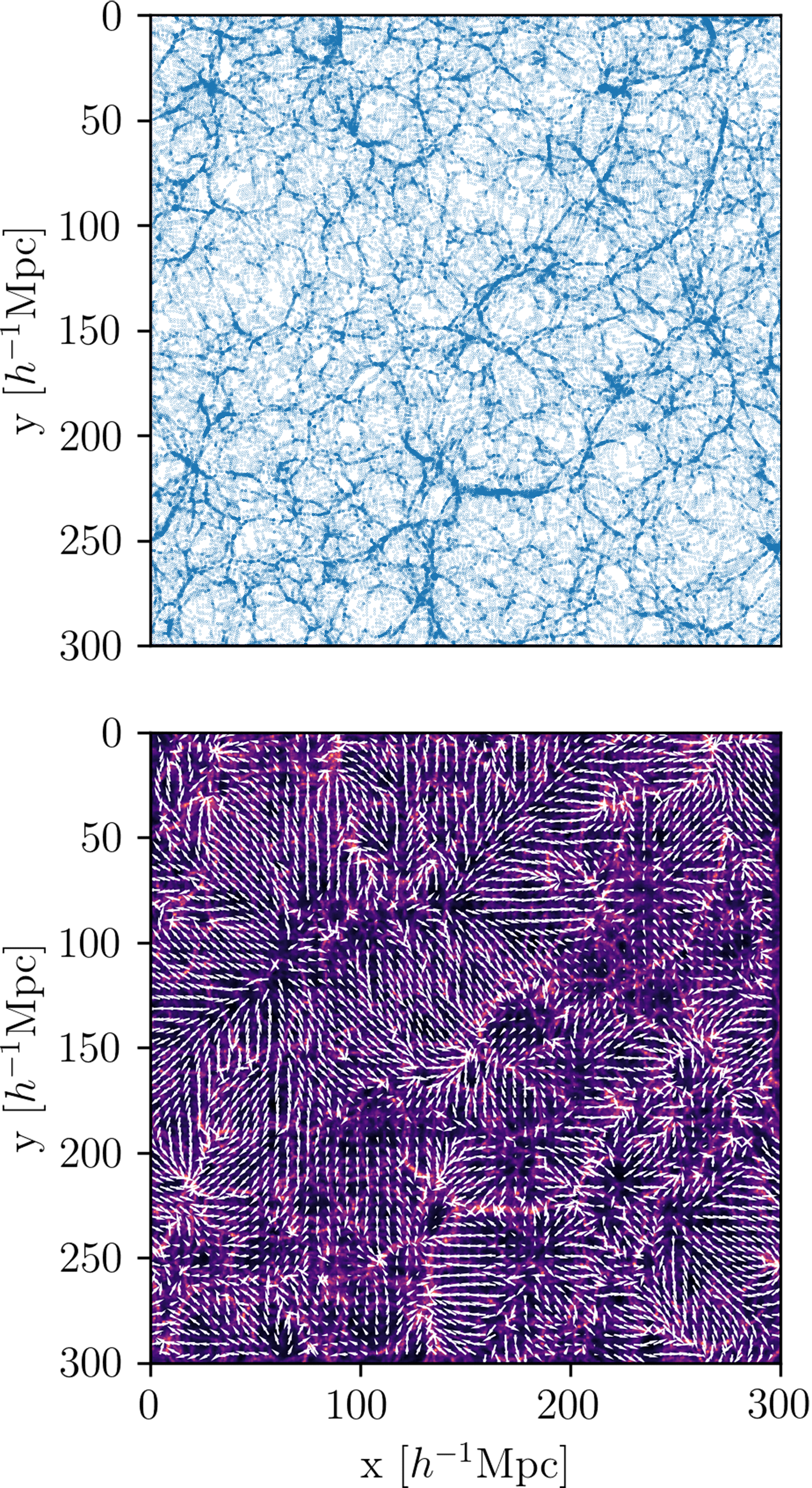}
  \caption{Cosmic Mass distribution and force field. Top panel: the particle distribution, at $z=0$ in a $0.59~h^{-1}\text{Mpc}$
    thick slice from a LCDM N-body simulation in a  $300~h^{-1}\text{Mpc}$ by $300~h^{-1}\text{Mpc}$ box.
    Bottom panel: the corresponding
    gravity force vector plot in the same slice. }
    \label{fig:nbody}
\end{figure*}

\begin{figure*}
  \includegraphics[width=.9\textwidth]{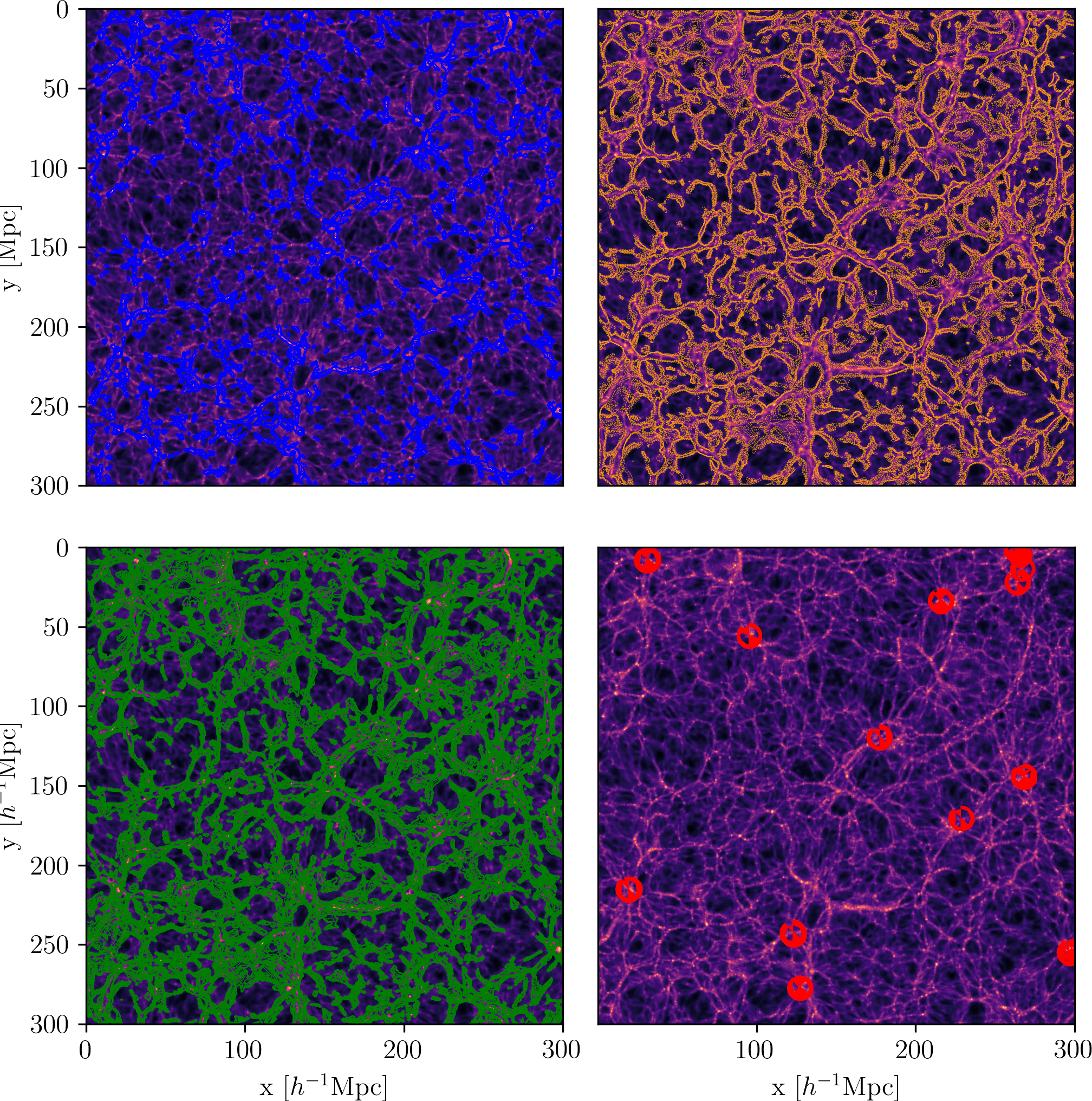}
  \caption{Cosmic web components identified by Nexus+: filaments, voids, walls and cluster nodes. Each panel shows a representation of a $300~h^{-1}\text{Mpc}$ by $300~h^{-1}\text{Mpc}$ by $0.59~h^{-1}\text{Mpc}$ slice from the N-body simulation used. Top left panel: Filaments. The \nexus{} filament contours, in blue, are superimposed on the log density field. Top right panel: Voids. The \nexus{} void boundary contours, in orange, are superimposed on the log density field. Bottom left panel: Walls. the \nexus{} wall contours, in green, are superimposed on the log density field. Bottom right panel: Cluster nodes. The cluster nodes, indicated by red symbols, are superimposed on the log density field. }
    \label{fig:denMMF}
\end{figure*}

\subsection{DTFE Density and Velocity maps}
To translate the spatial particle distribution into density and velocity maps, we convert the particle locations and velocities into a space-filling DTFE density and velocity maps. The density $\Delta$,
\begin{equation}
    \Delta = \frac{\rho}{\rho_u}=\delta+1
\end{equation}
\noindent and velocity field $\mathbf{v}_{\text{DTFE}}$ are sampled on a $512^3$ grid. It corresponds to a resolution of $0.59h^{-1}$Mpc
for each grid cell.  

The Delaunay tessellation field estimator (DTFE) \citep{schaap2000,weyschaap2009,NEXUS_MAIN}, uses the Voronoi and Delaunay tessellation of the spatial particle distribution to get an unbiased volume-weighted estimate of the local density including a natural interpolation over the entire volume using the dual Delaunay tessellation as natural interpolation grid. Application of DTFE to the velocity field, traced by the simulation particles, not only yields a volume-weighted space covering reconstruction of the velocity field but also yields maps of the first order gradient of the flow field, ie. the divergence and shear (and even vorticity) of the flow field \citep{bernwey1997,bernwey1998,romanowey2007,weyschaap2009}. THE DTFE density and velocity fields are sampled on a $512^3$ grid. 

\subsection{MMF/NEXUS+}\label{sec:NEXUS}
To dissect the cosmic matter distribution into the various morphological components of the cosmic web we use the MMF/NEXUS pipeline, specifically its \nexus{} version \citep{NEXUS_MAIN,NEXUS_CW_EVO}. It is the highest dynamic range version of the \Nexus{} library of routines for pattern classification based on the MMF/Nexus Multiscale Morphology Filter formalism \citep[][]{MMFb,MMFa,MMF3,NEXUS_MAIN,NEXUS_CW_EVO}. For a short review see appendix~\ref{app:mmfnexus}, as well as \cite{Libeskind2018}. We specifically chose to use the \nexus{} version as it is optimally suited to take into account the multiscale nature and the wide dynamic range of cosmic web density field \citep{NEXUS_CW_EVO}. Compared with the other NEXUS methods it provides the sharpest rendering of the cosmic web, including the more tenuous walls and filaments. 

Instrumental and unique for the class of MMF cosmic web identification methods is that it simultaneously pays heed to two principal aspects characterising the web-like cosmic mass distribution. The first aspect is that of the different geometric shapes of the various structural components of the cosmic web, in particular the anisotropic shapes of filaments and walls. The Hessian of the corresponding fields translates into the local geometry and local anisotropy on the basis of the ratios of its eigenvalues. That is, it determines whether the mass distribution has a roundish, flattened or elongated geometry. The second aspect is
that of the multiscale character of the cosmic mass distribution, the product of the hierarchical evolution and buildup of structure
in the Universe. To this end, MMF/Nexus invokes a \textit{Scale-Space} analysis to enable it to probe the scale dependence of the
local geometry of the mass (or velocity or gravity) distribution. It allows the detection, and complete and unbiased characterisation, of features present at all scales, from the prominent structures present in overdense regions to the tenuous networks pervading the
cosmic voids.  

The MMF/Nexus formalism results in a scale adaptive framework that classifies the matter distribution on the basis of local spatial
variations in either the density field, velocity field or gravity field. Their geometry and anisotropy are encoded in the Hessian
matrix determined at each of the spatial scales at which the fields are assessed. Subsequently, a set of morphological filters is
used to classify the spatial matter distribution into three basic components, the clusters, filaments and walls that constitute the
cosmic web. It produces a map in which for each location in the analysed volume the morphological identity is specified. The end
product of the pipeline is a multiscale identification of the cosmic web into its structural components. To this end, each location and mass element of the matter distribution is assigned a unique morphological identity as either belonging to a filament, cluster node, wall or void.

For more details on the MMF/Nexus formalism we refer to appendix~\ref{app:mmfnexus}.

\begin{table}
\centering
\caption{Mass and volume fractions of the different components of the cosmic web as identified by \nexus{}.}
\begin{tabular}{l r r}
 \hline
    Component & Mass fraction [\%] & Volume fraction [\%]\\ \hline
    Filaments & 46.2 & 7.25\\
    Voids & 21.4 & 72.09\\ 
    Nodes & 4.6 & 20.65 \\ 
    Walls & 27.8  & 0.01\\
    
\hline    
\end{tabular}
    \label{tab:fractions}
\end{table}

\subsubsection{NEXUS+: practical implementation}
While in principle the \Nexus{} \textit{Scale Space} formalism involves an infinite number of scales, in our practical
implementation we use a finite number of filter scales, restricted to the range of $[1.0,4.0] h^{-1}\text{Mpc}$.
We tested a few other filter ranges, and concluded that the used range provides a robust population of filaments. 
This includes the major dynamically dominant filaments, along with a fair fraction of the tenuous tendrils. We also
assessed this for the higher redshift simulation snapshots, to find that for $z>3.8$ most filaments are to be
found in this scale range. 

The morphological signature is evaluated on a $512^3$ grid, ie. for each of the gridpoints on this grid \nexus{}
provides us with the identity of the morphological features to which it belongs. 

\section{Force and Strain Field: Procedure}
For the computation of the diverse gravitational force and tidal fields, we use a brute force approach. At each grid location
we compute the contributions by the various morphological elements of the cosmic web by summing the gravitational and tidal force
contributions from the grid points located in the regions that have been identified with this morphology. In short, the force
and tidal influence of filaments is the sum of the mass elements residing in filaments, of voids the sum of mass elements
residing in voids, etc. The total (peculiar) gravitational force and tidal strain at each location is the
sum of the contributions by all mass elements in the volume.

\bigskip
The stated force and tidal force calculations involve the following practical issues:
\begin{itemize}
\item[1.] The gravitational force field and tidal strain field are evaluated on the same $512^3$ grid as the one on which we
  have computed the DTFE density field, and assessed the \nexus{} identity (see previous section).
\item[2.] The forces and tidal strains are computed by brute force, i.e. we compute these at a given location by summing
  the force and tidal contribution.
\item[3.] Following a sheer brute force summation, our computational resources allow us to carry out the computation for
  a restricted number of locations, instead of for all $512^3$ gridpoints.
\item[4.] For the brute force computation, we assume that a spherical surrounding volume of radius $150 h^{-1}\text{Mpc}$,
  corresponding to the largest sphere that fits inside the simulation box, is
  sufficient to include all gravitationally relevant influences. In other words, force and tide contributions are considered
  to be negligible beyond a distance of $r=150h^{1-}$Mpc. This is certainly true for the tidal contributions, and turns out to
  be valid to very good approximation for the gravitational force itself. 
\item[5.] A more efficient Fourier space based formalism is under development, but has not yet been applied to the analysis of
  the simulation in this study. A Fourier code would automatically guarantee periodic boundary conditions pertaining to the
  simulation box, and hence by implication include the force and tidal influences throughout the entire simulation volume. 
\end{itemize}

Because the brute force calculations are computationally intensive, we restricted the force and tidal evaluations to a few 
slices of the simulation box, yielding a total of $512^2$ force evaluations on the corresponding $512^2$ grid. The force 
at each point is computed using the full three dimensional grid, yielding a fully three dimensional force and tidal tensor
at each gridpoint. For the statistical results presented in our study, we established that the force and tides on a
$512^2$ two-dimensional grid is representative and hence sufficient for our discussion. Figure~\ref{fig:nbody} shows the
resulting total force field. Figure~\ref{fig:denMMF} shows the NEXUS+ morphological identifications in the same slice. The volume and mass fractions for the different components of the cosmic web are given in Table~\ref{tab:fractions}.

\subsection{Gravitational Force field}\label{sec:force_field}
The peculiar gravitational force field for a continuous matter distribution field, specified in terms of the density perturbations
$\delta_m(\mathbf{x})$,
\begin{equation}
\Delta_m(\mathbf{x})\,=\,\frac{\rho(\mathbf{x},t) - \rho_u(t)}{\rho_u(t)}\,,
\end{equation}
The resulting peculiar gravity field, following \cite{Peebles1980}, is given by
\begin{equation}
    \mathbf{ g} (\mathbf{ x}) = \frac{3\Omega_mH_0^2}{8\pi}\int \text{d} \mathbf{ x'} \Delta_m (\mathbf{ x'}) \frac{\mathbf{ x'}-\mathbf{ x}}{|\mathbf{ x'}-\mathbf{ x}|^3}.
\end{equation}
As we represent this field on a discrete grid, the force integral expression is converted into a sum over the mass-weighted contribution to the gravitational force by each gridpoint. Evaluating the force field at the gridpoint $i$, at location $\mathbf{x_i}$,
the corresponding grid expression for the full force field $\mathbf{g}_{i}(\mathbf{ x_i})$ is the sum over all $N$ gridpoints $j$, at
locations $\mathbf{x'}_j$,
\begin{equation}
  \mathbf{g}(\mathbf{x_i}) = \frac{3\Omega_mH_0^2}{8\pi} \sum_j^N\,\Delta_m(\mathbf{ x'}_j)\,\frac{(\mathbf{ x_i} - \mathbf{ x'}_j)}
         {|\mathbf{ x}_i -\mathbf{ x'}_j|^3}\,.
\end{equation}
The summation is done up to a distance of 150$h^{-1}$Mpc. 

The gravitational force induced by the different cosmic web morphological components is obtained by the straightforward mass-weighted summation over all gridpoints that are located within the \nexus{} identified regions, ie. the regions identified by \nexus{}
with either cluster node, filament, wall or void. The four cosmic web morphological components are assigned by index
${\rm CWM}=node,fil,wall,void$. The (peculiar) gravitational force induced by the mass residing in any of these four
morphological components is then the sum of the gravitational pull induced by the $M$ mass elements residing in the
corresponding areas,
\begin{equation}
  \mathbf{g}^{\text{CWM}}(\mathbf{x_i}) = \frac{3\Omega_mH_0^2}{8\pi} \sum_k^{\text M}\,\Delta_m(\mathbf{ x'}_k)\,\frac{(\mathbf{ x_i} - \mathbf{ x'}_k)} {|\mathbf{ x}_i -\mathbf{ x'}_k|^3}\,.
\end{equation}
where $\mathbf{ x}_i$ is the location for which the force is calculated, $\mathbf{ x}_k$ one of the locations identified as
belonging to morphological feature {\rm CWM} and M the total number of grid-points identified with that morphological feature. 

The net result of the force calculation is a representation of the force field $\mathbf{ g} (\mathbf{ x})$ at every gridpoint
by five force vectors, the full gravitational force $\mathbf{g}(\mathbf{x_i})$, and the four morphological contributions,  
$\mathbf{g}^{\text{node}}(\mathbf{x_i})$, $\mathbf{g}^{\text{fil}}(\mathbf{x_i})$, $\mathbf{g}^{\text{wall}}(\mathbf{x_i})$ and
$\mathbf{g}^{\text{void}}(\mathbf{x_i})$. Evidently, the total force $\mathbf{g}(\mathbf{x_i})$ is the sum of the four morphological contributions,
\begin{equation}
\mathbf{g}(\mathbf{x_i})\,=\,\mathbf{g}^{\text{node}}(\mathbf{x_i})\,+\,\mathbf{g}^{\text{fil}}(\mathbf{x_i})\,+\,\mathbf{g}^{\text{wall}}(\mathbf{x_i})\,+\,\mathbf{g}^{\text{void}}(\mathbf{x_i})\,.
\end{equation}

\subsubsection{Gravitational Force field: visualization}\label{sec:visual}
One of the principal aspects of the present study is to study and assess the gravity, velocity and tidal tensor fields. Before any
quantitative and statistical assessment of these fields, the most direct impression of their characteristics and spatial properties
is obtained by visual inspection.

To obtain an impression of the spatial structure and properties of the gravity and velocity fields, we invoke a few different
visualisations:
\begin{itemize}
\item{} {\it gravitational force amplitude} maps\\
  of the amplitude of the gravitational force, and of the amplitude contributions by the various
  individual cosmic web morphological components. 
\item{} {\it gravity vector field} maps\\
  This includes maps of the full gravity vector field, as well as the
  vector plots of the gravity contributions by the various individual cosmic web morphologies.\\ 
  In the various gravity vector plots, we depict the component of the gravity vector in the plotted (2-D) box slice. The vectors are
  oriented towards the direction of the gravity vector component in the box slice. The length of the gravity vectors is normalized in
  one of two different options:
\item[-] {\it total} gravitational force: \\
  vector length proportional to the amplitude of the total gravitational force
  $\mathbf{g}(\mathbf{x})$, in units of the mean gravitational force amplitude, ie. in units of the dispersion $\sigma(|\mathbf{g}|)$. 
\begin{equation}
  \mathbf{g}^{\text{CWM}}_{\text{tot}}(\mathbf{x}) = \frac{\mathbf{g}^{\text{CWM}}(\mathbf{x})}{\sigma(|\mathbf{g}|)}\,.
\end{equation}
\item[-] {\it relative} gravitational force: \\
  local relative contribution to the gravitational force by each of the cosmic web components,
  by expressing the amplitude of the gravitational force in terms of the (local) total gravitational force $\mathbf{g}(\mathbf{x})$, 
\begin{equation}
    \mathbf{g}^{\text{CWM}}_{\text{rel}}(\mathbf{x}) = \frac{\mathbf{g}^{\text{CWM}}(\mathbf{x})}{|\mathbf{g}(\mathbf{x})|} = \frac{\mathbf{g}^{\text{CWM}}_{\text{tot}}(\mathbf{x})}{|\mathbf{g}_{tot}(\mathbf{x})|}\,.
\end{equation}
In a sense, $\mathbf{g}^{\text{CWM}}_{\text{rel}}(\mathbf{x})$ adheres to the concept of the fractional contribution by the
cosmic web components CWM to the local gravitational force. 
\item{} {\it velocity  streamline} map. The velocity streamline field provides a transparent view of the spatial structure of the velocity field, and allows  the identification of regions dominated by divergence, ie. inflow or outflow, and those by
  shear flow. To this end, the gravity field is transformed into an \textit{equivalent} linear velocity field \citep{Peebles1980},
  \begin{equation}
    \mathbf{v}_{\text{lin}} = \frac{2f}{3H_0\Omega_m}\mathbf{ g}\,,
    \label{eq:linv}
   \end{equation}
  in which the linear growth rate
  \begin{equation}
    f(\Omega_m)\,=\,\frac{a}{D} \frac{dD}{da}\,\approx\,\Omega_m^{\gamma}\,,\\
    \gamma=0.55+0.05(1+w)\,,
  \end{equation}
  Streamlines are everywhere tangent to the velocity vectors in the flow field, and represent the direction of velocity at each point in the flow. On the basis of the linear velocity field, the streamline map is inferred using the \texttt{matplotlib} \citep{matplotlib} function \texttt{pyplot.streamplot}.
\end{itemize}

\subsection{Tidal field}\label{sec:tidalfield}
To calculate the tidal field at $\mathbf{r}=(x_1,x_2,x_3)$ we use the traceless tidal tensor defined as \citep{WeygaertBond2005}
\begin{equation}
   \begin{aligned}
   T_{ij}(\mathbf{ r}) = \frac{3\Omega_mH_0^2}{8\pi} \int \text{d}\mathbf{ r'} \Delta_m(\mathbf{ r'}) \frac{3(x_i' - x_i)(x_j' - x_j) - |\mathbf{ r}' - \mathbf{ r}|^2 \delta_{ij}}{|\mathbf{ r} - \mathbf{ r'}|^5} 
   \\ - \frac{1}{2}\Omega H^2 \Delta_m(\mathbf{ r})\delta_{ij}\,,
\end{aligned} 
\end{equation}
In order to calculate this numerically we turn the integral into a discrete sum over the grid with the DTFE density field values. In our evaluations, we smooth the density field with a Gaussian on a scale of 5~$h^{-1}$Mpc. This is done to smooth out the regions in which nonlinear behaviour would be dominant. Representing this field on a discrete grid, the tidal force integral expression is converted into a sum over the mass-weighted contribution to the tidal force by each gridpoint. Evaluating the tidal field at the gridpoint $(\mathbf{ r})$, the corresponding grid expression for the full tidal field ${\tilde T}_{ij}(\mathbf{ r})$ is the sum over all $N$ gridpoints $(\mathbf{r}')$ 
\begin{equation}
    {\tilde T}_{ij}(\mathbf{ r}) = \frac{3\Omega_m H_0^2}{8\pi} \sum_{k}^{N} \Delta_m(\mathbf{r}') \frac{3(x_i' - x_i)(x_j' - x_j) - |\mathbf{ r}' - \mathbf{ r}|^2 \delta_{ij}} {|\mathbf{ r} - \mathbf{r'}|^5}\,.
\end{equation}
In order to separate the tidal field contribution by the different components CWM - (node,filament, wall or void) - of the cosmic web we sum only over the $M$ gridpoints that are located within a region identified as of morphology CWM. Only the mass allocated at those gridpoints ia taken along in the tidal force calculation. 
\begin{equation}
    {\tilde T}_{ij}^{\text{CWM}}(\mathbf{ r}) = \frac{3\Omega_m H_0^2}{8\pi} \sum_{k}^{M} \Delta_m(\mathbf{r}') \frac{3(x_i' - x_i)(x_j' - x_j) - |\mathbf{ r}' - \mathbf{ r}|^2 \delta_{ij}} {|\mathbf{ r} - \mathbf{r'}|^5}\,.
\end{equation}
The sum uses all cells up to a radius of 150h$^{-1}$Mpc. The traceless tidal tensor $T_{ij}$ follows from 
\begin{equation}
    T_{ij} = {\tilde T}_{ij} - \frac{1}{3}({\tilde T}_{11}+{\tilde T}_{22}+{\tilde T}_{32})\delta_{ij}\,. 
\end{equation}
The eigenvalues of tensor $T_{ij}$ are
\begin{align}
    T_1 &> T_2 > T_3 \nonumber \\
    |T| &= \sqrt{T_1^2 + T_2^2 + T_3^2} \label{eq:tidal_tot}
\end{align} 
where $T_1$, $T_2$ and $T_3$ are the sorted eigenvalues of the $T_{ij}$ matrix and |T| signifies the strength of the tidal field. The accompanying eigenvectors $\hat T_1, \hat T_2, \hat T_3$ are sorted accordingly. The value of the different eigenvalues signify the level of elongation or contraction that the corresponding mass element undergoes due to tidal forces. 

\subsubsection{Tidal field: normalization \& visualization}\label{sec:visualtidal}
In the analysis of the tidal field, we use two different normalization options:
\begin{itemize}
\item[-] {\it total} tidal force: \\
  the tidal force induced by cosmic web component ${\text{CWM}}$, $T_{ij}^{\text{CWM}}(\mathbf{x})$, in units of the mean tidal force amplitude, ie.
  the dispersion $\sigma(|T|)$
\begin{equation}
  T_{ij,\text{tot}}^{\text{CWM}}(\mathbf{x}) = \frac{T_{ij}^{\text{CWM}}(\mathbf{x})}{\sigma(|T|)}\,.
\end{equation}
\item[-] {\it relative} tidal force: \\
  local relative contribution to the tidal force by each of the cosmic components, 
  by expressing the tidal tensor components and amplitude in terms of the local total
  gravitational force $|T|_{\text{tot}}(\mathbf{x})$, 
\begin{equation}
    T_{ij,\text{rel}}^{\text{CWM}}(\mathbf{x}) = \frac{T_{ij}^{\text{CWM}}(\mathbf{x})}{|T|(\mathbf{x})|}\,.
\end{equation}
In a sense, $T_{ij,\text{rel}}^{\text{CWM}}(\mathbf{x})$ adheres to the concept of the fractional contribution by the
cosmic web components ${\text{CWM}}$ to the local tidal field. 
\end{itemize}

\bigskip
To visualize orientation, coherence and strength of the tidal force field, we use two different kinds of maps:
\begin{itemize}
\item{} {\it tidal force amplitude} maps\\
  maps of the amplitude of the tidal field and tidal field contributions by the various cosmic web morphological components.
  For the amplitude of the tidal field we use $|T|$ as defined in Eq.~\ref{eq:tidal_tot}.
\item{} {\it tidal bar} maps\\
  Tidal bars indicate the direction of the principal orientation of the tidal force field, in combination with their strength. They
  have the orientation of the eigenvectors of the tidal force, with a length proportional to the absolute value of the corresponding eigenvector,
\begin{equation}
    T_{i,\text{plot}} = \hat{T}_{i} |T_{i}|.
\end{equation}
In the tidal bar maps in the present study we depict the compressional component of the tidal force.
\end{itemize}
From these maps we may hence deduce the directions along which the mass elements get compressed.

\begin{figure*}
    \centering
    \includegraphics[width=\textwidth]{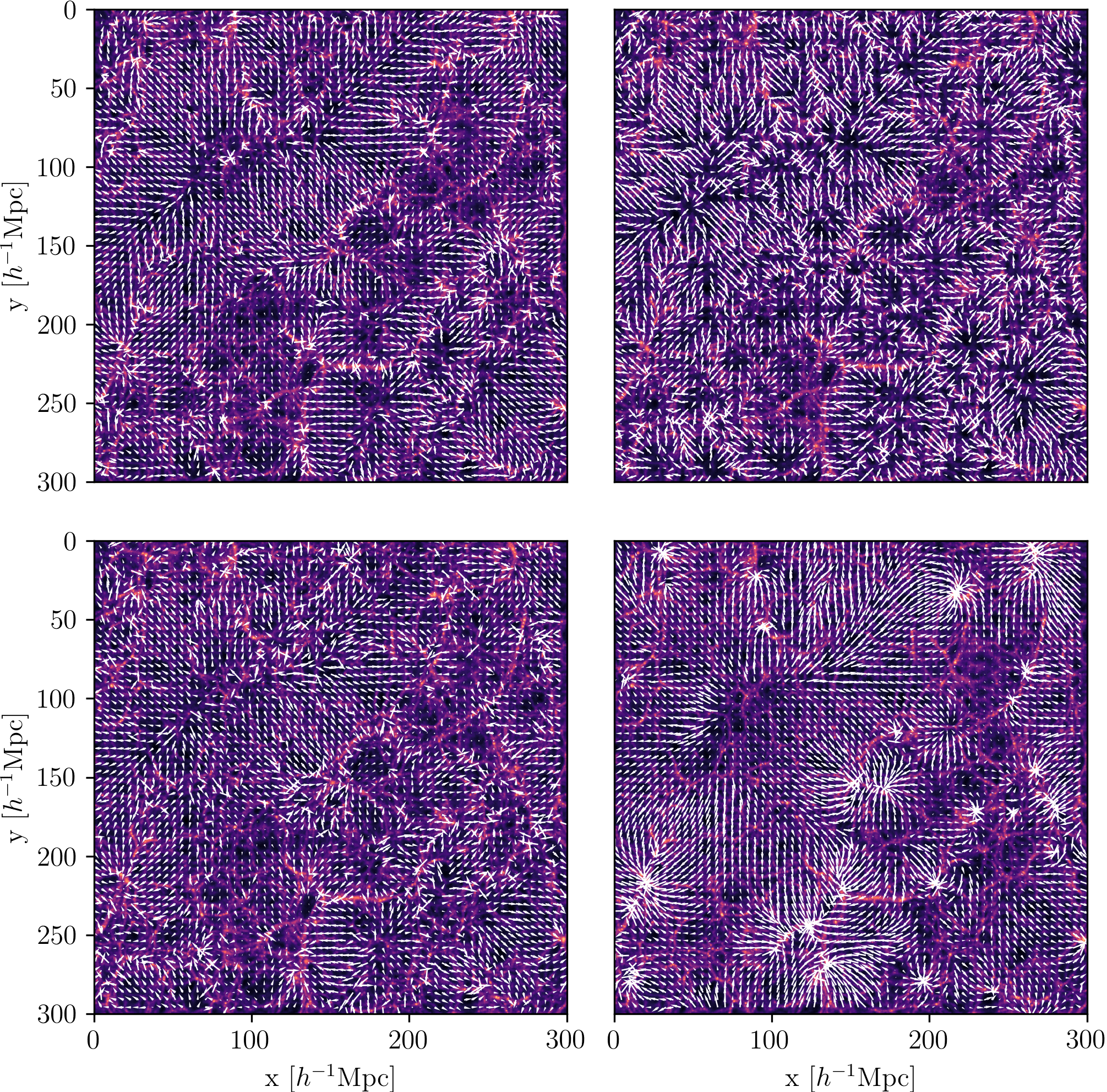}
    \caption{Gravitational Force field by Cosmic Web component: vector field.  The force fields originating of the different components for the complete field, normalised by the total field (see Section~\ref{sec:visual}). Top left has filaments, Top right has voids, bottom left has walls and bottom right has nodes. \textbf{Note:} For this figure, filaments are rescaled with respect to the other components. In comparison the filament arrows should be longer. Voids, walls and nodes are one the same scale.}
    \label{fig:totarrs}
\end{figure*}
\begin{figure*}
    \centering
    \includegraphics[width=\textwidth]{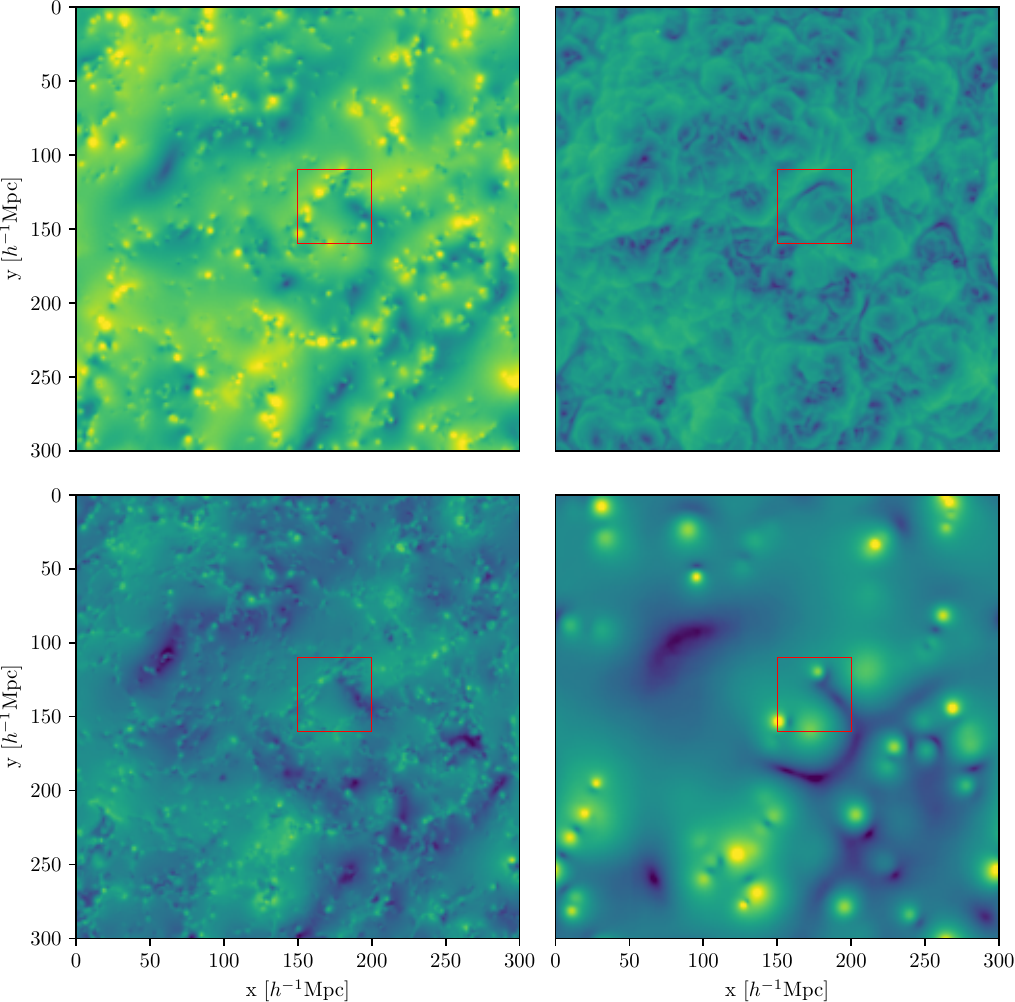}
    \caption{Gravitational Force field by Cosmic Web component: amplitude field.
      The amplitude of the peculiar force field $\log_{10}|\mathbf{ g}|$ for the different components of the cosmic web for a $300~h^{-1}\text{Mpc}$ by $300~h^{-1}\text{Mpc}$ by $0.59~h^{-1}\text{Mpc}$ slice. The top left panel shows the component originating from filaments, the top right panel shows the component originating from voids, the bottom left panel shows the component originating from walls and the bottom right panel shows the component originating from nodes. The red box indicates the region used for the zoom in Figure~\ref{fig:voidarrs}.}
    \label{fig:totg}
\end{figure*}

\section{Gravitational Force Field}\label{sec:totforce}
The gravity field contributions by the various cosmic web components, for the specific case of the slice in figure~\ref{fig:nbody}, are shown in figure~\ref{fig:totarrs} and figure~\ref{fig:totg}. The four panels in figure~\ref{fig:totarrs} show the gravity vectors for the combined force impact by filaments (top lefthand frame), voids (top righthand frame), walls (bottom lefthand frame) and cluster nodes (bottom righthand panel). We should note that the vector arrows for the filament force field are rescaled, as they are by far the most substantial contribution to the gravity force field (which should perhaps not surprise us, given filaments contain more than 50\% of the cosmic matter content, see \cite{NEXUS_CW_EVO}). With the gravity vector plot providing insight into the structural patterns of the induced gravitational force fields, maps of the strength of the gravitational force yield the needed information on the relative strength of the cosmic web morphological components and the spatial extent of their influence. Figure~\ref{fig:totg} contains the four corresponding panels with force amplitude level maps for the filament population (top lefthand), void population (top righthand), wall population (bottom lefthand) and cluster nodes (bottom righthand). 

One immediate observation is the significant differences between the gravitational influence of the different morphological components of the cosmic web. Each contributes uniquely to the total force field, yielding force fields that differ substantially in shape, coherence, scale and amplitude. Most outstanding is the dominant gravitational influence of the filamentary network. Their overarching strength can be directly inferred from the gravity amplitude maps in figure~\ref{fig:totg}. The towering influence of filaments is expressed in the fact that throughout the simulation box the filament force is factors higher than that exerted by voids and walls. Only cluster nodes reach comparable strengths, but this is only restricted to the immediate vicinity of the cluster nodes. Perhaps the most interesting finding is that of the significant and coherent gravitational impact of voids, marked by
a typical signature in terms of characteristic bubble shaped regions in and around minima in the matter distribution, the
manifestation of a typical - effectively repulsive - force field pattern. Moreover, comparison between the cluster node and
void force field reveals the perhaps surprising conclusion that voids are the more dominant component over far larger regions
of a cosmic volume. 

\subsection{Filaments}
From the filament force vector field map (top lefthand panel, fig.~\ref{fig:totarrs}), we see that the gravitational
influence of filaments acts coherently over a large range of distances. In fact, its overall structure and pattern
is quite close to that of the total gravitational force field (bottom panel, fig.~\ref{fig:nbody}, a telling
testimony of the fact that most of the large scale gravitational force field is due to the combined influence
of the filamentary network. We should also note that, as expected, the force field only traces the largest structures.
Small scale features hardly induce a noticeable effect. The overpowering dynamical influence of the filamentary
network should not come as a surprise: the filamentary network of the cosmic web represents more than 50\% of the
cosmic matter content of the universe \citep{NEXUS_CW_EVO}. As important is the pervasive spatial character of the multiscale
filamentary network, with filaments spreading throughout cosmic volumes and their tendrils branching out into even the
most remote and desolate realms (voids).

Inspection of the force amplitude map shows that over nearly the entire cosmic volume filaments are responsible for
high force amplitudes, their overdense nature guaranteeing a consistent and coherent cumulative attractive force
contribution (top lefthand panel, fig.~\ref{fig:totg}). We find that over almost $97\%$ of the cosmic volume, the filament
induced forces dominate over
that induced by any of the other morphological components (table~\ref{tab:results}, 3rd column). 
The largest and densest filaments create coherent force fields on scales up to $\sim50~h^{-1}$Mpc. In
fact, the coherence scale of the force field is much larger than that of the weblike structure of the underlying
matter distribution. 

\subsection{Voids}
The picture is considerably different for the void force field. The top righthand panel of figure~\ref{fig:totarrs}
shows a spatial pattern marked by individual divergent patches, indicating individual expanding void regions. Inside
these individual void regions, the force field increases monotonically - and usually almost linearly - from the center of
a void to their boundary of surrounding walls and filaments. It produced the characteristic superhubble expanding
void regions \citep[see e.g.][]{icke1984,weykamp1993}. Due to the rapidly rising mass density near their boundary,
the strength of the void force field drops at the void edge, resulting in the conspicuous bubble-like configurations
that stand out in the void force field in figure~\ref{fig:totarrs}.
 
Overall the void force field appears to be one in which individual superhubble expanding voids \citep{icke1984} produce a
segmented volume with clearly distinguishable individual void regions. Indicative for the latter are the small
voids near the center, whose influence appears to be mainly limited to their own interior. Nonetheless, we may
observe assemblies of voids that seem to operate collectively in pushing out the surrounding mass, such as the
the void agglomeration at the top left of the field. Its force fields adds up to create an expanding region on a scale
of $\sim50~h^{-1}$Mpc.

In all, it suggests that in general voids hold sway over in particular their local environment, in which
large scale effects are significant but less prominent than those seen in the filament force field. On large scales voids
do produce significant residual large scale effects emanating due to their spatial clustering, be it of a lower amplitude than
that characteristic for the forces in individual voids. It confirms the finding by \cite{platen2008}, who showed that voids
even induce noticeable tidal effects over scales in excess of $\sim30~h^{-1}$Mpc).

\subsection{Walls}
The force field induced by the walls is the least conspicuous one of the web morphologies. Throughout the cosmic
volume it is very weak, as testified by the map of its force strength (see bottom lefthand panel fig.~\ref{fig:totg}.
The walls do not contribute much to the overall force field. Also, as they populate moderate density realms as well as
the interior of voids \citep[see e.g.][]{NEXUS_CW_EVO}, their effect may be effectively either be attractive or repulsive. To some extent, it leads to the moderate and low density walls compensating each others force contributions. It results in a force field
that only faintly reflects the cosmic web pattern seen in the filament and void force fields.  

Globally, the spatial structure of the wall force map bears resemblance to that of the filament strength map
(top lefthand panel fig.~\ref{fig:totg}. It appears to be a faint reflection of the filament strength map.  
One may recognise only a part of the most prominent features that are seen in the filament force field, others are
conspicuously absent: e.g. the strong filament force field at the top lefthand corner of the box does not seem to have a wall
equivalent. Also the overall large scale pattern of the wall force vector map does resemble that of the filament
force vector map. However, the wall force map appears far less coherent and organised, marked by a considerable
number of randomly scattered small scale patches of irregular oriented force vectors. 

\subsection{Cluster Nodes}
A rather surprising finding is the fact that cluster nodes only appear as highly localised force centers, and they are far from the dominant gravitational component of the cosmic web that we had expected them to be. 

In the force strength map (bottom righthand panel, fig.~\ref{fig:totg}) cluster nodes stand out
as compact high amplitude peaks in the force field. They act as a set of randomly clustered monopole attractors. 
The cluster nodes are only dominant over the rather small scales of their immediate environment. At this range,
they overshadow all other force contributions. In most cases, this is only over distances of a few $h^{-1}$Mpc,
in an occasional exception out to $\sim20h^{-1}$Mpc. The latter concerns an agglomerate, ``superclusters", such as
that at the bottom centre of the map (bottom righthand panel, fig.~\ref{fig:totg}). 

Their spatial connection to the cosmic web is less clear, and it is not easy to recognize the global weblike mass
distribution in their own spatial arrangement or the direct dynamical relation between the cluster nodes and the
spatial intricacies of the cosmic web. At large scales there are substantial regions were the influence of multiple
nodes is one in which they effectively cancel each other, resulting in a contribution that is less than - or at best
comparable to - that by the voids. 

\subsection{Gravity Field strength: statistical analysis}
To quantify the visual impressions of the force field contributions by the different cosmic web components, discussed extensively above, we assess the statistical (volume-weighted) distribution of the fractional gravitational force contributions by each of the
cosmic web components $CWM$. The fractional gravitational force ${\cal F}({\mathbf r})$ at each location ${\mathbf r}$ is
defined by the ratio,
\begin{equation}
    {\cal F}^{\text{CWM}}({\mathbf r}) = \frac{|\mathbf{g}^{\text{CWM}}(\mathbf{r})|}{|\mathbf{g}_{\text{tot}}(\mathbf{r})|},
    \label{eq:ratiomap}
\end{equation}
where $\mathbf{g}^{\text{CWM}}({\mathbf r})$ is the gravitational force exerted by the components CWM - i.e. filaments, voids,
walls and cluster nodes - at location ${\mathbf r}$. The total sum of these is 
\begin{equation}
  \mathbf{ g}_{\text{tot}}(\mathbf{r})=\sum_{\text{CWM}}^{} \mathbf{g}^{\text{CWM}}(\mathbf{r})\,. 
\end{equation}
Note that the ratio is independent of direction, while the sum of the forces does take into account the orientation of
each force contribution. As a result, the fractional force ${\cal F}({\mathbf r})$ may have a value larger than unity.
The average ratios for the fractional force contributions are listed in table~\ref{tab:results}. There is a clear
hierarchy of force field contributions: filaments $\gg$ voids $>$ nodes $>$ walls. Averaged over all volume elements,
the filament impact is on average no less than $60\%$, with voids as a surprising second at on average contributing
$\sim 20\%$ of the local gravitational force. 

\begin{table}
\centering
\caption{Gravitational Force Statistics.
  The second column shows the mean force ratio for the different components of the cosmic web, Note the direction dependence in the ratio $|\mathbf{ g}|^{\text{CWM}}/|\mathbf{ g}|^{\text{tot}}$. The third column shows the percentage of the total volume where each component has a contribution that is larger than the other components.}
\begin{tabular}{r l l}
 \hline
    Component & Mean [\%] & Largest in [\%]\\ \hline
    Filaments & 60.4 & 97.2\\
    Voids & 20.3 & 0.9\\ 
    Nodes & 17.0 & 1.5 \\ 
    Walls & 13.7  & 0.4\\
    
\hline    
\end{tabular}
    \label{tab:results}
\end{table}

\begin{figure*}
    \centering
    \includegraphics[width=0.7\textwidth]{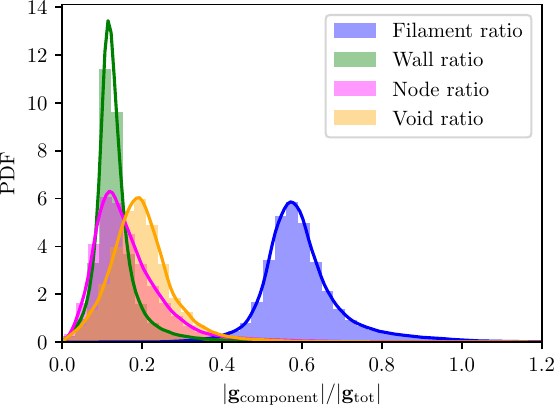}
    \vskip 0.5cm
    \includegraphics[width=0.7\textwidth]{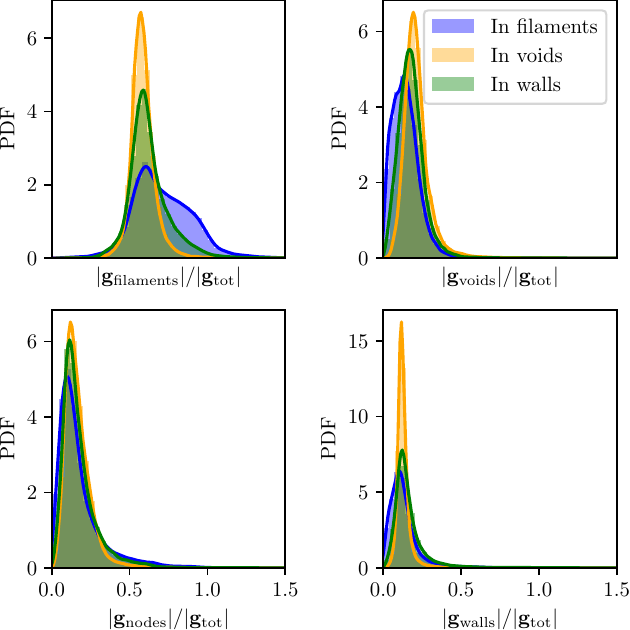}
    \caption{Cosmic Web Gravitational Force Field: Strength Inventory. Top panel: The distribution of $|\mathbf{ g}^{\text{CWM}}|/|\mathbf{g}_{\text{tot}}|$ for the different components of the cosmic web. Each using the complete field. Bottom four panels: the distribution of $|\mathbf{g}^{\text{CWM}}|/|\mathbf{g}_{\text{tot}}|$ for the different components of the cosmic web. Per panel the
      pdf for cosmic web component CWM is plotted, within three different environments: filaments (blue), voids (yellow), walls (green). Centre left: filament induced force field. Centre right: void induced force field. Bottom left: wall induced force field. Bottom right: cluster induced force field.}
    \label{fig:totratio}
\end{figure*}
\begin{figure*}
    \includegraphics[width=0.7\textwidth]{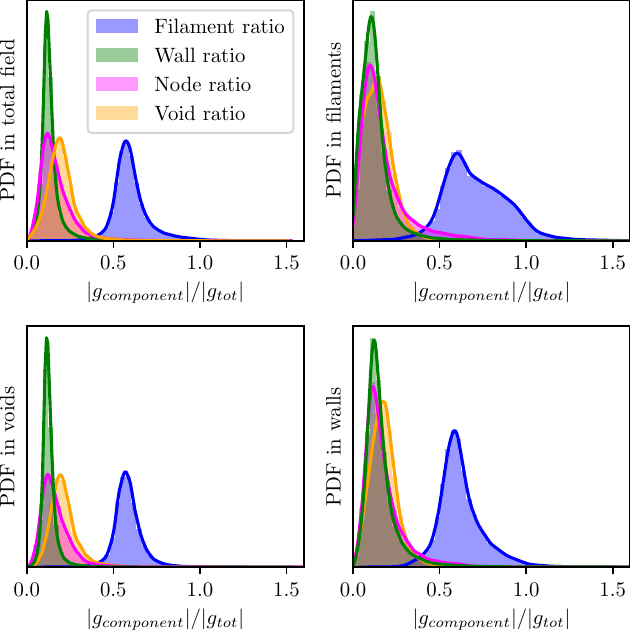}
    \caption{Cosmic Web Gravitational Force Field: Strength Inventory by Cosmic Web environment. The panels show the pdf of the
      induced gravitational force field $|\mathbf{ g}^{\text{CWM}}|/|\mathbf{g}_{\text{tot}}|$ by each of the four cosmic web
      components CWM in the different cosmic web environments. The different colours represent the different components:
      filament induced force field (blue), void induced force field (yellow), wall induced force field (green) and cluster node
      induced force field (magenta). Top left panel: total gravitational field. Top right panel: gravitational field in filaments.
      Bottom left panel: gravitational field in voids. Bottom right panel: gravitational field in walls.}
    \label{fig:totratio2}
  \end{figure*}

\bigskip
The distribution for the complete force field is shown in the top panel of Figure~\ref{fig:totratio}. The bottom four
panels show the same distribution, but split up by morphology. They show for each of the four cosmic web environments,
the fractional force contribution by each of the morphologies. In other words, these tell you what the relative impact
is of filaments in voids, of voids in filaments, or of cluster nodes in filaments. 

The top panel of Figure \ref{fig:totratio} reveals the major differences in force impact by the different cosmic web components. 
Filaments stand out as by far the most dominant element of the cosmic web. Their force impact ranges over a wide range of values
${\cal F}^{fil}({\mathbf r})$, centering around $60\%$ of the total gravitational force. At the vast majority of locations the
filamentary cosmic web is the leading gravitational influence.

The role of voids is one of the most interesting aspects of the gravitational force inventory. They appear to
be the second most dominant source of gravitational force. They account for more than $20\%$ of the local force at far more
locations than e.g. walls, and even than cluster nodes. At many locations voids assume an even stronger fraction of the force budget. 
The fact that the void distribution function has a long high end tail implies that voids are the dominant
influence at various regions of the cosmic web. Nearly without exception this concerns the interior of large voids. A similar long
high end tail is found for the cluster node gravitational influence ${\cal F}^{node}({\mathbf r})$, reflecting the overpowering
gravitational influence of nodes in and around their own location. We find that the cluster node
distribution has a mode near $10-12\%$ of the total force, substantially less than that of voids. It means that over most
of the cosmic volume, voids have a stronger gravitational influence than cluster nodes !

As expected, the distribution of the wall force fraction ${\cal F}^{walls}({\mathbf r})$ expresses their weak role in the
cosmic web force budget. Not only does the distribution peak around $10\%$, quite similar to the mode for cluster nodes, but
it also concerns a narrow distribution without an outstanding tail towards higher values. It shows that walls, over nearly
the entire cosmic volume, are only a minor fiddle in the force concert.

\bigskip
Assessing the force impact differentiated by cosmic web environment provides additional information on the nature of the 
dynamical impact of filaments, voids, walls and nodes. These are provided by the bottom panels of
figure~\ref{fig:totratio} and by the panels of figure~\ref{fig:totratio2}. 

Differentiating between the impact of morphological components in different environments, the four bottom panels of
figure~\ref{fig:totratio} reveal a few interesting aspects. Filaments are clearly the dominant
dynamical component over the entire cosmic volume, and this is true in voids, walls as well as in filaments
themselves (voids: bottom lefthand panel fig.~\ref{fig:totratio2}, walls: bottom righthand panel fig.~\ref{fig:totratio2},
filaments: top righthand panel fig.~\ref{fig:totratio2}). In the case of the filament interior, the shoulder of  
the filament pdf reveals that it is in particular inside filaments themselves
that they display the strongest impact: filaments hold sway inside filaments. A minor detail is that we
see that ${\cal F}^{fil}({\mathbf r})$ is higher in walls than in voids. This surely is a manifestation
of the spatial proximity of filaments to these structures: voids are large, and in the central interior
the filament's force may be somewhat weaker than in and around the walls to which they are connected.

For the impact of voids in different cosmic web environments, we see that they hold their strongest influence 
over the void regions themselves, while they are relatively stronger in walls than in the interior
of filaments (centre righthand panel fig.~\ref{fig:totratio}). Inside void interiors, the force inventory  
quantified in the bottom lefthand panel of fig.~\ref{fig:totratio2} shows how important
external influences are in the dynamics and evolution of voids. The filament forces dominate the force field
of voids. This is certainly true for the outer regions of voids, but may even be so for their inner regions.
The latter will be most evident for small voids, where these external forces may even induce their collapse,
an essential process - called void-in-cloud - in the buildup of the void population \citep{shethwey2004}.
However, the filaments represents the principal gravitational influence for even the large voids. This
implies that any analysis of void dynamics should include the mass distribution surrounding the void
\citep[see e.g][for a recent review]{weygaert2016}, while any exploitation of the characteristics of the
void population for cosmological purposes cannot ignore this fact and base their analysis on simplistic 
isolated void dynamics \citep[see e.g.][]{pisani2019}. A recent study of the dynamics of a sample of
voids from SDSS, including velocity flow information from the Cosmicflows-4 galaxy peculiar velocity
survey, did also indicate this finding in the observational reality \citep{courtois2023}.

\begin{figure*}
    \centering
    \includegraphics[width=\textwidth]{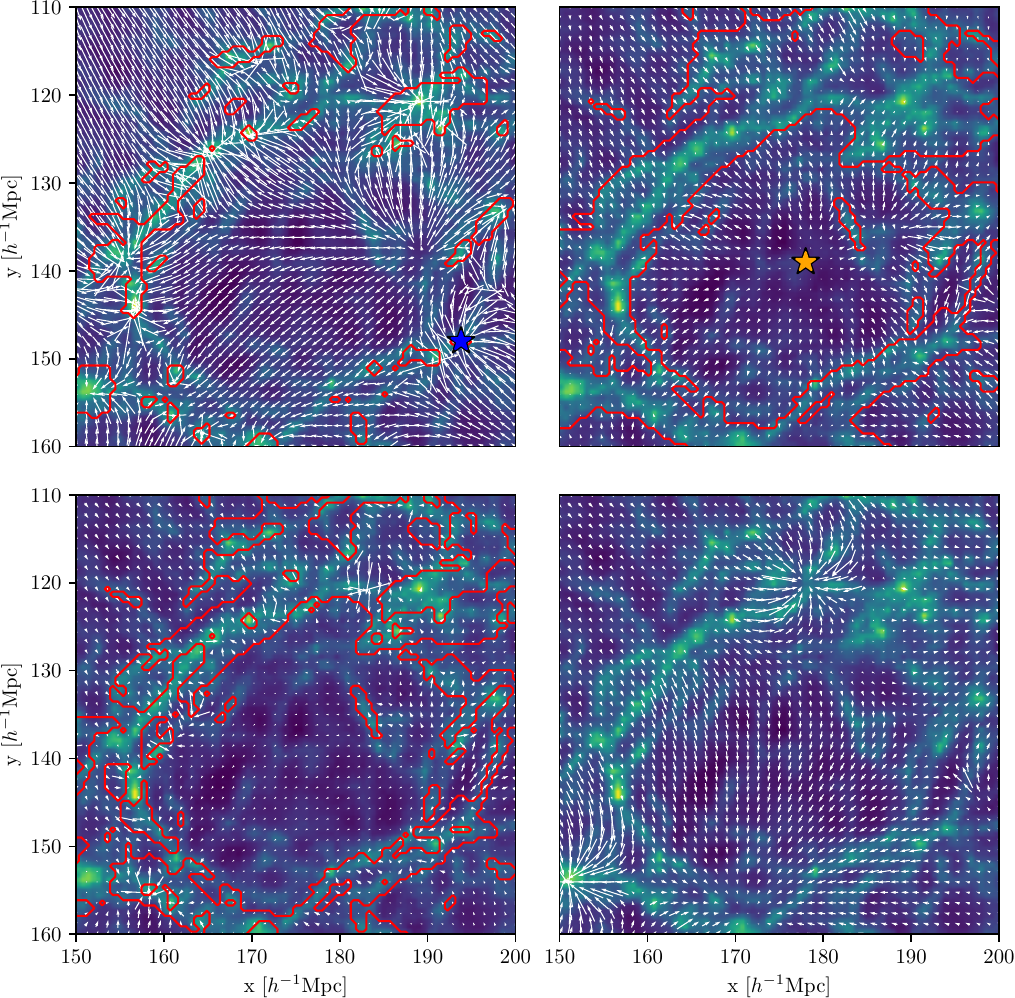}
    \caption{Gravitational force vector field in and around void region. The region is a $50~h^{-2}$Mpc$^2$ region centered
      on a void (void center: orange star, top lefthand panel). The panels represent the force vector field in this region
      generated by the different morphological components of the cosmic web. Top left panel: filament induced force field;
      top right panel: void induced force field; bottom left panel: wall induced panel; borrom right panel: cluster node
      induced field. The arrows show the magnitude and direction of the corresponding (normalized) gravitational force, $\mathbf{ g}^{\text{CWM}}_{tot}$. The vector field is superimposed on the (log) density field map. In the different
    panels the red contours indicate the regions belonging to the corresponding NEXUS+ identified cosmic web component.
    The orange star (top righthand panel) indicates the void center, the blue star (top left panel) the filament location,
    around which the radial force profiles in figures~\ref{fig:voidratvoid} and \ref{fig:voidratfila} are determined.}
    \label{fig:voidarrs}
\end{figure*}

Meanwhile, less outstanding are the observations with respect to the gravitational influence of nodes and walls. Their
impact appears relatively indifferent to the environment. The node force fraction decreases slightly going from
voids to walls to filaments. This might be somewhat counter intuitive, but is an expression of the fact that
filaments are so much stronger than walls and voids. Relatively speaking, nodes will then assume a smaller
force fraction ${\cal F}^{node}({\mathbf r})$ in a filament environment. 

\section{Gravitational Force Field:\\ \ \ \ \ individual structures}\label{sec:zooms}
In this section we zoom in on the force structure in and around a few representative individual features. This with the intention 
to develop an intuitive and visual image of the role and gravitational influence of the different cosmic web components with
respect to the various large scale environments, and to appreciate the complex interplay between them. We concentrate on the force field
in and around a single void, around a filament, and one in and around a cluster node. In each case we zoom in on $50\times50 h^{-1}$ Mpc
segment. They confirm the observations and conclusions that we reached in the previous section on the basis of its structural and statistical
assessment analysis of the cosmic web gravitational force field.

\subsection{Case study 1:\ \ \ \ \ void force field}
Our first study case is the force field in and around a void. Against the backdrop of the corresponding density field,
with the red contours indicating the NEXUS+ identified region, the four panels in figure~\ref{fig:voidarrs} show 
the decomposition of the gravitational force field in and around the void into its individual morphological contributions. 
The force field is decomposed into the filament induced force field (top lefthand panel), that by voids themselves (top righthand panel)
by walls (bottom lefthand panel) and by cluster nodes (bottom righthand panel).

Evidently, the filament induced force field dominates over nearly the entire region. Two of the most characteristic aspects of the
filament induced force field is its large scale reach and coherence, and the easily recognisable signature shear pattern in the
gravity vector field. The latter involves the splitting of the gravity vector field in opposite directions at
saddle points of the corresponding gravitational potential. Two of the most outstanding ones are the one near righthand edge of
the void (at $(x,y)\approx(190,140)h^{-1}$Mpc, and the one near $(x,y)\approx(170,120)h^{-1}$. The latter is situated in between
two dense concentrations within the filamentary network. These are responsible for stronger attractive cores along the filaments.
The two dense mass concentrations along the filament turn out to be massive filamentary branches connecting
to a cluster node, forming the connections of the cluster with the filamentary skeleton of the cosmic web. Over this
specific void, the cluster node represents a sizeable external influence over the entire realm of the void. The
cluster induced force field is comparable in magnitude to that of the void force field itself. 

The coherence of the filament induced gravity field over the entire reach of the central void is the source for a large scale bulk flow
(eq.~\ref{eq:linv}), superimposed on the iconic superhubble divergent flow induced by the void(s) themselves. The induced force field
by the voids themselves is easily recognisable, consisting of an almost pure divergent vector field centered around
the density minimum (top righthand panel fig.~\ref{fig:voidarrs}. To first approximation, the superhubble force field in the
interior of the void involves a radially outward directed force that increases with distance to the center
\citep[see eg.][]{icke1984,weygaert2016}. It reflects the almost uniform (underdense) mass distribution in the interior of the
void \citep[see e.g][]{weykamp1993,shethwey2004}. In reality, also voids display a substantial level of substructure. The
presence of a small expanding subvoid superimposed on the large central void reflects the multiscale structure of both
the interior void density and velocity field. It is the product of the hierarchical buildup of the void population
\citep{shethwey2004,aragon2013}.

\begin{figure}
  \centering
    \vskip 0.1cm
    \includegraphics[width=\columnwidth]{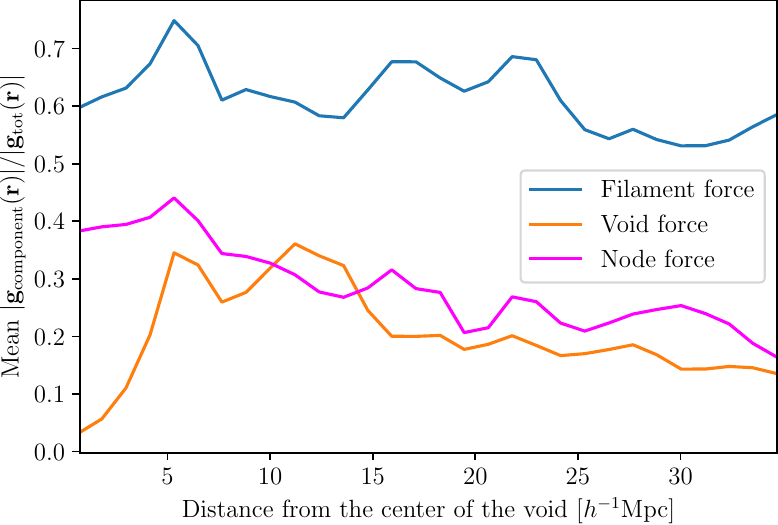}
    \vskip -0.25cm
    \caption{Radial gravitational force (amplitude) profile for the cosmic web components CWM, around the (orange) void center
      (fig.\ref{fig:voidarrs}~, top righthand panel). Plotted is the force ratio
      $|\mathbf{ g}^{\text{CWM}}|/|\mathbf{g}_{\text{tot}}|$ for the CWM induced gravitational force amplitude, as
      a function of distance from the void centre. Blue: filament force amplitude; yellow: void force amplitude;
    magenta: cluster node force amplitude.}
    \vskip 0.05cm
    \label{fig:voidratvoid}
    \centering
    \includegraphics[width=\columnwidth]{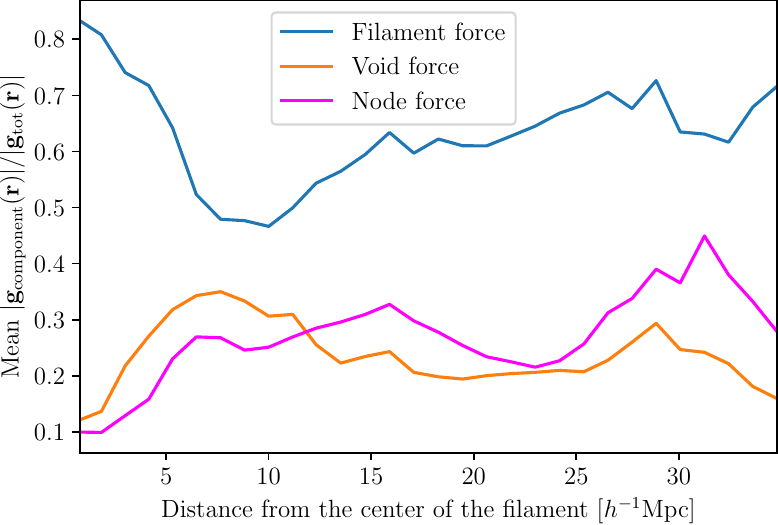}
    \vskip -0.25cm
    \caption{Radial gravitational force (amplitude) profile for the cosmic web components CWM, around the (blue) filament location
      (fig.\ref{fig:voidarrs}~, top lefthand panel). Plotted is the force ratio
      $|\mathbf{ g}^{\text{CWM}}|/|\mathbf{g}_{\text{tot}}|$ for the CWM induced gravitational force amplitude, as
      a function of distance from the central filament location. Blue: filament force amplitude; yellow: void force amplitude;
    magenta: cluster node force amplitude.}
    \vskip 0.05cm
    \label{fig:voidratfila}
    \centering
    \includegraphics[width=\columnwidth]{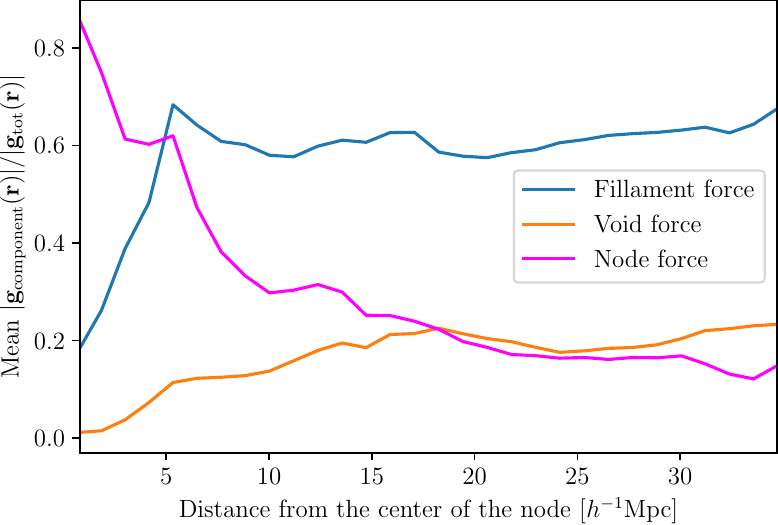}
    \vskip -0.25cm
    \caption{Radial gravitational force (amplitude) profile for the cosmic web components CWM, around the prominent massive cluster
      in fig.~\ref{fig:nodestr}. Plotted is the force ratio
      $|\mathbf{ g}^{\text{CWM}}|/|\mathbf{g}_{\text{tot}}|$ for the CWM induced gravitational force amplitude, as
      a function of distance from the cluster node location. Blue: filament force amplitude; yellow: void force amplitude;
    magenta: cluster node force amplitude. }
    \vskip 0.05cm
    \label{fig:noderatio}
\end{figure}

\begin{figure*}
    \centering
    \includegraphics[width=\textwidth]{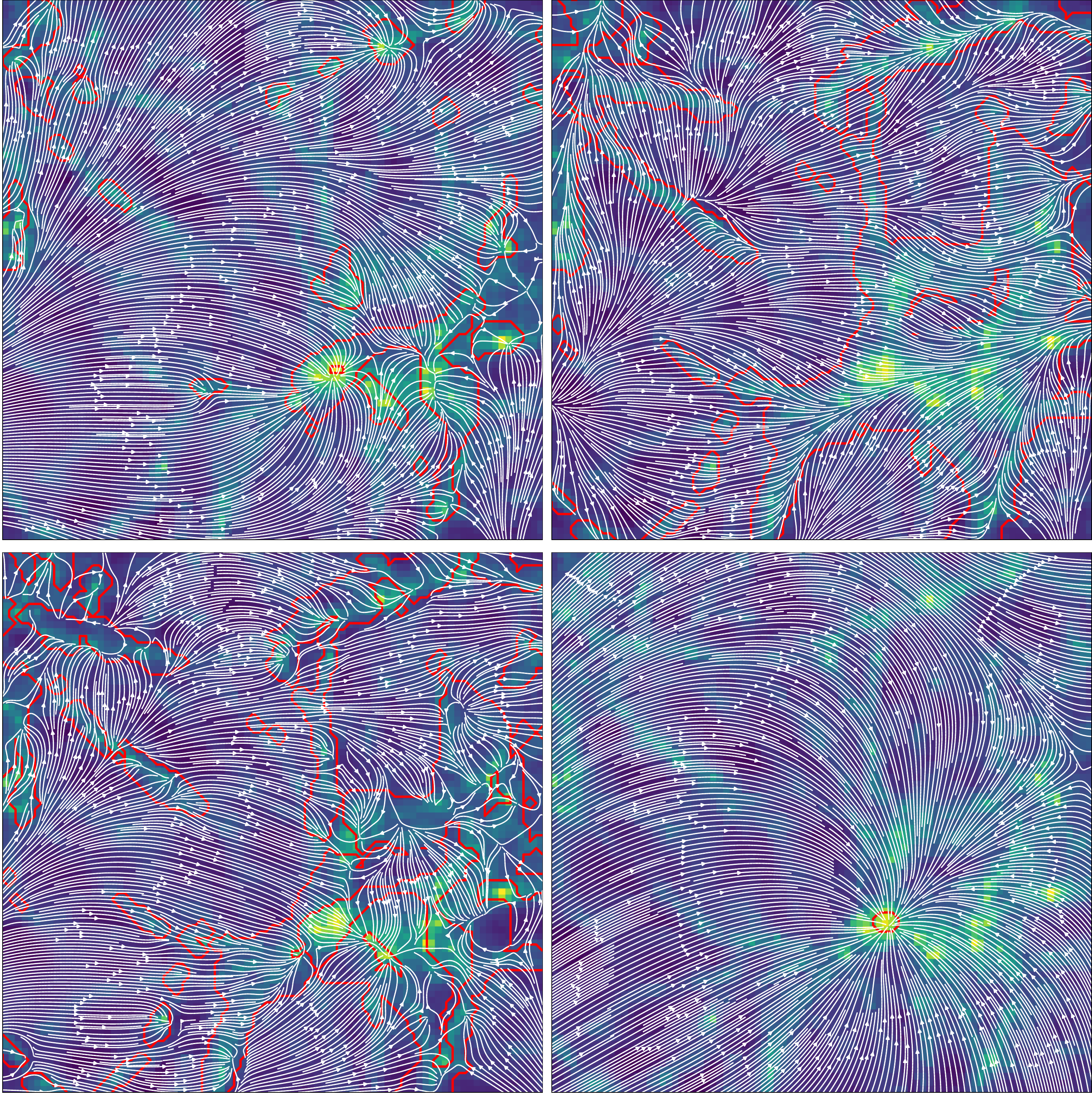}
    \caption{Velocity flow field in and around cluster node. The region is a $50~h^{-2}$Mpc$^2$ region, containing a massive
      cluster node (at bottom righthand corner). The velocity vector field is represented by the corresponding streamlines,
      highlighting the overall spatial structure of the flowfield. The four panels show the streamline field for the
      induced velocity flow field by each of the cosmic web morphological components. The velocity field streamlines are
      superimposed on the (log) density field map. In the different panels the red contours indicate the regions belonging
      to the corresponding NEXUS+ identified cosmic web component. 
      Top left panel: filament induced velocity field; top right panel: void induced velocity field;
      bottom left panel: wall induced velocity field; borrom right panel: cluster node induced velocity field. In this
      indivual case, the velocity field is clearly dominated by the cluster.}
    \label{fig:nodestr}
\end{figure*}

While the filamentary force field displays a coherence over nearly the entire volume, the void induced field appears to be
more restricted in spatial extent. The void gravitational force increases radially outward up to its boundary, at
a radius of around $r \approx 10-15 h*{-1}$Mpc. At the edge it plummets to an almost negligible influence, as the
gravity by the higher density outer realms of the voids kick in to compensate for the influence underdense interiors.
Minor residual void forces may be traced in the interstitial boundary regions, yielding the suggestion of the voids
pushing matter into the surrounding weblike filamentary and planar structures. Similar patterns are seen to emanate from
the surrounding voids. As a result, we get the impression of a void force field segmented into separate void regions. This
was also seen on the more global scale of the simulation volume in the void force field in figure~\ref{fig:totarrs} (top righthand
panel). In all, we may conclude that while voids may have some minor effect on large scale flows, their main impact seems to be
that of the local pushing around of matter into the interstitial elements of the cosmic web.

Quantitatively the above impressions are supported by the radial profile of the different force amplitudes around the center
of the void. The top panel of Figure~\ref{fig:noderatio} plots the radially averaged force amplitude as seen from the center of the
void. It confirms the dominance of the filament induced gravity (blue line), over the entire interior of the void (and beyond).
The void induced force field (orange line) reveals the expected characteristic (almost) superhubble expansion. Near the
edge of the void, the outward directed void force represents a substantial fraction of the complete gravitational force, but never
more than around 40\% of the filament force. A full account of the force field for this particular voids also includes a substantial
influence from the massive cluster node on its north side, whose magnitude is comparable to that of the void itself. The
corresponding force profile (magenta line) confirms the visual impression offered by the bottom lefthand panel of fig.~\ref{fig:voidarrs}. 

Seen from the perspective of the filament, a complementary view of the sizeable void influence is obtained. The middle panel of Figure~\ref{fig:noderatio}
shows the radial force profile around the dense filament location at the bottom right of the region (marked by a blue star,
top left panel fig.~\ref{fig:voidarrs}). At small distances from this location, still inside the filament, the filament induced
force field is overwhelmingly dominant, responsible for more than 80\% of the full force field. As we move towards the
edge of the filament, and enter into the void's realm, we observe a rapid decline in the filament's influence, going along with
a corresponding rise of the void's influence. Further afield, as the influence of overdense surroundings at the other side of the
void becomes noticeable, we see that the force contribution by the void diminishes while that of filaments remains more or less
constant. 

\subsection{Case study 2:\ \ \ \ \  cluster node force field \& flowlines}
The second case study concerns the force field around an isolated cluster node. To get a visual appreciation of its impact on
the environment, we assess the induced flowfield in and around the node. To this end, we use the linear relation between
gravity and velocity (Eq. \ref{eq:linv}). Formally, it is only valid in the linear regime, but to reasonable approximation
may also be used in the quasi-linear configurations we are investigating. It yields the implied velocity fields induced
by the four different cosmic web components, filaments, voids, walls and cluster nodes. These velocity fields are
visualized by means of the corresponding streamlines. It yields a flowline representation of the velocity field that provides
a direct insight into the structure of the cosmic migration flows. Figure~\ref{fig:nodestr} shows the streamline fields
for the induced flowfield by filaments (top lefthand panel), the voids (top righthand panel), the walls (bottom lefthand panel)
and the cluster nodes (bottom righthand panel). 

The most prominent aspect of the flowfield is the cluster node induced flowfield (bottom righthand panel). The node induced
a massive inflow pattern over the entire entire region. The dominant impact of the cluster on its environment is confirmed
by the radial force profile centered on the node (bottom panel of fig.~\ref{fig:noderatio}). In the inner sphere of $\approx 5h^{-1}$Mpc around
the cluster node, it dictates the gravitational force field, taking care of even more than $80\%$ at its center. Outside of
this immediate vicinity, its sway rapidly declines, taken over by the surrounding filamentary web. To a considerable extent, the
filament streamlines offer a similar spatial flow pattern, augmented by the presence of shear signature. It concerns the impact
of the massive filamentary branches connecting to the node, and as such strongly related to this peak in the mass distribution. 

By contrast, the void induced flow field offer a more localized pattern. In the overall pattern we recognize the typical
segmented nature of the field, carrying the imprint of individual voids. More so than we have seen in the force vector
field, the flow field also reveals the tendency for a large void scale induced bulk flow, running from the lefthand to
the righthand side of the region. It is a manifestation of the multiscale nature of the void population, a direct consequence 
of its hierarchical buildup \citep{shethwey2004,aragon2013}. Voids tend to be embedded in larger (less) underdense regions,
whose dynamical influence shows up in the superposition of the individual void outflows as a residual large scale
(bulk) outflow \citep[see][for telling illustrations]{aragon2013}. As seen from the center of the cluster node, the
void induced flow field becomes noticeable at a distance of around $\sim16h^{-1}$Mpc, beyond which distance it represents
a stronger contribution than that of the cluster itself. 

\begin{figure*}
    \centering
    \includegraphics[width=\textwidth]{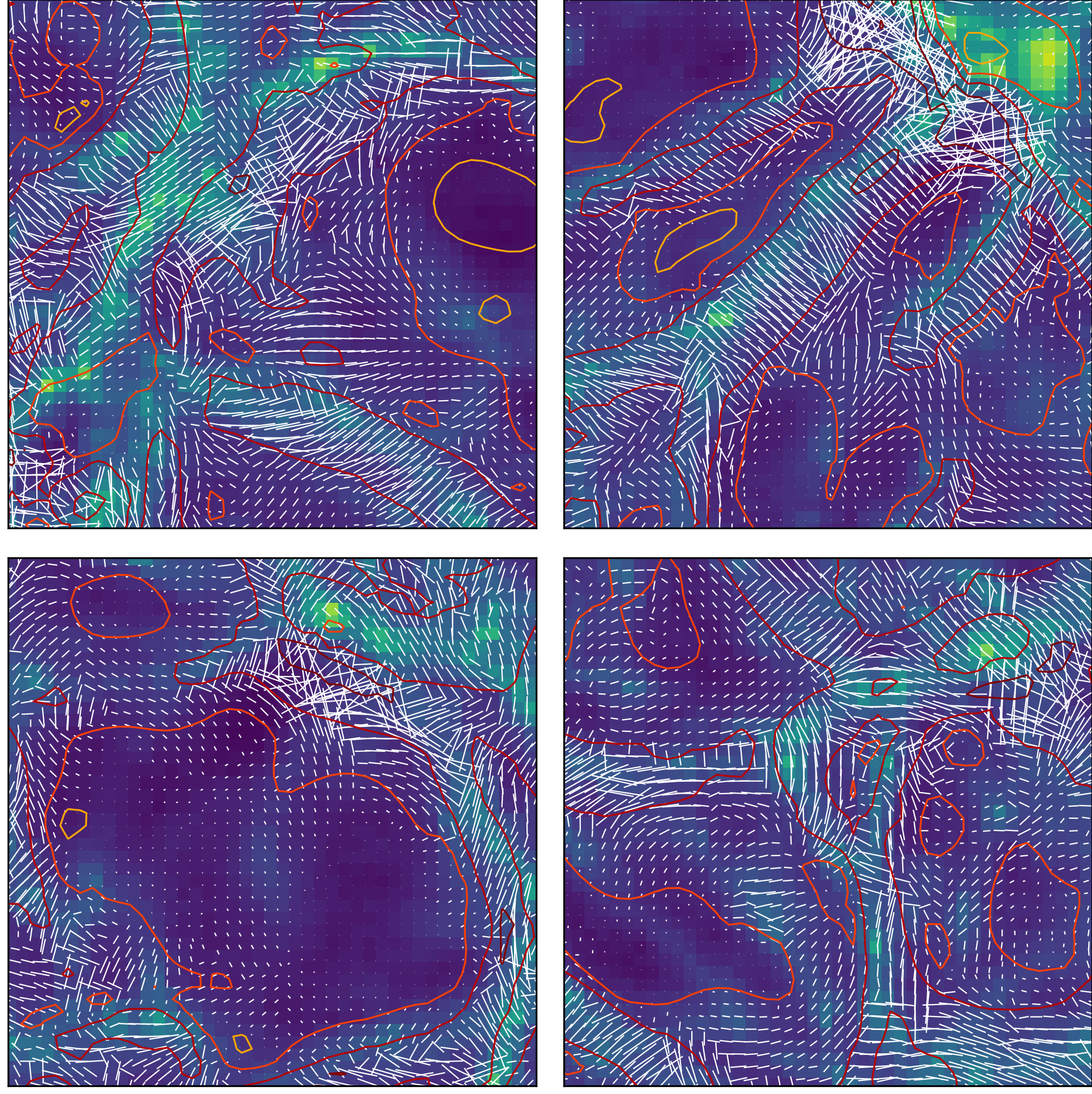}
    \caption{Tidal field induced by voids, in four different $50~h^{-2}$Mpc$^2$ regions. The tidal force field is represented
      by tidal bars and contours. All four panels show the tidal field induced \textbf{exclusively} by the void population
      in the simulation box. The tidal bars and contours are superimposed on the (log) density field map. The tidal bars indicate
      strength and direction of the compressional component of the tidal field. The definition and construction of the tidal bars is
      explained in section~\ref{sec:visual}.}
    \label{fig:zoomtid}
\end{figure*}

\bigskip
\section{Tidal field}\label{sec:Tidalfield}
To understand the dynamics underlying the formation and shaping of the cosmic web, tidal gravitational forces and the induced
deformation of mass elements play an instrumental role \citep[see e.g.][]{zeldovich1970,bond1996,weybond2008,hahn2010,Feldbrugge2023,Feldbrugge2024}. The present study therefore needs to complement the inventory of the gravitational force in the previous section by
an analysis of the corresponding tidal force field. 

\begin{figure*}
    \centering
    \includegraphics[width=\textwidth]{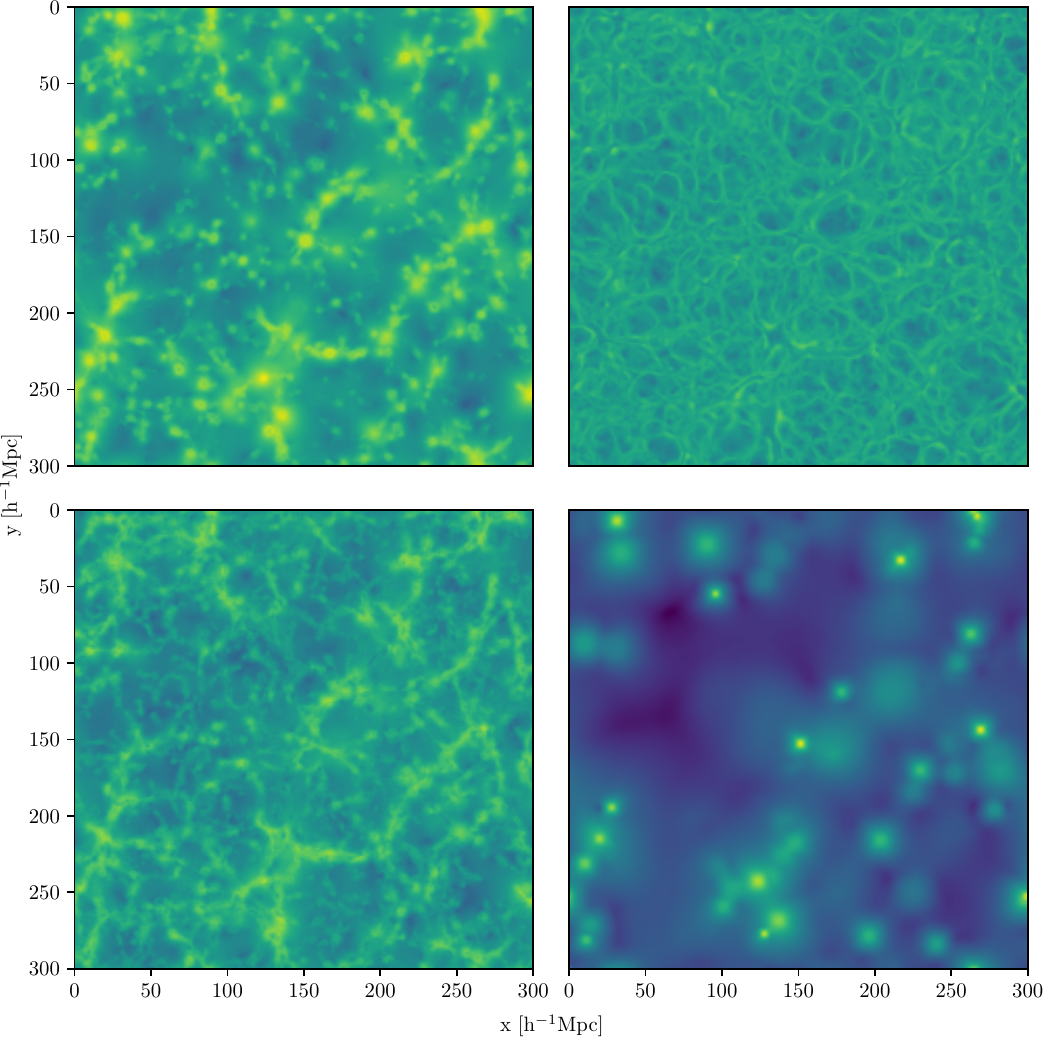}
    \caption{Tidal field by Cosmic Web component: amplitude field. 
      The amplitude of the tidal field $\log_{10}|T|$ for the different components of the cosmic web for a $300~h^{-1}\text{Mpc}$ by $300~h^{-1}\text{Mpc}$ by $0.59~h^{-1}\text{Mpc}$ slice. The top left panel shows the component originating from filaments, the top right panel shows the component originating from voids, the bottom left panel shows the component originating from walls and the bottom right panel shows the component originating from nodes.}
    \label{fig:tidaltot}
\end{figure*}

Following the definitions in section~\ref{sec:tidalfield}, in the present study we first investigate the tidal force field generated by
filaments, voids, walls and cluster nodes by means of tidal force amplitude maps, i.e. maps of the (traceless) tidal field amplitude $|T|$, 
\begin{equation}
    |T| = \sqrt{T_1^2 + T_2^2 + T_3^2} \label{eq:tidal_tot_def}\,,
\end{equation}
with $T_1 > T_2 > T_3$ the tidal field eigenvalues. Figure~\ref{fig:tidaltot} presents the tidal amplitude maps for the tidal fields generated by the filament population (top lefthand panel), the void population (top righthand panel), the wall population (bottom lefthand panel) and cluster nodes (bottom righthand panel). The first direct observation from the four frames is the difference between the tidal field morphology and patterns to that of the force field. The tidal force field is intimately linked to the weblike spatial pattern of the cosmic mass distribution, and also reflects more closely its smaller scale structure than the force field. Recent work has also revealed the close similarity between the
structure of the primordial tidal field, specifically its eigenvalues, and that of the emerging structure of the cosmic web 
\citep[see e.g][]{Feldbrugge2023,Feldbrugge2024}. 

\bigskip
\subsection{Tidal Field illustrated: tidal impact voids}
To appreciate the intricacies of the cosmic web tidal force field, it is most insightful to focus on the component induced by the
voids and the void population. To this end, we first zoom in on a few interesting and telling regions. Figure~\ref{fig:zoomtid} shows zoom-ins on four different regions of size $25\times25$h$^{-1}$Mpc. The four panels show the tidal bars of the compressional components of the tidal field induced by the void population in the matter distribution. The direction of the bars is along the corresponding tidal tensor eigenvectors, their size proportional to that of the corresponding eigenvalues.

The panels illustrate the close connection between the weblike spatial pattern of the matter distribution, and its multiscale nature, and the tidal force field induced by voids, which we find - as discussed in more detail below - to be a surprisingly striking aspect of the large scale tidal field. Even more so than the filament induced tidal force field, we find that the deformation of the matter distribution is strongly correlated to the spatial pattern and connectivity seen in the void induced tidal force field. The void tidal field traces out structures at all scales and over a wide range of densities. It causes compression along even the smaller, more tenuous,
structures seen in the density field. 

The clearest example of a void's tidal impact is offered by the bottom lefthand panel, centered at the interior of a void. It has
the typical characteristics
of a tidal force field expected in and around voids.  A typical aspect of the void induced tidal force field is that it is weak in the interior of voids. It steeply increases in amplitude as it enters the overdense boundaries, a reflection of the strong differential
gravitational force between interior and surroundings. In this sense, it adheres closely to that expected for isolated spherical
voids. It induces a compressional component $T_{rr}$ along the radial direction, while the dilational components $T_{tt}$ are oriented
along the tangential direction. At a radial distance $r$ from the void center, for a void with density profile $\Delta(r)$,
\begin{align}
  T_{rr}&\,=\,\Omega H^2\,\big[\Delta(r)-\langle \Delta(r) \rangle\big] \,,\nonumber\\
  T_{tt}&\,=\,-\frac{1}{2} T_{rr}(r)\,,
  \label{eq:voidtide}
  \end{align}
\noindent in which $\langle \Delta \rangle(r)$ is the mean interior density of the void. Hence, in the near uniform - bucket shape - mass
distribution in the interior of voids \citep[see e.g.][]{weygaert2016} the tidal force field will be negligible. Meanwhile, the
fact that $\Delta(r) > \langle \Delta(r) \rangle$ as we enter the boundary of voids implies into a compressional
tidal force along the radial direction of the void. It translates into the compression of mass elements in the radial
direction of the void, reflecting the formation of an overdense boundary around the void. 

The variety and extent of the tidal impact of voids on their environment can be appreciated from the other three panels in
figure~\ref{fig:zoomtid}.  The top righthand panel is a telling illustration of their influence on the dynamics of filaments: the
tidal bars trace the three filamentary structures that are connecting to the node at the top righthand corner of the panel. Evidently,
the implication is that voids are instrumental in effecting further compression of the filaments. Overall, we see that the void
induced tidal force field follows the underlying mass distribution in meticulous detail, which we find to be an exclusive property
of the void population. Even on scales smaller then a few Mpc the voids are able to closely trace out the underlying matter
distribution. The tidal fields of the other components do not display this level of detail. 

\subsection{Tidal Field Structure}
For comparison between the tidal influence of the various cosmic web components, figure~\ref{fig:tidaltot} shows the
tidal amplitude maps for each individual component of the cosmic web. The top left panel shows the filament tidal field,
the top right panel shows the void tidal field, the bottom right panel shows the wall tidal field and the bottom right panel shows the tidal field for nodes. The first global impression offered by these maps is that the differences are starker than for the
corresponding force fields. 

\begin{figure*}
    \centering
    \includegraphics[width=0.7\textwidth]{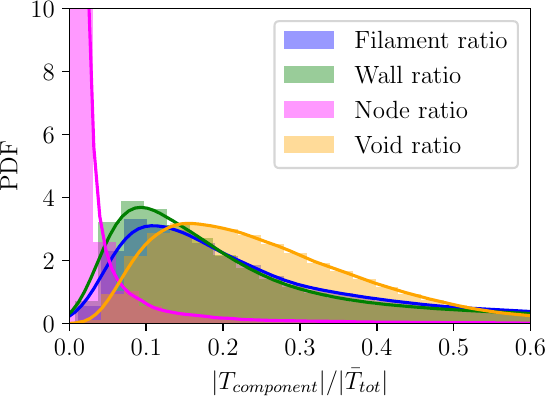}
    \vskip 0.5cm
    \includegraphics[width=0.7\textwidth]{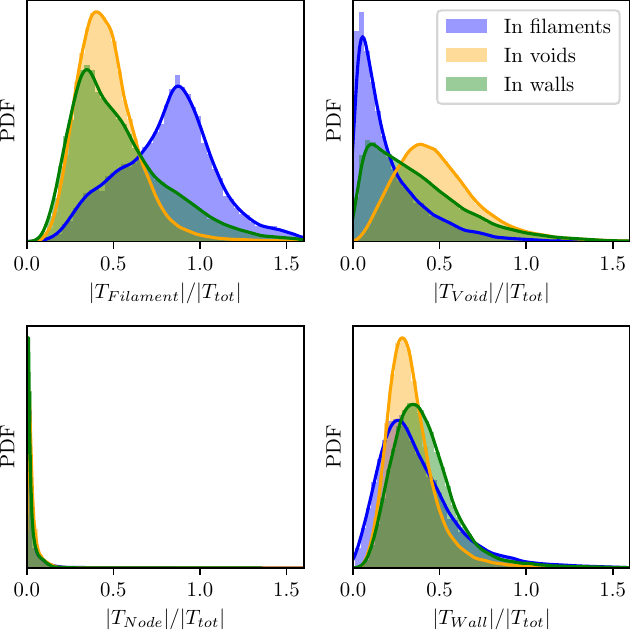}
    \caption{Cosmic Web Tidal Field: Strength Inventory. Top panel: The distribution of $|T|^{\text{CWM}}|/|T|_{\text{tot}}|$ for the different components of the cosmic web. This is the distribution for the entire field. Bottom four panels: the distribution of
    $|T|^{\text{CWM}}|/|T|_{\text{tot}}|$ for the different components of the cosmic web. Per panel the
      pdf for cosmic web component CWM is plotted, within three different environments: filaments (blue), voids (yellow), walls (green). Centre left: filament induced tidal field. Centre right: void induced tidal field. Bottom left: wall induced tidal field. Bottom right: cluster induced tidal field.}
    \label{fig:tidratio}
\end{figure*}

\begin{figure*}
      \includegraphics[width=0.7\textwidth]{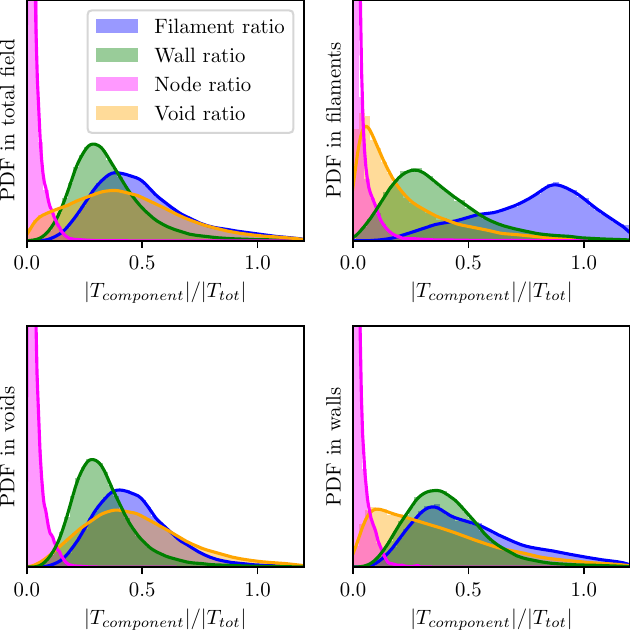}
      \caption{Cosmic Web Tidal Field: Strength Inventory by Cosmic Web environment. The panels show the pdf of the
      induced tidal field $|T|^{\text{CWM}}/|T|_{\text{tot}}$ by each of the four cosmic web
      components CWM in the different cosmic web environments. The different colours represent the different components:
      filament tidal field (blue), void induced tidal field (yellow), wall induced tidal field (green) and cluster node
      induced tidal field (magenta). Top left panel: total tidal field. Top right panel: tidal field in filaments.
      Bottom left panel: tidal field in voids. Bottom right panel: tidal field in walls. }
      \label{fig:tidcomp}
  \end{figure*}

On large scales, the amplitude maps for the filaments and walls share a similar pattern, be it that the tides generated by
filaments is substantially stronger than that for walls. The fact that they share a similar large scale spatial structure is
an expression of the fact that these anisotropic features are intimately related aspects of the weblike network of which they
are the principal constituent elements. A clear difference with the corresponding force amplitude maps (fig.~\ref{fig:totg})
is that the tidal fields rapidly drop to a negligible level directly outside the filaments and walls, a direct reflection
of the more localised nature of the tidal field. In terms of the relative amplitudes the filament induced tidal forces are still
dominant (see fig.~\ref{fig:tidratio}), in particular within the filament and wall network of the cosmic web, but far less so than in the
gravitational force field. Within the interior of the filaments, the filament induced tides stand out as by far the strongest
influence. Wall induced tides are marginally stronger inside walls, somewhat stronger than those induced by (nearby) filaments
and voids (see fig.~\ref{fig:tidcomp}).

The filament tidal amplitude field delineates a large scale pattern of massive elongated filamentary structures. With the elongated shape
of the filaments is an outstanding fundamental aspect of the filament tidal field, this goes along with a rather substantial level
of inhomogeneity along the filaments, with high amplitude peaks marking the immediate high-density environment of massive clusters
or the branching connections with other filaments in the weblike network. It is a direct reflection of the wide range of densities
of filaments \citep[see e.g.][]{NEXUS_CW_EVO} and of the large density variations along the ridges of filaments in
the cosmic web\citep[see][]{NEXUS_CW_EVO}, which manifest themselves strongly in the more localised nature of the
tidal field (as opposed to the more large-scale nature of the force field). 

The wall tidal amplitude map shares the dominant large scale features seen in the filament map, but also includes several
different and significant characteristics. While the same large scale structures in the wall tidal field can be seen,
their contrast is substantially less than that seen in the filament induced tidal field. Also the tidal strength
induced by the walls shows far less internal variation, yielding a more coherent tidal field within the outlined
structures. This is a direct consequence of the mass density in walls being more uniform and spanning a much narrower
range than that of the filament population \citep[see e.g.][]{NEXUS_CW_EVO}. By far the most outstanding difference to
the filament tidal field, in addition to the degree of coherence, is the presence of far more small-scale structure filling
up the space between the dominant large scale structures. Throughout the entire volume, it outlines a more intricate weblike
network marked by the tidal footprint of small walls. Often they surround or connect with each other to form the boundaries of
small void regions. In implies the multiscale nature of the cosmic web to be more readily visible in the wall induced
tidal field than that in the overpowering filament tidal field, the latter more dominated by the major arteries of the
cosmic web. 


It is the void induced tidal field that reveals a surprisingly different pattern than that seen in the filament tidal amplitude
map. In a sense it extrapolates the trend seen in the wall tidal amplitude maps of the dominant presence of small scale
weblike structure. It yields a rich spatial pattern marked by small scale voidlike regions, with their boundaries
connected into a pervasive network, yielding an ordered assembly of small-scale voids surrounded by wall-like boundaries.
The pattern is also coherent and has a rather uniform amplitude, a direct reflection of the narrow density range of
these cosmic underdensities \citep{NEXUS_CW_EVO}. It suggests that voids are largely responsible for defining and outlining
the small-scale structure of the cosmic web, and appears to imply that voids play a crucial role in the formation, development
and spatial organisation of the multiscale cosmic web.

By contrast, the node tidal field offers the least significant contribution to the tidal field. Even more so than in the
case of the corresponding force field, the node induced tides define a spatial distribution that is highly localised
in the immediate vicinity of the cluster nodes. The field does not contain a clearly defined structure, and has the
appearance of a random set of tidal monopoles, with a field strength falling off like $r^{-3}$. 

\subsection{Tidal field strength: statistical analysis}
For a quantitative judgement over the relative and absolute role of the induced tidal fields in the cosmic web, we turn
to a similar assessment as in the case of the cosmic web force field, in terms of the probability functions for the
total and relative tidal force field. Overall, we find a similar hierarchy in tidal force dominance as that in
the case of the force field. However, filaments are far less dominant in the tidal field than in the force field. Walls, and
in particular voids, have a considerably larger impact in setting the tides in the Universe. By contrast, nodes do not have any influence outside their
immediate vicinity they , despite their high density contrast sometimes in excess of a thousand.

The top panel of figure~\ref{fig:tidratio} compares the pdf of the amplitude of the tides generated by the individual
cosmic web components. The amplitude $|T|_{tot}^{\text{CWM}}$ is given in units of the average amplitude of the total field,
$\sigma(|T|)$. The distribution of the amplitude of the induced tidal fields shows a substantial difference between the
various components. Voids are by far the most ubiquitous components in the medium range of tidal values, in between
$0.15 < |T|_{norm} < 0.5$. This is a clear reflection of their substantial tidal imprint, as we have already noticed
in the amplitude maps, and the fact that they occupy a major fraction of the cosmic volume. 

Walls and filaments have a major presence at low tidal amplitudes, with a mode near $T_{norm} \approx 0.1$, a reflection
of the rapid falloff towards low tidal strengths outside of their own realm. However, both have a long tail towards high tidal values,
ie. for $T_{norm} > 0.5$. It even leads to an average value substantially higher than that for voids, as may be seen in
table~\ref{tab:resultstid}. The filament tail reaches the highest tidal values, confirming the visual impression
of figure~\ref{fig:tidaltot}, with the walls remaining at more moderate tidal values. The high tidal values
for these structures are a direct reflection and manifestation of the corresponding mass concentration in lower
dimensionless geometric structures, i.e. of their elongated and flattened geometry. In turn, these are the
result of their formation driven by the gravitational contraction induced by these anisotropic forces. 

Telling is also the conclusion that quantitatively cluster nodes play only a minor role in setting the overall
tidal force field. In most of the cosmic volume the nodes take care of only minor tidal values. The fact that their tidal
pdf has a very long but low tail hints at their dominance in the minute regions of their immediate neighbourhood, but nowhere
else. 

\begin{table}
\centering
\caption{Tidal field statistics. The middle two columns show the mean tidal force ratio for the different components of the cosmic web using either global or local normalisation. Note the direction dependence in the ratio $|T|^{\text{CWM}}/|T|^{\text{tot}}$ and $|T|^{\text{CWM}}/|T|^{\text{tot}}$. The fourth column shows the percentage of the volume where each component is larger than all other components.}
\begin{tabular}{r l l l}
 \hline
    Component & Global [\%] & Local [\%] & Largest in [\%]\\ \hline
    Filaments & 65.3 & 51.5 & 48.5\\ 
    Voids & 26.9 & 45.2 & 41.0\\ 
    Nodes & 3.5  & 3.0 & 0.1 \\
    Walls & 34.9 & 36.2 & 10.4 \\ 
\hline    
\end{tabular}
    \label{tab:resultstid}
\end{table}

\bigskip
Upon assessing in which environments the various component tidal influences hold sway, we notice substantial differences
in the case of voids and filaments, far less so for walls and cluster nodes. Overall, the wall and node contributions to
the tidal inventory of the cosmic web appear to be far less dependent on location than that of filaments and voids. For
the nodes this is largely because of their spatially severely limited influence. 

Evidently, the relative filament and void
influence are highly dependent on their location in the cosmic web. The four panels in figure~\ref{fig:tidratio}
show the distribution of the relative filament, void, cluster and wall tidal contribution in either filaments (purple),
void (orange) or wall (green) environments. The top righthand panel reveals the dominance of the filamentary induced
tides within the realm of the filaments themselves, with the related panels for voids and walls indicating only a minor
influence of these structures within filaments. Meanwhile, the tidal influence of voids appears to be largely restricted
to that in voids themselves. Walls hardly ever are a dominant tidal source, even not within walls themselves,
although they always appear to represent an omnipresent minor influence. 

The latter implies that the dynamics of filaments is largely propelled
by the filamentary network itself. This is emphasised in the accompanying panels in figure~\ref{fig:tidcomp}, which
for each of the cosmic web environments shows the distribution of the relative tidal impact by filaments, walls, voids and
cluster nodes. Within filaments (top righthand panel), tidal forces are mostly due to the influence of filaments themselves,
with a moderate but substantially weaker influence by walls, and far less by voids and cluster nodes. The moderate
influence by walls is an expression of the close spatial and geometric relation between the anisotropic filamentary
and wall-like elements in the cosmic web.

Void regions show a more chequered tidal image. From the bottom lefthand panel of figure~\ref{fig:tidcomp}, we see
that within voids, the tidal forces induced by the voids themselves represent a major influence, usually taking care
of up to $40-50\%$ of the tidal force field. Nonetheless, the impact of the surrounding filaments remains important
and often is even dominant (the long tail of the pdf). To a lesser extent this is also true for the walls. This often
concerns the outer parts and boundary regions
of the voids, where the dynamical influence of the higher density of filaments rapidly takes over the gravitational
influence of the underdense voids. It emphasises the observation seen earlier with respect to the gravitational
force field, the fact that the dynamics and dynamical evolution of voids cannot be understood without taking
into account the external influence by in particular filaments. 

\subsection{Tidal field alignment}
Given the notion that the elongated filamentary ridges and flattened walls in the cosmic web are the result of
the tidally induced deformation of primordial matter concentrations, we should recognise this in the existence
of an alignment of filaments and walls with respect to the tidal force field. To assess to what extent
structures are aligned, we include a rough appraisal of the orientation of structures in
the cosmic web with the tidal eigenvectors.

\bigskip
At each location we measure the alignment between the local geometry of the mass distribution and the tidal force. 
To this end we determine the orientation between the eigenvectors of the local tidal tensor and the eigenvectors of
local inertia tensor. The latter is represented by the Hessian ${\cal H}(\mathbf x)$ of the density field
\begin{equation}
  H_{ij}(\mathbf x)\,=\,\frac{\partial^2 \Delta(\mathbf x)}{\partial x_i\partial x_j}\,.
\end{equation}
The eigenvectors of the Hessian are ${\hat e}_1$, ${\hat e}_2$ and ${\hat e}_3$, with corresponding eigenvalues  
\begin{align}
    e_1 &> e_2 > e_3 \\
    |e| &= \sqrt{e_1^2 + e_2^2 + e_3^2}\,.
\end{align}

\begin{figure}
    \centering
    \includegraphics[width=0.5\textwidth]{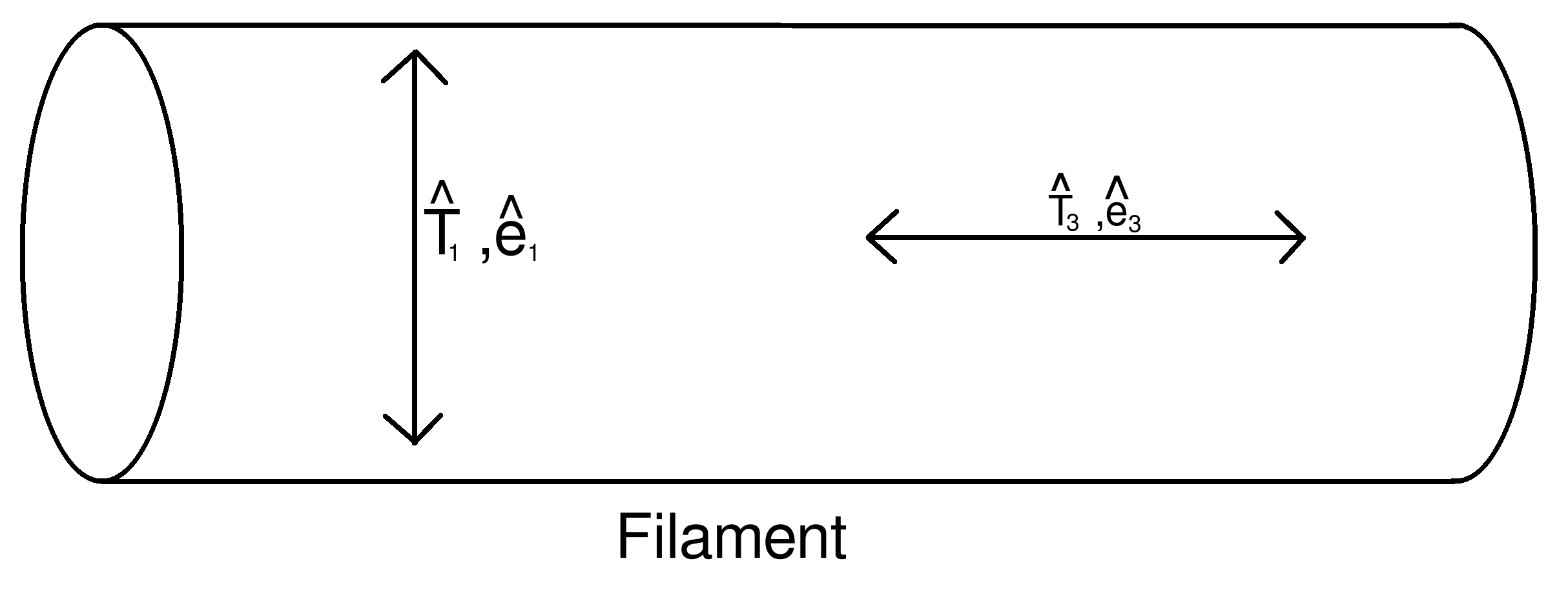}
    \caption{Tidal field \& Cosmic Web aligment.  Diagram illustration of (idealized) orientations inertial (Hessian density field) and tidal eigenvectors with respect to an elongated filament. Indicated is the orientation of the larges tidal eigenvector ($T_1$) and
    that of the smallest inertia eigenvector ${\hat e}_3$, which is aligned along the principal ridge of the filament.}
    \label{fig:draw}
    \vskip 0.5cm
    \centering
    \includegraphics[width=0.5\textwidth]{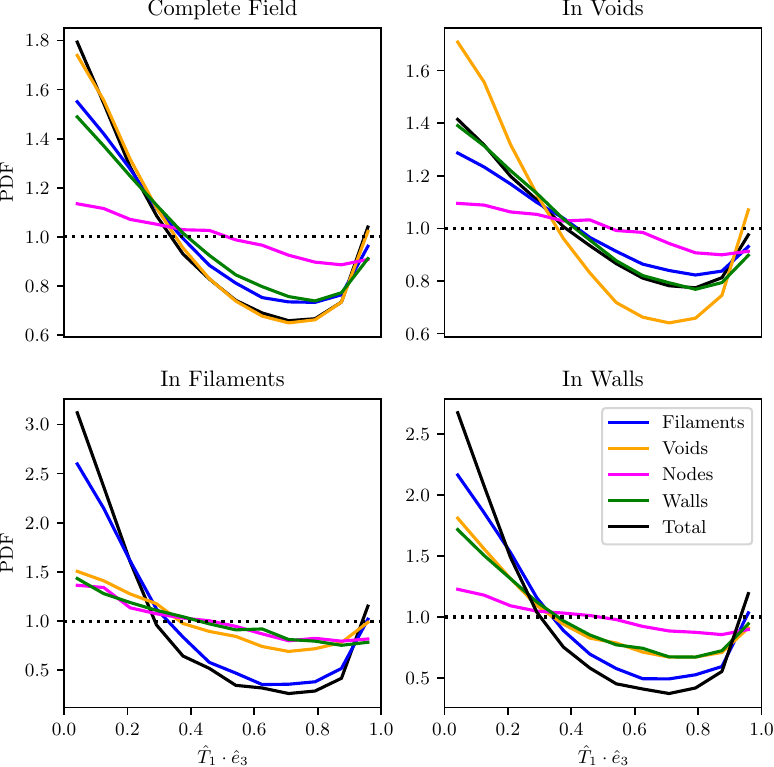}
    \caption{Tidal field \& Cosmic Web aligment. The pdf of the alignment angle $\cos \theta=\hat T_1 \cdot \hat e_3$ between the
      largest (compressional) tidal eigenvector and the inertia tensor direction along the filament ridge, ${\hat e}_3$. The
      four panels show the pdf curves for the tidal field induced by the entire mass distribution and the specific tidal
      fields induced by the four morphological components $CWM$ in four different cosmic web environments.
      Black curve: tidal field induced by entire mass distribution; blue curve: filament induced tidal field; orange curve:
      void induced tidal field; green curve: wall induced tidal field; magenta curve: cluster node tidal field.
      The environments in the four panels: entire field (top lefthand panel); void regions (top righthand panel); filament regions
      (bottom lefthand panel); wall region (bottom righthand panel). Note the different range for the vertical axis of the
      four panels.}
    \label{fig:tidangd}
\end{figure}

A schematic diagram indicating the eigenvectors, of both tidal and inertia tensor, with respect to an elongated filament is
shown in figure~\ref{fig:draw}. Our alignment analysis investigates the orientation between the largest - compressional -
tidal eigenvector, ${\hat T}_1$ and the inertia eigenvector along the ridge of the filament, ie. the smallest density
Hessian, ${\hat e}_3$. In a perfectly aligned setting, the compressional tidal eigenvector ${\hat T}_1$ would be expected
to be perpendicular to the filament's ridge, and hence to ${\hat e}_3$, implying an inproduct
\begin{equation}
\cos(\theta)={\hat T}_1 \cdot {\hat e}_3=0\,. 
\end{equation}
On the other hand, in the absence of any correlation between the shape of the structure and the tidal field, the
orientation $\theta$ between ${\hat T}_1$ and ${\hat e}_3$ would be random and the distribution of
$\mu=\cos{\theta})$ entirely uniform (flat).

While the diagram in figure~\ref{fig:draw} relates to filaments, similar geometric considerations
also hold for walls and voids. In the case of walls, ${\hat e}_3$ is one of the two eigenvectors directed
along the plane of a wall, while ${\hat T}_1$ is the compressional tidal component, ideally directed
perpendicular to the wall. For voids, the compressional tidal component ${\hat T}_1$ is expected to be
directed along the radial direction of the void (see eqn.~\ref{eq:voidtide}). To first approximation,
inside voids the density hardly varies along radial shells, so that the density Hessian
eigenvector ${\hat e}_3$ is oriented along the transverse direction of the void.

\begin{figure*}
    \centering
    \includegraphics[width=\textwidth]{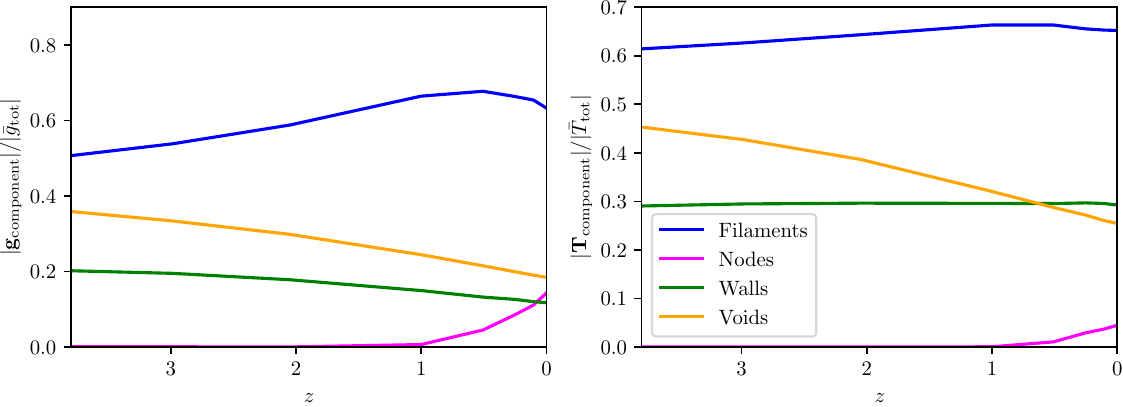}
    \caption{Redshift Evolution Forces and Strains in the Cosmic Web. Left frame: redshift evolution of the
      global average gravitational force fraction ${\cal F}^{\text{CWM}}\,=\,|{\mathbf{g}}|^{\text{CWM}}/|\mathbf{g}|^{\text{total}}$ for the force induced by cosmic web component \text{CWM}. 
      Right frame: redshift evolution of the global average tidal field fraction
      ${\cal F}^{\text{CWM}}\,=\,|T|^{\text{CWM}}/|T|^{\text{total}}$ for the tide induced by cosmic web component \text{CWM}.}
    \label{fig:zevolution}
\end{figure*}

The distribution of the orientation angle $\cos(\theta)$ between ${\hat T}_1$ and ${\hat e}_3$ for
the different cosmic web environments is shown in the four panels of figure~\ref{fig:tidangd}. For
the entire field, as well as for the void, filament and wall environments, the panels depict the
orientation pdf for the compressional component of the complete tidal force ${\hat T}_1$,
as well as for the corresponding compressional components for each of the cosmic web components
${\text{CWM}}$, ${\hat T}_1$.

All panels reveal a substantial level of alignment between the compressional tidal force and the
local geometry of the environment. The distribution functions peak strongly towards
$\mu=\cos(\theta)=0$, ie. we find  a strongly perpendicular tendency. The one major exception is that for the node induced tidal force, which in all environments shows a mere weak alignment. It is a manifestation of the weak cluster node induced tidal field throughout the entire volume. With respect to filaments and voids the following observations are made:
  
  \begin{itemize}
  \item{} {\textit Filaments}\\
    Inside filaments, the alignment between tide and geometry can be almost exclusively ascribed to the the
    filament induced tidal force. To a lesser extent, the filament induced tide is responsible for
    a major share of alignment in walls, confirming yet again the strong structural bond between
    filaments and walls. Hence, the filament induced tides are not only dominant with respect to
    their amplitude and strength, but also with respect to their orientation and dynamical impact.
    On the other hand, we also see that even within filamentary environments, the filament induced tides
    do not account for all alignments. It hints at the still significant influence by the other
    cosmic web components, in particular voids and walls.
  \item{} {\textit Voids} \\
    A most interesting finding is that the mass distribution inside voids is more strongly aligned
    with the tidal field than that in the overall field (top righthand frame fig:tidangd). To a large
    extent this is to be ascribed to the strong alignment with the tidal force induced by the void
    itself. Also, we find that the latter has only a weak alignment with the ridge of filaments
    (bottom righthand frame fig.~\ref{fig:tidangd}), and a mere moderate alignment with the plane
    of walls (top righthand frame fig:tidangd). In all, it confirms the earlier impression of
    the voids tidally dominating the more tenuous parts of the cosmic web.
    \end{itemize}

\section{Cosmic Web Force \& Tidal Evolution}
The present study has concentrated predominantly on the dynamical structure of the cosmic web at the
current cosmic epoch, $z=0$. In order to
develop more definitive conclusions on the buildup of structure by the force and tidal influence of
the various cosmic web components, it will be imperative not only to study the dynamics at the
current time, but also to investigate their behaviour over time. While in an upcoming study we address the
dynamical evolution of the cosmic web in detail, we may get a global impression by evaluating global averages
of the force and tidal amplitudes as a function of redshift $z$.

We follow the evolution of the mean (normalized) force and tide contribution
by the various cosmic web components - filaments, walls, voids and cluster nodes. To this end, for
each redshift $z$, at each location ${\mathbf x}$ we determine the fractional contribution ${\cal F}^{\text{CWM}}({\mathbf x},z)$ by a
component ${\text{CWM}}$ to the total force or tide amplitude ${\cal A}^{\text{total}}({\mathbf x},z)$,
\begin{equation}
  {\cal F}^{\text{CWM}}({\mathbf x},z)\,=\,\frac{{\cal A}^{\text{CWM}}({\mathbf x},z)}{{\cal A}^{\text{total}}({\mathbf x},z)}\,.
\end{equation}
By normalizing with respect to the average (total) force or tide amplitude at redshift $z$, we compensate for
the time evolution of the general force and tidal fields. We average over all locations ${\mathbf x}$ to obtain a
global fractional average ${\tilde {\cal F}}(z)$ at redshift $z$.

Figure~\ref{fig:zevolution} plots the redshift evolution of the fractional force contributions by cosmic web
components ${\text{CWM}}$ (lefthand panel) and corresponding fractional tidal amplitude contributions (righthand panel),
  over a redshift range $0<z<4$.  On the basis of these rudimentary global averages, a few generic conclusions may be drawn
  with respect to the evolution of cosmic web dynamics.

  Over the entire redshift range, filaments are consistently by far the most dominant cosmic web component, with respect to
  both force and tidal fields. Up to a redshift $z \approx 0.5-1$, the influence of filaments is increasing, 
  partially the reflection of the growth of the filament population and mass as cosmic structure develops
  \citep[see e.g.][]{NEXUS_CW_EVO}. The influence of filaments stabilises after this redshift, and may even
  decrease somewhat. It is tantalising to identify this with the universe transiting from a matter-dominated
  to a dark energy dominated regime, and the corresponding global stemming of structure growth as the Universe
  enters an exponential de Sitter expansion phase. Another factor may be that of the steep emergence and growth
  of massive cluster nodes, whose fractional force and tidal influence we see rising steeply after $z \approx 1$
  (see fig.~\ref{fig:zevolution}, magenta curve). This goes along with the accretion of large amounts of
  mass along dense filamentary branches connecting the filamentary arteries with the cluster nodes, and
  the subsequent increasing force influence of the nodes. At high redshifts the influence of cluster nodes
  is almost negligible, as their hierarchical buildup has not yet formed virialized clusters of comparable mass.
  
  Arguably the most interesting evolutionary trend is seen for the void population. The force and tidal field
  influence of voids turns out to be even more prominent at higher redshifts than it is today. To some extent
  this may seem surprising. Because voids are smaller and less empty at higher redshifts, at first one might
  expect them to have a weaker influence over the cosmic mass distribution. On the other hand, the filamentary
  and wall-like skeleton of the cosmic web is also developing rapidly at high redshifts, and represents
  less mass at those redshifts. Relatively speaking, voids may therefore be more prominent at high redshifts.
  The details of the dynamical influence of voids, of the induced mass migration out of their interior, within
  the context of the hierarchically evolving void populations \citep{shethwey2004,aragon2013}, therefore
  need further investigation. This is the subject of a related upcoming study. 

  Finally, walls offer a chequered view of force and tidal evolution. Their tidal influence appears to remain
  almost constant throughout time. Their force evolution, however, appears to follow a similar downward trend
  as that seen for the void population. It may reflect a proportionally higher mass content at higher redshift,
  a consequence of the flow of mass from void troughs, via walls towards filaments.

\section{Summary and Discussion}\label{sec:conc}
The cosmic web emerges at the transition between the linearly evolving cosmic matter distribution on large
cosmological scales, and the highly nonlinear realm on small scales where we find matter assembled in
virialized halos\citep{bond1996,weybond2008,NEXUS_CW_EVO,Libeskind2018}. At this transition scale, we see the
emergence of complex spatial structure in the shape  
of an intricate weblike pattern. Long elongated filaments and flattened tenuous planar walls form the
boundaries of near-empty void regions and defined a pervasive interconnected network, whose dense
compact nodes are the sites where we find massive virialized halos. 

With the cosmic web forming the manifestation of this key dynamical transition in the organisation of the
cosmic matter distribution, it is essential to develop a more profound insight and understanding of the
dynamics driving the buildup of cosmic structure, of the induced migration flows, and the gravitational
influence of the various constituents of the cosmic web on these processes. This will bear upon a range
of major cosmological issues. Prime is a more fundamental understanding of the formation and assembly of
the cosmic web itself, and its relation to the primordial mass distribution \citep[see e.g.][]{Feldbrugge2018,Feldbrugge2024}.
Of major current interest is the influence of the large scale environment on the formation and evolution
of galaxies, of which the most widely recognised factor is that of the generation of angular momentum of
galaxies by tidal torqueing by the same tidal forces that shape the filaments and walls in the cosmic web
\citep[e.g.][]{Hoyle1949,White1984ttt,porciani2002a,porciani2002b,Jones2009TTT,codis2012,codis2015,Punya2019TTT}.
Of increasing interest is also the potential information content of the intricate cosmic web structure
on the value of cosmological parameters and the properties of dark matter and dark energy. Over the recent
years in particular the structure and kinematics of cosmic voids have received ample attention for
the potentially high precision with which they may reflect the character and value of dark energy
\citep[see e.g.][]{leepark2009,weygaert2011,lavaux2012,bos2012,weygaert2016,pisani2019}. 

\bigskip
The present study is the first stage in a systematic investigation of the detailed dynamics involved
in the emergence and assembly of the cosmic web and its various morphological components. 
We assess the force field and tidal field induced by filaments, walls, cluster nodes and voids, and investigate in
how far they contribute, and dominate, the gravity and tides in the various regions of the
cosmic web. It facilitates an in-depth investigation of the role of the different morphological
components of the cosmic web, and its environmental dependence, in driving the gravitational evolution driven
formation and evolution of the largest structures in the universe.

The dynamical inventory study is based on the redshift $z=0$ structure in a $512^3$ large $\Lambda$CDM dark matter
only N-body simulation in a $(300h^{-1}\text{Mpc})$ box. At each location in the simulation box the gravitational
force and tidal tensor is computed for the entire mass content in the simulation volume, as well as separate for
the force and tides generated by filament regions, walls, void regions and cluster nodes separately. In the
current study this is based on a brute-force direct summation over the mass elements residing in the
various identified morphological structures. Given the morphological identity of each mass element, we are able
to do the gravitational force and tidal field analysis separately for the various cosmic web regions. Hence, it allows
us to study in how far the dynamics of voids is influenced by the surrounding filaments or, reversely, in how
far the gravitational force and tidal forces in filaments are feeling the presence of the surrounding voids.
In addition, it enables an evaluation of the reach of void induced forces and tides, the scales over which
filaments dominate, and a wide range of related issues. 

For the identification of the cosmic web structures, we invoke the density field \nexus{} multiscale morphological
formalism \citep{MMFa,MMFb,MMF3,NEXUS_MAIN,NEXUS_CW_EVO}, which delineate the filament, void, wall and node regions.
\nexus{} is the highest dynamic range version of the Scale Space MMF/Nexus pipeline, unique in its combination of
geometric and dynamical criteria for the
morphological classification and that of taking into account of the multiscale nature of the cosmic
matter distribution. It enables us to identify structures over a wide spectrum of scales and densities, and hence
assures an optimal assessment of the multiscale aspects of cosmic web dynamics. The \Nexus{} formalism assigns
to each location in the simulation volume the appropriate morphology - filaments, walls, void or cluster node -
following a parameter-free evaluation at the proper physical scale of the locally dominant feature. 

\bigskip
The systematic inventory of cosmic web dynamics yielded insights that confirm existing assumptions. More tantalising
are new and surprising insights with respect to the role of voids and cluster nodes. Amongst the conclusions based
on the presented force and tide inventory are:

\begin{itemize}
\item The gravitational force fields generated by the distinct cosmic web components differ significantly from each other.
  The gravitational fields induced by filaments, voids, walls and cluster nodes have markedly and systematically dissimilar
  characters. The overall amplitude of the gravitational force, their range, and spatial structure and pattern are
  quite dissimilar. In terms of the amplitude of the induced forces, there is a clear hierarchy: filaments are by far the
  most dominant and powerful force, followed by voids, only then - surprisingly - by cluster nodes, followed by walls.
\item At nearly every location, the total gravitational force is the combination of the contributions by several
  cosmic web components. The important implication is that it is hardly possible to describe the dynamics and development of a
  structure as an isolated object. In general the gravitational influence is the result of the interplay between the
  several contributions. The only exceptions may be the force field in the interior of the most prominent filaments and
  in and around the immediate vicinity of massive cluster nodes. Yet, over most of the filamentary network also the influence
  of nearby voids should be taken into account, while the dynamical evolution of walls cannot be understood at all without
  taking into account the impact by the nearly filaments and voids. It certainly is true without any exception for all void
  regions, whose dynamics cannot be understood on the basis of their underdense interior but should also take into account
  the even more dominant exterior influence by the nearby filaments. 
\item Filaments dominate the gravitational force field over nearly the entire cosmic volume, with the exception of the immediate
  vicinity of the massive cluster nodes. On average, they are responsible for more than $50\%$ of the exerted gravitational
  force at any location. Filaments' exert their influence over a large spatial range.
\item While voids turn out to represent a sizeable gravitational influence over nearly the entire cosmic volume, they
  hardly dominate anywhere, even within their own interior. Instead, at a large fraction of space they are the second
  gravitational influence after filaments. 
\item  Voids have a limited spatial range over which they wield their gravitational influence. The spatial pattern defined by
  the void generated gravitational field is one of a segmented volume, each segment surrounding a locally strong repulsive
  density trough (with radii up to $50~h^{-1}$Mpc. Zooming in on individual void regions shows their superhubble expanding effect
  over their interior and up to their overdense boundary. Within the interstitial regions between voids, i.e. for their filamentary
  and planar boundaries, their gravitational impact declines rapidly to very low levels.
\item Despite the mostly segmented nature of the void induced force field, still we find a residual large scale void
  force field. This is a manifestation of the multiscale nature of the void population
  \citep[see][]{shethwey2004,aragon2013}, a direct result of its hierarchical evolution and which has recently been
  detected in the observational Cosmicflows-3 peculiar velocity field \citep{courtois2023}. 
\item A complete dynamical description of voids in terms of isolated expanding underdense regions is not representative, The
  evolution of voids depends strongly on the exterior mass distribution. Void dynamics is predominantly a combined effect Of
  the induced superhubble expansion by voids themselves and the force and tidal influence of the the surrounding filaments
  (and, if in their near vicinity, the nearby cluster nodes). It shows that any consideration of void dynamics in terms of
  isolated almost spherically expanding regions needs a fundamental revision: the dynamics of voids cannot be understood
  without taking into account the gravitational and tidal influence of the environment. 
\item A surprising finding is that of the rather low dynamical influence by massive cluster nodes. Both the gravitational
  force induced by cluster nodes, as well as the corresponding tidal force, are only significant in and in the immediate
  surroundings of the cluster nodes, out to a radius of $\approx 5h^{-1}$Mpc. Because of their rather limited gravitational
  influence, they only wield a minor impact on the structure of the cosmic web.
\end{itemize}

\bigskip
\begin{itemize}
\item With respect to the corresponding gravitational tides, the situation in terms of strength is less outspoken
  than in the case of the gravitational force. Still, filaments are the strongest source of gravitational tides. However,
  voids, and even walls, yield a significant additional and alternative tidal influence. While the tidal impact of filaments
  has a predominantly large scale character, strongly correlated with the main elongated arteries in the cosmic web,
  walls and in particular voids add an essential small-scale aspect of the tidal field that is to be identified with the
  multiscale nature and structure of the cosmic web.
  \item The tidal field generated by filaments is
  largely a large-scale phenomenon, reflecting the outline of the most massive and prominent arteries in the
  cosmic web. The multiscale structure of the cosmic web can hardly be recognised in the filament induced tidal field. 
  Filaments are the predominant source for the tidal forces inside the filaments themselves, as a result of their
  overwhelming strength within the filamentary ridges of the cosmic web.  
\item Along the filamentary ridges, the tidal field generated by filaments is
  rather inhomogeneous and patchy, marked by a fluctuating trend between high amplitude and low amplitude regions along a
  filament's ridge. This is a direct manifestation of the rather strong mass density variations along a filament's
  ridge \citep[see e.g.][]{NEXUS_CW_EVO}. With respect to the orientation of the tidal force field along filaments,
  we see a coherent pattern of aligned compressional tidal bars along the ridge of a filament. 
\item Voids are the dominant source of tidal influence within their interior. As voids represent the major share of
  a cosmic volume, void assume a major organising role in the shaping of the cosmic matter distribution.
\item An interesting finding is the remarkable spatial pattern in the void induced tidal amplitude map: the void tidal field
  defines a coherent rather uniform cellular pattern over the entire volume, dominated by the small-scale voids in
  the multiscale buildup of the cosmic web. The void tidal field appears to trace the cosmic web down to even the most
  tenuous structures found inside voids \citep[also see][with respect to a similar observation wrt. the interior void
  velocity field]{aragon2013}.
\item The finding that the spatial organisation of the cosmic web is most readily expressed in the void tidal field
  suggests voids to be seen as the organisers of the cosmic web. It is confirmed by the orientation of the void induced
  tidal field (see the case studies in fig.~\ref{fig:zoomtid}): the compressional component appears to trace the more
  tenuous parts of the cosmic web, suggesting them to be responsible for the shaping of the tenuous walls and filaments
  that form the boundaries of the small voids. Analysis of the alignment between tidal field and mass distribution
  underlines this conclusion, revealing a stronger alignment within the interior of voids than in the overall cosmic web. 
\item The tidal force field induced by the wall population is more noticeable than its gravitational force field,
  arguably the reflection of their intimate relationship with the filamentary spine of the cosmic web. Both are
  the product of the anisotropic force field emanating from the inhomogeneous mass distribution, i.e. from the
  tidal force field. The wall tidal field follows the large scale spatial outline seen in the filament tidal field,
  tracing the heavy arteries and connecting sheets in the cosmic web. Yet, it also reflects - more faintly - the
  small-scale cellular pattern seen in the void induced tidal field.
 \end{itemize}

\bigskip
While most of our analysis concerns the current cosmic matter distribution, a global statistical overview shows that
the overall conclusions concerning strength of the gravitational force and tidal forces generated by the various
cosmic web components pertains over a rather wide range of redshifts, at least from $z=4$ down to $z=0$. 

In summary, we find that filaments are the main drivers of the dynamical evolution of the cosmic web. This may not
be surprising, given filaments represent at least $\approx 50\%$ of the mass in the universe, concentrated in
rather compact elongated ridges pervading the entire universe. More revealing is the finding of the prominent role
of voids in the dynamics of the cosmic web. When looking at the cause of anisotropic structure, of which tidal fields
are the main agent, we find that voids play an important role. In other words, they can be seen as the main spatial organisers
of the mulitscale cosmic web.

\bigskip
This work has the intention to open the window on to the intricate interplay of the various gravitational influences
in the cosmic web. It is the starting point of a more extensive program. Several immediate and evident practical
and computational improvements of the current study are foreseen. These include a more efficient force and tidal tensor
computation that would enable a more detailed statistical study and the extension of our study to large cosmological
simulations including gas dynamics, baryonic physics and galaxy formation. More fundamental will be the astrophysical
and cosmological issues that should be addressed on the basis of the new insights gained from the present
inventory of cosmic web dynamics. Amongst a wide range of related issues, two prominent aspects are the
following. 

The most outstanding issue to be addressed is that of the dynamics and evolution of cosmic voids. In a follow-up
study the multiscale structure of the void force and tidal field, as well as the generated mass flows, will
be analysed armed with the insights obtained in the current study. The low-density environment in voids imply
the major influence of external gravitational and tidal influences. As demonstrated in the present study, for  
most void configurations it is essential to include the role of the surrounding filamentary cosmic web if we
seek to understand and exploit their structure and flow fields. This will alter our understanding of galaxy
formation and evolution within void environments, given for example the impact on matter accretion. It will
substantially impact on the use of voids as cosmological probes, given the fact that their internal structure
and dynamics will even more highly altered by the surrounding cosmic web structure than by cosmological
factors.

Another important observable effect of the cosmic web is that on the shape and rotation of galaxies. While it
has long been recognised that especially the strong tidal forces exerted by (proto)filaments are the principal source
of rotation for many galaxies, many recent studies have uncovered additional processes such as the
anisotropic accretion along filaments that may seriously affect the outcome. In a slightly different context,
the tidal forces induced by the cosmic web yield intrinsic alignments of gravitationally lensed background
galaxy images and their interpretation within the cosmological context. Given the current study's finding that the tidal
force field is highly sensitive to the small-scale aspects of the cosmic web, highlighting the presence of
small voids and walls that are observationally difficult to detect, it may be necessary to focus in more
detail on the intricate dynamical structure of the cosmic web and its effect on to compensate for its impact.

\section*{Acknowledgements}
We are grateful to Marius Cautun and Bernard Jones for their willingness to share the Nexus code
for this project, and to Joop Schaye for helpful suggestions. Many decades ago Vincent Icke taught RvdW about the importance of voids, the present study is a token of gratitude for this wise lesson. RvdW owes Dick Bond for emphasizing the key role of tidal forces to understand the formation of the cosmic web. Finally, RvdW also acknowledges Simon White for a highly motivating remark that inspired a major share of the current study. This work is partly funded by research programme Athena 184.034.002 from
the Dutch Research Council (NWO).

\section*{Data Availability}

All data presented in this paper will be shared upon reasonable request to the corresponding author.




\bibliographystyle{mnras}
\bibliography{bibliography} 

\begin{thebibliography}{}
\makeatletter
\relax
\def\mn@urlcharsother{\let\do\@makeother \do\$\do\&\do\#\do\^\do\_\do\%\do\~}
\def\mn@doi{\begingroup\mn@urlcharsother \@ifnextchar [ {\mn@doi@}
  {\mn@doi@[]}}
\def\mn@doi@[#1]#2{\def\@tempa{#1}\ifx\@tempa\@empty \href
  {http://dx.doi.org/#2} {doi:#2}\else \href {http://dx.doi.org/#2} {#1}\fi
  \endgroup}
\def\mn@eprint#1#2{\mn@eprint@#1:#2::\@nil}
\def\mn@eprint@arXiv#1{\href {http://arxiv.org/abs/#1} {{\tt arXiv:#1}}}
\def\mn@eprint@dblp#1{\href {http://dblp.uni-trier.de/rec/bibtex/#1.xml}
  {dblp:#1}}
\def\mn@eprint@#1:#2:#3:#4\@nil{\def\@tempa {#1}\def\@tempb {#2}\def\@tempc
  {#3}\ifx \@tempc \@empty \let \@tempc \@tempb \let \@tempb \@tempa \fi \ifx
  \@tempb \@empty \def\@tempb {arXiv}\fi \@ifundefined
  {mn@eprint@\@tempb}{\@tempb:\@tempc}{\expandafter \expandafter \csname
  mn@eprint@\@tempb\endcsname \expandafter{\@tempc}}}

\bibitem[\protect\citeauthoryear{{Abel}, {Hahn}  \& {Kaehler}}{{Abel}
  et~al.}{2012}]{Abel2012}
{Abel} T.,  {Hahn} O.,   {Kaehler} R.,  2012, \mn@doi [\mnras]
  {10.1111/j.1365-2966.2012.21754.x}, \href
  {https://ui.adsabs.harvard.edu/abs/2012MNRAS.427...61A} {427, 61}

\bibitem[\protect\citeauthoryear{{Alam}, {Paranjape}  \& {Peacock}}{{Alam}
  et~al.}{2024}]{alam2024}
{Alam} S.,  {Paranjape} A.,   {Peacock} J.~A.,  2024, \mn@doi [\mnras]
  {10.1093/mnras/stad3423}, \href
  {https://ui.adsabs.harvard.edu/abs/2024MNRAS.527.3771A} {527, 3771}

\bibitem[\protect\citeauthoryear{{Alpaslan} et~al.,}{{Alpaslan}
  et~al.}{2014}]{Alpaslan2014}
{Alpaslan} M.,  et~al., 2014, \mn@doi [\mnras] {10.1093/mnras/stt2136}, \href
  {https://ui.adsabs.harvard.edu/abs/2014MNRAS.438..177A} {438, 177}

\bibitem[\protect\citeauthoryear{{Angulo}, {Zennaro}, {Contreras}, {Aric{\`o}},
  {Pellejero-Iba{\~n}ez}  \& {St{\"u}cker}}{{Angulo} et~al.}{2021}]{Angulo2021}
{Angulo} R.~E.,  {Zennaro} M.,  {Contreras} S.,  {Aric{\`o}} G.,
  {Pellejero-Iba{\~n}ez} M.,   {St{\"u}cker} J.,  2021, \mn@doi [\mnras]
  {10.1093/mnras/stab2018}, \href
  {https://ui.adsabs.harvard.edu/abs/2021MNRAS.507.5869A} {507, 5869}

\bibitem[\protect\citeauthoryear{{Aragon-Calvo} \& {Szalay}}{{Aragon-Calvo} \&
  {Szalay}}{2013}]{aragon2013}
{Aragon-Calvo} M.~A.,  {Szalay} A.~S.,  2013, \mn@doi [\mnras]
  {10.1093/mnras/sts281}, \href
  {https://ui.adsabs.harvard.edu/abs/2013MNRAS.428.3409A} {428, 3409}

\bibitem[\protect\citeauthoryear{{Aragon-Calvo} \& {Yang}}{{Aragon-Calvo} \&
  {Yang}}{2014}]{Aragon2014}
{Aragon-Calvo} M.~A.,  {Yang} L.~F.,  2014, \mn@doi [\mnras]
  {10.1093/mnrasl/slu009}, \href
  {https://ui.adsabs.harvard.edu/abs/2014MNRAS.440L..46A} {440, L46}

\bibitem[\protect\citeauthoryear{{Arag{\'o}n-Calvo}, {Jones}, {van de Weygaert}
   \& {van der Hulst}}{{Arag{\'o}n-Calvo} et~al.}{2007a}]{MMFb}
{Arag{\'o}n-Calvo} M.~A.,  {Jones} B.~J.~T.,  {van de Weygaert} R.,   {van der
  Hulst} J.~M.,  2007a, \mn@doi [\aap] {10.1051/0004-6361:20077880}, \href
  {https://ui.adsabs.harvard.edu/abs/2007A&A...474..315A} {474, 315}

\bibitem[\protect\citeauthoryear{{Arag{\'o}n-Calvo}, {van de Weygaert}, {Jones}
   \& {van der Hulst}}{{Arag{\'o}n-Calvo} et~al.}{2007b}]{MMFa}
{Arag{\'o}n-Calvo} M.~A.,  {van de Weygaert} R.,  {Jones} B. J.~T.,   {van der
  Hulst} J.~M.,  2007b, \mn@doi [\apjl] {10.1086/511633}, \href
  {https://ui.adsabs.harvard.edu/abs/2007ApJ...655L...5A} {655, L5}

\bibitem[\protect\citeauthoryear{{Arag{\'o}n-Calvo}, {van de Weygaert}  \&
  {Jones}}{{Arag{\'o}n-Calvo} et~al.}{2010}]{MMF3}
{Arag{\'o}n-Calvo} M.~A.,  {van de Weygaert} R.,   {Jones} B. J.~T.,  2010,
  \mn@doi [\mnras] {10.1111/j.1365-2966.2010.17263.x}, \href
  {https://ui.adsabs.harvard.edu/abs/2010MNRAS.408.2163A} {408, 2163}

\bibitem[\protect\citeauthoryear{{Awad} et~al.,}{{Awad}
  et~al.}{2023}]{Awad2023}
{Awad} P.,  et~al., 2023, \mn@doi [\mnras] {10.1093/mnras/stad428}, \href
  {https://ui.adsabs.harvard.edu/abs/2023MNRAS.520.4517A} {520, 4517}

\bibitem[\protect\citeauthoryear{{Bernardeau}, {van de Weygaert}, {Hivon}  \&
  {Bouchet}}{{Bernardeau} et~al.}{1997}]{bernwey1997}
{Bernardeau} F.,  {van de Weygaert} R.,  {Hivon} E.,   {Bouchet} F.~R.,  1997,
  \mn@doi [\mnras] {10.1093/mnras/290.3.566}, \href
  {https://ui.adsabs.harvard.edu/abs/1997MNRAS.290..566B} {290, 566}

\bibitem[\protect\citeauthoryear{{Bocquet}, {Heitmann}, {Habib}, {Lawrence},
  {Uram}, {Frontiere}, {Pope}  \& {Finkel}}{{Bocquet}
  et~al.}{2020}]{Miratitan2020}
{Bocquet} S.,  {Heitmann} K.,  {Habib} S.,  {Lawrence} E.,  {Uram} T.,
  {Frontiere} N.,  {Pope} A.,   {Finkel} H.,  2020, \mn@doi [\apj]
  {10.3847/1538-4357/abac5c}, \href
  {https://ui.adsabs.harvard.edu/abs/2020ApJ...901....5B} {901, 5}

\bibitem[\protect\citeauthoryear{{Bond}, {Kofman}  \& {Pogosyan}}{{Bond}
  et~al.}{1996}]{bond1996}
{Bond} J.~R.,  {Kofman} L.,   {Pogosyan} D.,  1996, \mn@doi [\nat]
  {10.1038/380603a0}, \href
  {https://ui.adsabs.harvard.edu/abs/1996Natur.380..603B} {380, 603}

\bibitem[\protect\citeauthoryear{{Bonjean}, {Aghanim}, {Salom{\'e}}, {Douspis}
  \& {Beelen}}{{Bonjean} et~al.}{2018}]{bonjean2018}
{Bonjean} V.,  {Aghanim} N.,  {Salom{\'e}} P.,  {Douspis} M.,   {Beelen} A.,
  2018, \mn@doi [\aap] {10.1051/0004-6361/201731699}, \href
  {https://ui.adsabs.harvard.edu/abs/2018A&A...609A..49B} {609, A49}

\bibitem[\protect\citeauthoryear{Bos}{Bos}{2016}]{patrickthesis}
Bos E.,  2016, PhD thesis, University of Groningen

\bibitem[\protect\citeauthoryear{{Bos}, {van de Weygaert}, {Dolag}  \&
  {Pettorino}}{{Bos} et~al.}{2012}]{bos2012}
{Bos} E.~G.~P.,  {van de Weygaert} R.,  {Dolag} K.,   {Pettorino} V.,  2012,
  \mn@doi [\mnras] {10.1111/j.1365-2966.2012.21478.x}, \href
  {https://ui.adsabs.harvard.edu/abs/2012MNRAS.426..440B} {426, 440}

\bibitem[\protect\citeauthoryear{{Cautun} \& {van de Weygaert}}{{Cautun} \&
  {van de Weygaert}}{2011}]{Cautun2011}
{Cautun} M.~C.,  {van de Weygaert} R.,  2011, {The DTFE public software: The
  Delaunay Tessellation Field Estimator code}

\bibitem[\protect\citeauthoryear{{Cautun}, {van de Weygaert}  \&
  {Jones}}{{Cautun} et~al.}{2013}]{NEXUS_MAIN}
{Cautun} M.,  {van de Weygaert} R.,   {Jones} B. J.~T.,  2013, \mn@doi [\mnras]
  {10.1093/mnras/sts416}, \href
  {https://ui.adsabs.harvard.edu/abs/2013MNRAS.429.1286C} {429, 1286}

\bibitem[\protect\citeauthoryear{{Cautun}, {van de Weygaert}, {Jones}  \&
  {Frenk}}{{Cautun} et~al.}{2014}]{NEXUS_CW_EVO}
{Cautun} M.,  {van de Weygaert} R.,  {Jones} B. J.~T.,   {Frenk} C.~S.,  2014,
  \mn@doi [\mnras] {10.1093/mnras/stu768}, \href
  {https://ui.adsabs.harvard.edu/abs/2014MNRAS.441.2923C} {441, 2923}

\bibitem[\protect\citeauthoryear{{Cen}}{{Cen}}{1997}]{cen1997}
{Cen} R.,  1997, \mn@doi [\apjl] {10.1086/310587}, \href
  {https://ui.adsabs.harvard.edu/abs/1997ApJ...479L..85C} {479, L85}

\bibitem[\protect\citeauthoryear{{Codis}, {Pichon}, {Devriendt}, {Slyz},
  {Pogosyan}, {Dubois}  \& {Sousbie}}{{Codis} et~al.}{2012}]{codis2012}
{Codis} S.,  {Pichon} C.,  {Devriendt} J.,  {Slyz} A.,  {Pogosyan} D.,
  {Dubois} Y.,   {Sousbie} T.,  2012, \mn@doi [\mnras]
  {10.1111/j.1365-2966.2012.21636.x}, \href
  {https://ui.adsabs.harvard.edu/abs/2012MNRAS.427.3320C} {427, 3320}

\bibitem[\protect\citeauthoryear{{Codis}, {Pichon}  \& {Pogosyan}}{{Codis}
  et~al.}{2015}]{codis2015}
{Codis} S.,  {Pichon} C.,   {Pogosyan} D.,  2015, \mn@doi [\mnras]
  {10.1093/mnras/stv1570}, \href
  {https://ui.adsabs.harvard.edu/abs/2015MNRAS.452.3369C} {452, 3369}

\bibitem[\protect\citeauthoryear{{Colless} et~al.,}{{Colless}
  et~al.}{2003}]{colless2003}
{Colless} M.,  et~al., 2003, \mn@doi [arXiv e-prints]
  {10.48550/arXiv.astro-ph/0306581}, \href
  {https://ui.adsabs.harvard.edu/abs/2003astro.ph..6581C} {pp
  astro--ph/0306581}

\bibitem[\protect\citeauthoryear{{Courtois} et~al.,}{{Courtois}
  et~al.}{2023}]{courtois2023}
{Courtois} H.~M.,  et~al., 2023, \mn@doi [\aap] {10.1051/0004-6361/202245578},
  \href {https://ui.adsabs.harvard.edu/abs/2023A&A...673A..38C} {673, A38}

\bibitem[\protect\citeauthoryear{{Dietrich}, {Werner}, {Clowe}, {Finoguenov},
  {Kitching}, {Miller}  \& {Simionescu}}{{Dietrich}
  et~al.}{2012}]{dietrich2012}
{Dietrich} J.~P.,  {Werner} N.,  {Clowe} D.,  {Finoguenov} A.,  {Kitching} T.,
  {Miller} L.,   {Simionescu} A.,  2012, \mn@doi [\nat] {10.1038/nature11224},
  \href {https://ui.adsabs.harvard.edu/abs/2012Natur.487..202D} {487, 202}

\bibitem[\protect\citeauthoryear{{Efstathiou} \& {Jones}}{{Efstathiou} \&
  {Jones}}{1980}]{efstathiou1980}
{Efstathiou} G.,  {Jones} B.~J.~T.,  1980, Comments on Astrophysics, \href
  {https://ui.adsabs.harvard.edu/abs/1980ComAp...8..169E} {8, 169}

\bibitem[\protect\citeauthoryear{{Einasto}}{{Einasto}}{1977}]{einasto1977}
{Einasto} J.,  1977, in Problems of Observational and Theoretical Astronomy. pp
  26--43

\bibitem[\protect\citeauthoryear{{Elek}, {Burchett}, {Prochaska}  \&
  {Forbes}}{{Elek} et~al.}{2020}]{Elek2020}
{Elek} O.,  {Burchett} J.~N.,  {Prochaska} J.~X.,   {Forbes} A.~G.,  2020,
  \mn@doi [arXiv e-prints] {10.48550/arXiv.2009.02441}, \href
  {https://ui.adsabs.harvard.edu/abs/2020arXiv200902441E} {p. arXiv:2009.02441}

\bibitem[\protect\citeauthoryear{{Elek}, {Burchett}, {Prochaska}  \&
  {Forbes}}{{Elek} et~al.}{2022}]{Elek2022}
{Elek} O.,  {Burchett} J.~N.,  {Prochaska} J.~X.,   {Forbes} A.~G.,  2022,
  \mn@doi [arXiv e-prints] {10.48550/arXiv.2204.01256}, \href
  {https://ui.adsabs.harvard.edu/abs/2022arXiv220401256E} {p. arXiv:2204.01256}

\bibitem[\protect\citeauthoryear{{Feldbrugge} \& {van de
  Weygaert}}{{Feldbrugge} \& {van de Weygaert}}{2023}]{Feldbrugge2023}
{Feldbrugge} J.,  {van de Weygaert} R.,  2023, \mn@doi [\jcap]
  {10.1088/1475-7516/2023/02/058}, \href
  {https://ui.adsabs.harvard.edu/abs/2023JCAP...02..058F} {2023, 058}

\bibitem[\protect\citeauthoryear{{Feldbrugge} \& {van de
  Weygaert}}{{Feldbrugge} \& {van de Weygaert}}{2024}]{Feldbrugge2024}
{Feldbrugge} J.,  {van de Weygaert} R.,  2024, \mn@doi [arXiv e-prints]
  {10.48550/arXiv.2405.20475}, \href
  {https://ui.adsabs.harvard.edu/abs/2024arXiv240520475F} {p. arXiv:2405.20475}

\bibitem[\protect\citeauthoryear{{Feldbrugge}, {van de Weygaert}, {Hidding}  \&
  {Feldbrugge}}{{Feldbrugge} et~al.}{2018b}]{Feldbrugge2018}
{Feldbrugge} J.,  {van de Weygaert} R.,  {Hidding} J.,   {Feldbrugge} J.,
  2018b, \mn@doi [\jcap] {10.1088/1475-7516/2018/05/027}, \href
  {https://ui.adsabs.harvard.edu/abs/2018JCAP...05..027F} {2018, 027}

\bibitem[\protect\citeauthoryear{{Feldbrugge}, {van de Weygaert}, {Hidding}  \&
  {Feldbrugge}}{{Feldbrugge} et~al.}{2018a}]{job2018Caustics}
{Feldbrugge} J.,  {van de Weygaert} R.,  {Hidding} J.,   {Feldbrugge} J.,
  2018a, \mn@doi [\jcap] {10.1088/1475-7516/2018/05/027}, \href
  {https://ui.adsabs.harvard.edu/abs/2018JCAP...05..027F} {2018, 027}

\bibitem[\protect\citeauthoryear{{Feldbrugge}, {Yan}  \& {van de
  Weygaert}}{{Feldbrugge} et~al.}{2023}]{Feldbrugge2023b}
{Feldbrugge} J.,  {Yan} Y.,   {van de Weygaert} R.,  2023, \mn@doi [\mnras]
  {10.1093/mnras/stad2777}, \href
  {https://ui.adsabs.harvard.edu/abs/2023MNRAS.526.5031F} {526, 5031}

\bibitem[\protect\citeauthoryear{{Florack}, {ter Haar Romeny}, {Koenderink}  \&
  {Viergever}}{{Florack} et~al.}{1992}]{florack1992}
{Florack} L.,  {ter Haar Romeny} B.,  {Koenderink} J.,   {Viergever} M.,  1992,
  Image and Vision Computing, 10, 376

\bibitem[\protect\citeauthoryear{{Forero-Romero}, {Hoffman}, {Gottl{\"o}ber},
  {Klypin}  \& {Yepes}}{{Forero-Romero} et~al.}{2009}]{foreroromero2009}
{Forero-Romero} J.~E.,  {Hoffman} Y.,  {Gottl{\"o}ber} S.,  {Klypin} A.,
  {Yepes} G.,  2009, \mn@doi [\mnras] {10.1111/j.1365-2966.2009.14885.x}, \href
  {https://ui.adsabs.harvard.edu/abs/2009MNRAS.396.1815F} {396, 1815}

\bibitem[\protect\citeauthoryear{{Frieman}, {Turner}  \& {Huterer}}{{Frieman}
  et~al.}{2008}]{DEReview2008}
{Frieman} J.~A.,  {Turner} M.~S.,   {Huterer} D.,  2008, \mn@doi [\araa]
  {10.1146/annurev.astro.46.060407.145243}, \href
  {https://ui.adsabs.harvard.edu/abs/2008ARA&A..46..385F} {46, 385}

\bibitem[\protect\citeauthoryear{{Ganeshaiah Veena}, {Cautun}, {van de
  Weygaert}, {Tempel}, {Jones}, {Rieder}  \& {Frenk}}{{Ganeshaiah Veena}
  et~al.}{2018}]{Punya2018}
{Ganeshaiah Veena} P.,  {Cautun} M.,  {van de Weygaert} R.,  {Tempel} E.,
  {Jones} B. J.~T.,  {Rieder} S.,   {Frenk} C.~S.,  2018, \mn@doi [\mnras]
  {10.1093/mnras/sty2270}, \href
  {https://ui.adsabs.harvard.edu/abs/2018MNRAS.481..414G} {481, 414}

\bibitem[\protect\citeauthoryear{{Ganeshaiah Veena}, {Cautun}, {Tempel}, {van
  de Weygaert}  \& {Frenk}}{{Ganeshaiah Veena} et~al.}{2019}]{Punya2019TTT}
{Ganeshaiah Veena} P.,  {Cautun} M.,  {Tempel} E.,  {van de Weygaert} R.,
  {Frenk} C.~S.,  2019, \mn@doi [\mnras] {10.1093/mnras/stz1343}, \href
  {https://ui.adsabs.harvard.edu/abs/2019MNRAS.487.1607G} {487, 1607}

\bibitem[\protect\citeauthoryear{{Ganeshaiah Veena}, {Cautun}, {van de
  Weygaert}, {Tempel}  \& {Frenk}}{{Ganeshaiah Veena}
  et~al.}{2021}]{ganeshaiah2021}
{Ganeshaiah Veena} P.,  {Cautun} M.,  {van de Weygaert} R.,  {Tempel} E.,
  {Frenk} C.~S.,  2021, \mn@doi [\mnras] {10.1093/mnras/stab411}, \href
  {https://ui.adsabs.harvard.edu/abs/2021MNRAS.503.2280G} {503, 2280}

\bibitem[\protect\citeauthoryear{{Hahn}, {Carollo}, {Porciani}  \&
  {Dekel}}{{Hahn} et~al.}{2007}]{hahn2007}
{Hahn} O.,  {Carollo} C.~M.,  {Porciani} C.,   {Dekel} A.,  2007, \mn@doi
  [\mnras] {10.1111/j.1365-2966.2007.12249.x}, \href
  {https://ui.adsabs.harvard.edu/abs/2007MNRAS.381...41H} {381, 41}

\bibitem[\protect\citeauthoryear{{Hahn}, {Teyssier}  \& {Carollo}}{{Hahn}
  et~al.}{2010}]{hahn2010}
{Hahn} O.,  {Teyssier} R.,   {Carollo} C.~M.,  2010, \mn@doi [\mnras]
  {10.1111/j.1365-2966.2010.16494.x}, \href
  {https://ui.adsabs.harvard.edu/abs/2010MNRAS.405..274H} {405, 274}

\bibitem[\protect\citeauthoryear{{Hidding}, {van de Weygaert}, {Vegter},
  {Jones}  \& {Teillaud}}{{Hidding} et~al.}{2012}]{hidding2012Adhesion}
{Hidding} J.,  {van de Weygaert} R.,  {Vegter} G.,  {Jones} B. J.~T.,
  {Teillaud} M.,  2012, arXiv e-prints, \href
  {https://ui.adsabs.harvard.edu/abs/2012arXiv1205.1669H} {p. arXiv:1205.1669}

\bibitem[\protect\citeauthoryear{{Hidding}, {van de Weygaert}  \&
  {Shandarin}}{{Hidding} et~al.}{2016}]{hidding2016Adhesion}
{Hidding} J.,  {van de Weygaert} R.,   {Shandarin} S.,  2016, in {van de
  Weygaert} R.,  {Shandarin} S.,  {Saar} E.,   {Einasto} J.,  eds,  IAU
  Symposium Vol. 308, The Zeldovich Universe: Genesis and Growth of the Cosmic
  Web. pp 69--76 (\mn@eprint {arXiv} {1611.01221}),
  \mn@doi{10.1017/S1743921316009650}

\bibitem[\protect\citeauthoryear{{Hirschmann}, {Dolag}, {Saro}, {Bachmann},
  {Borgani}  \& {Burkert}}{{Hirschmann} et~al.}{2014}]{Magneticum2014}
{Hirschmann} M.,  {Dolag} K.,  {Saro} A.,  {Bachmann} L.,  {Borgani} S.,
  {Burkert} A.,  2014, \mn@doi [\mnras] {10.1093/mnras/stu1023}, \href
  {https://ui.adsabs.harvard.edu/abs/2014MNRAS.442.2304H} {442, 2304}

\bibitem[\protect\citeauthoryear{{Hoffman}, {Metuki}, {Yepes}, {Gottl{\"o}ber},
  {Forero-Romero}, {Libeskind}  \& {Knebe}}{{Hoffman}
  et~al.}{2012}]{hoffman2012}
{Hoffman} Y.,  {Metuki} O.,  {Yepes} G.,  {Gottl{\"o}ber} S.,  {Forero-Romero}
  J.~E.,  {Libeskind} N.~I.,   {Knebe} A.,  2012, \mn@doi [\mnras]
  {10.1111/j.1365-2966.2012.21553.x}, \href
  {https://ui.adsabs.harvard.edu/abs/2012MNRAS.425.2049H} {425, 2049}

\bibitem[\protect\citeauthoryear{{Hoyle}}{{Hoyle}}{1949}]{Hoyle1949}
{Hoyle} F.,  1949.
eds. {{Burgers}, J.M. and {van de Hulst}, H.C.} (Central Air Documents Office,
  Dayton)

\bibitem[\protect\citeauthoryear{Huchra et~al.,}{Huchra
  et~al.}{2012}]{huchra20122mass}
Huchra J.~P.,  et~al., 2012, \mn@doi [The Astrophysical Journal Supplement
  Series] {10.1088/0067-0049/199/2/26}, 199, 26

\bibitem[\protect\citeauthoryear{Hunter}{Hunter}{2007}]{matplotlib}
Hunter J.~D.,  2007, \mn@doi [Computing in Science \& Engineering]
  {10.1109/MCSE.2007.55}, 9, 90

\bibitem[\protect\citeauthoryear{{Icke}}{{Icke}}{1984}]{icke1984}
{Icke} V.,  1984, \mn@doi [\mnras] {10.1093/mnras/206.1.1P}, \href
  {https://ui.adsabs.harvard.edu/abs/1984MNRAS.206P...1I} {206, 1P}

\bibitem[\protect\citeauthoryear{{Jones} \& {van de Weygaert}}{{Jones} \& {van
  de Weygaert}}{2009}]{Jones2009TTT}
{Jones} B.,  {van de Weygaert} R.,  2009, \mn@doi [Astrophysics and Space
  Science Proceedings] {10.1007/978-3-540-75826-6_48}, \href
  {https://ui.adsabs.harvard.edu/abs/2009ASSP....8..467J} {8, 467}

\bibitem[\protect\citeauthoryear{{Jones}, {van de Weygaert}  \&
  {Arag{\'o}n-Calvo}}{{Jones} et~al.}{2010}]{Jones2010_allign}
{Jones} B. J.~T.,  {van de Weygaert} R.,   {Arag{\'o}n-Calvo} M.~A.,  2010,
  \mn@doi [\mnras] {10.1111/j.1365-2966.2010.17202.x}, \href
  {https://ui.adsabs.harvard.edu/abs/2010MNRAS.408..897J} {408, 897}

\bibitem[\protect\citeauthoryear{{Kitaura}, {Angulo}, {Hoffman}  \&
  {Gottl{\"o}ber}}{{Kitaura} et~al.}{2012a}]{kitaura2012b}
{Kitaura} F.-S.,  {Angulo} R.~E.,  {Hoffman} Y.,   {Gottl{\"o}ber} S.,  2012a,
  \mn@doi [\mnras] {10.1111/j.1365-2966.2012.21589.x}, \href
  {https://ui.adsabs.harvard.edu/abs/2012MNRAS.425.2422K} {425, 2422}

\bibitem[\protect\citeauthoryear{{Kitaura}, {Erdo{\v{g}}du}, {Nuza},
  {Khalatyan}, {Angulo}, {Hoffman}  \& {Gottl{\"o}ber}}{{Kitaura}
  et~al.}{2012b}]{kitaura2012}
{Kitaura} F.-S.,  {Erdo{\v{g}}du} P.,  {Nuza} S.~E.,  {Khalatyan} A.,  {Angulo}
  R.~E.,  {Hoffman} Y.,   {Gottl{\"o}ber} S.,  2012b, \mn@doi [\mnras]
  {10.1111/j.1745-3933.2012.01340.x}, \href
  {https://ui.adsabs.harvard.edu/abs/2012MNRAS.427L..35K} {427, L35}

\bibitem[\protect\citeauthoryear{{Kourkchi} et~al.,}{{Kourkchi}
  et~al.}{2020}]{Kourkchi2020}
{Kourkchi} E.,  et~al., 2020, \mn@doi [\apj] {10.3847/1538-4357/abb66b}, \href
  {https://ui.adsabs.harvard.edu/abs/2020ApJ...902..145K} {902, 145}

\bibitem[\protect\citeauthoryear{{Lavaux} \& {Wandelt}}{{Lavaux} \&
  {Wandelt}}{2012}]{lavaux2012}
{Lavaux} G.,  {Wandelt} B.~D.,  2012, \mn@doi [\apj]
  {10.1088/0004-637X/754/2/109}, \href
  {https://ui.adsabs.harvard.edu/abs/2012ApJ...754..109L} {754, 109}

\bibitem[\protect\citeauthoryear{{Lee} \& {Park}}{{Lee} \&
  {Park}}{2009}]{leepark2009}
{Lee} J.,  {Park} D.,  2009, \mn@doi [\apjl] {10.1088/0004-637X/696/1/L10},
  \href {https://ui.adsabs.harvard.edu/abs/2009ApJ...696L..10L} {696, L10}

\bibitem[\protect\citeauthoryear{{Lee} \& {Pen}}{{Lee} \&
  {Pen}}{2000}]{LeePen2000ttt}
{Lee} J.,  {Pen} U.-L.,  2000, \mn@doi [\apjl] {10.1086/312556}, \href
  {https://ui.adsabs.harvard.edu/abs/2000ApJ...532L...5L} {532, L5}

\bibitem[\protect\citeauthoryear{{Lee} \& {Springel}}{{Lee} \&
  {Springel}}{2010}]{lee2010}
{Lee} J.,  {Springel} V.,  2010, \mn@doi [\jcap]
  {10.1088/1475-7516/2010/05/031}, \href
  {https://ui.adsabs.harvard.edu/abs/2010JCAP...05..031L} {2010, 031}

\bibitem[\protect\citeauthoryear{{Lee}, {Hahn}  \& {Porciani}}{{Lee}
  et~al.}{2009}]{lee2009}
{Lee} J.,  {Hahn} O.,   {Porciani} C.,  2009, \mn@doi [\apj]
  {10.1088/0004-637X/705/2/1469}, \href
  {https://ui.adsabs.harvard.edu/abs/2009ApJ...705.1469L} {705, 1469}

\bibitem[\protect\citeauthoryear{{Lee} et~al.,}{{Lee}
  et~al.}{2018}]{khanlee2018}
{Lee} K.~G.,  et~al., 2018, VizieR Online Data Catalog, \href
  {https://ui.adsabs.harvard.edu/abs/2018yCat..22370031L} {p. J/ApJS/237/31}

\bibitem[\protect\citeauthoryear{{Libeskind} et~al.,}{{Libeskind}
  et~al.}{2018}]{Libeskind2018}
{Libeskind} N.~I.,  et~al., 2018, \mn@doi [\mnras] {10.1093/mnras/stx1976},
  \href {https://ui.adsabs.harvard.edu/abs/2018MNRAS.473.1195L} {473, 1195}

\bibitem[\protect\citeauthoryear{{Lindeberg}}{{Lindeberg}}{1994}]{Lindeberg1994}
{Lindeberg} T.,  1994, \mn@doi [Journal of Applied Statistics]
  {10.1080/757582976}, \href
  {https://ui.adsabs.harvard.edu/abs/1994JApSt..21..225L} {21, 225}

\bibitem[\protect\citeauthoryear{{Liske} et~al.,}{{Liske}
  et~al.}{2015}]{GAMA2015}
{Liske} J.,  et~al., 2015, \mn@doi [\mnras] {10.1093/mnras/stv1436}, \href
  {https://ui.adsabs.harvard.edu/abs/2015MNRAS.452.2087L} {452, 2087}

\bibitem[\protect\citeauthoryear{{L{\'o}pez}, {Cautun}, {Paz}, {Merch{\'a}n}
  \& {van de Weygaert}}{{L{\'o}pez} et~al.}{2021}]{lopez2021}
{L{\'o}pez} P.,  {Cautun} M.,  {Paz} D.,  {Merch{\'a}n} M.,   {van de Weygaert}
  R.,  2021, \mn@doi [\mnras] {10.1093/mnras/stab451}, \href
  {https://ui.adsabs.harvard.edu/abs/2021MNRAS.502.5528L} {502, 5528}

\bibitem[\protect\citeauthoryear{{Macquart} et~al.,}{{Macquart}
  et~al.}{2020}]{IGMNatureMacquart2020}
{Macquart} J.~P.,  et~al., 2020, \mn@doi [\nat] {10.1038/s41586-020-2300-2},
  \href {https://ui.adsabs.harvard.edu/abs/2020Natur.581..391M} {581, 391}

\bibitem[\protect\citeauthoryear{{McCarthy}, {Schaye}, {Bird}  \& {Le
  Brun}}{{McCarthy} et~al.}{2017}]{BAHAMAS2017}
{McCarthy} I.~G.,  {Schaye} J.,  {Bird} S.,   {Le Brun} A. M.~C.,  2017,
  \mn@doi [\mnras] {10.1093/mnras/stw2792}, \href
  {https://ui.adsabs.harvard.edu/abs/2017MNRAS.465.2936M} {465, 2936}

\bibitem[\protect\citeauthoryear{{Meiksin}}{{Meiksin}}{2009}]{meiksin2009}
{Meiksin} A.~A.,  2009, \mn@doi [Reviews of Modern Physics]
  {10.1103/RevModPhys.81.1405}, \href
  {https://ui.adsabs.harvard.edu/abs/2009RvMP...81.1405M} {81, 1405}

\bibitem[\protect\citeauthoryear{{Neyrinck} \& {Shandarin}}{{Neyrinck} \&
  {Shandarin}}{2012}]{Neyrinck2012}
{Neyrinck} M.~C.,  {Shandarin} S.~F.,  2012, \mn@doi [arXiv e-prints]
  {10.48550/arXiv.1207.4501}, \href
  {https://ui.adsabs.harvard.edu/abs/2012arXiv1207.4501N} {p. arXiv:1207.4501}

\bibitem[\protect\citeauthoryear{{Nicastro} et~al.,}{{Nicastro}
  et~al.}{2018}]{IGM2018}
{Nicastro} F.,  et~al., 2018, \mn@doi [\nat] {10.1038/s41586-018-0204-1}, \href
  {https://ui.adsabs.harvard.edu/abs/2018Natur.558..406N} {558, 406}

\bibitem[\protect\citeauthoryear{{Ostriker} \& {Cen}}{{Ostriker} \&
  {Cen}}{1996}]{cen1996}
{Ostriker} J.~P.,  {Cen} R.,  1996, \mn@doi [\apj] {10.1086/177297}, \href
  {https://ui.adsabs.harvard.edu/abs/1996ApJ...464...27O} {464, 27}

\bibitem[\protect\citeauthoryear{{Paranjape}}{{Paranjape}}{2021}]{paranjape2021}
{Paranjape} A.,  2021, \mn@doi [\mnras] {10.1093/mnras/stab359}, \href
  {https://ui.adsabs.harvard.edu/abs/2021MNRAS.502.5210P} {502, 5210}

\bibitem[\protect\citeauthoryear{{Park} \& {Lee}}{{Park} \&
  {Lee}}{2007}]{parklee2007}
{Park} D.,  {Lee} J.,  2007, \mn@doi [\prl] {10.1103/PhysRevLett.98.081301},
  \href {https://ui.adsabs.harvard.edu/abs/2007PhRvL..98h1301P} {98, 081301}

\bibitem[\protect\citeauthoryear{{Peebles}}{{Peebles}}{1969}]{Peebles69TTT}
{Peebles} P.~J.~E.,  1969, \mn@doi [\apj] {10.1086/149876}, \href
  {https://ui.adsabs.harvard.edu/abs/1969ApJ...155..393P} {155, 393}

\bibitem[\protect\citeauthoryear{{Peebles}}{{Peebles}}{1980}]{Peebles1980}
{Peebles} P.~J.~E.,  1980, {The large-scale structure of the universe}

\bibitem[\protect\citeauthoryear{{Pichon} \& {Bernardeau}}{{Pichon} \&
  {Bernardeau}}{1999}]{pichon2000}
{Pichon} C.,  {Bernardeau} F.,  1999, \mn@doi [\aap]
  {10.48550/arXiv.astro-ph/9902142}, \href
  {https://ui.adsabs.harvard.edu/abs/1999A&A...343..663P} {343, 663}

\bibitem[\protect\citeauthoryear{{Pillepich} et~al.,}{{Pillepich}
  et~al.}{2018}]{Illustris2018}
{Pillepich} A.,  et~al., 2018, \mn@doi [\mnras] {10.1093/mnras/stx2656}, \href
  {https://ui.adsabs.harvard.edu/abs/2018MNRAS.473.4077P} {473, 4077}

\bibitem[\protect\citeauthoryear{{Pisani} et~al.,}{{Pisani}
  et~al.}{2019}]{pisani2019}
{Pisani} A.,  et~al., 2019, \mn@doi [\baas] {10.48550/arXiv.1903.05161}, \href
  {https://ui.adsabs.harvard.edu/abs/2019BAAS...51c..40P} {51, 40}

\bibitem[\protect\citeauthoryear{{Platen}, {van de Weygaert}  \&
  {Jones}}{{Platen} et~al.}{2008}]{platen2008}
{Platen} E.,  {van de Weygaert} R.,   {Jones} B. J.~T.,  2008, \mn@doi [\mnras]
  {10.1111/j.1365-2966.2008.13019.x}, \href
  {https://ui.adsabs.harvard.edu/abs/2008MNRAS.387..128P} {387, 128}

\bibitem[\protect\citeauthoryear{{Porciani}, {Dekel}  \& {Hoffman}}{{Porciani}
  et~al.}{2002a}]{porciani2002a}
{Porciani} C.,  {Dekel} A.,   {Hoffman} Y.,  2002a, \mn@doi [\mnras]
  {10.1046/j.1365-8711.2002.05305.x}, \href
  {https://ui.adsabs.harvard.edu/abs/2002MNRAS.332..325P} {332, 325}

\bibitem[\protect\citeauthoryear{{Porciani}, {Dekel}  \& {Hoffman}}{{Porciani}
  et~al.}{2002b}]{porciani2002b}
{Porciani} C.,  {Dekel} A.,   {Hoffman} Y.,  2002b, \mn@doi [\mnras]
  {10.1046/j.1365-8711.2002.05306.x}, \href
  {https://ui.adsabs.harvard.edu/abs/2002MNRAS.332..339P} {332, 339}

\bibitem[\protect\citeauthoryear{{Romano-D{\'\i}az} \& {van de
  Weygaert}}{{Romano-D{\'\i}az} \& {van de Weygaert}}{2007}]{romanowey2007}
{Romano-D{\'\i}az} E.,  {van de Weygaert} R.,  2007, \mn@doi [\mnras]
  {10.1111/j.1365-2966.2007.12190.x}, \href
  {https://ui.adsabs.harvard.edu/abs/2007MNRAS.382....2R} {382, 2}

\bibitem[\protect\citeauthoryear{{Schaap} \& {van de Weygaert}}{{Schaap} \&
  {van de Weygaert}}{2000}]{schaap2000}
{Schaap} W.~E.,  {van de Weygaert} R.,  2000, \mn@doi [\aap]
  {10.48550/arXiv.astro-ph/0011007}, \href
  {https://ui.adsabs.harvard.edu/abs/2000A&A...363L..29S} {363, L29}

\bibitem[\protect\citeauthoryear{{Sch{\"a}fer}}{{Sch{\"a}fer}}{2009}]{schaeffer2008}
{Sch{\"a}fer} B.~M.,  2009, \mn@doi [International Journal of Modern Physics D]
  {10.1142/S0218271809014388}, \href
  {https://ui.adsabs.harvard.edu/abs/2009IJMPD..18..173S} {18, 173}

\bibitem[\protect\citeauthoryear{{Schaye} et~al.,}{{Schaye}
  et~al.}{2015}]{Eaglemain}
{Schaye} J.,  et~al., 2015, \mn@doi [\mnras] {10.1093/mnras/stu2058}, \href
  {https://ui.adsabs.harvard.edu/abs/2015MNRAS.446..521S} {446, 521}

\bibitem[\protect\citeauthoryear{{Schaye} et~al.,}{{Schaye}
  et~al.}{2023}]{FLAMINGO2023}
{Schaye} J.,  et~al., 2023, \mn@doi [arXiv e-prints]
  {10.48550/arXiv.2306.04024}, \href
  {https://ui.adsabs.harvard.edu/abs/2023arXiv230604024S} {p. arXiv:2306.04024}

\bibitem[\protect\citeauthoryear{{Schlafly} et~al.,}{{Schlafly}
  et~al.}{2023}]{DESI}
{Schlafly} E.~F.,  et~al., 2023, \mn@doi [arXiv e-prints]
  {10.48550/arXiv.2306.06309}, \href
  {https://ui.adsabs.harvard.edu/abs/2023arXiv230606309S} {p. arXiv:2306.06309}

\bibitem[\protect\citeauthoryear{{Shandarin}}{{Shandarin}}{2011}]{Shandarin2011}
{Shandarin} S.~F.,  2011, \mn@doi [\jcap] {10.1088/1475-7516/2011/05/015},
  \href {https://ui.adsabs.harvard.edu/abs/2011JCAP...05..015S} {2011, 015}

\bibitem[\protect\citeauthoryear{{Shandarin} \& {Zeldovich}}{{Shandarin} \&
  {Zeldovich}}{1989}]{shandarin1989}
{Shandarin} S.~F.,  {Zeldovich} Y.~B.,  1989, \mn@doi [Reviews of Modern
  Physics] {10.1103/RevModPhys.61.185}, \href
  {https://ui.adsabs.harvard.edu/abs/1989RvMP...61..185S} {61, 185}

\bibitem[\protect\citeauthoryear{{Shandarin}, {Habib}  \&
  {Heitmann}}{{Shandarin} et~al.}{2012}]{Shandarin2012}
{Shandarin} S.,  {Habib} S.,   {Heitmann} K.,  2012, \mn@doi [\prd]
  {10.1103/PhysRevD.85.083005}, \href
  {https://ui.adsabs.harvard.edu/abs/2012PhRvD..85h3005S} {85, 083005}

\bibitem[\protect\citeauthoryear{{Sheth} \& {van de Weygaert}}{{Sheth} \& {van
  de Weygaert}}{2004}]{shethwey2004}
{Sheth} R.~K.,  {van de Weygaert} R.,  2004, \mn@doi [\mnras]
  {10.1111/j.1365-2966.2004.07661.x}, \href
  {https://ui.adsabs.harvard.edu/abs/2004MNRAS.350..517S} {350, 517}

\bibitem[\protect\citeauthoryear{{Shim}, {Codis}, {Pichon}, {Pogosyan}  \&
  {Cadiou}}{{Shim} et~al.}{2021}]{codis2021}
{Shim} J.,  {Codis} S.,  {Pichon} C.,  {Pogosyan} D.,   {Cadiou} C.,  2021,
  \mn@doi [\mnras] {10.1093/mnras/stab263}, \href
  {https://ui.adsabs.harvard.edu/abs/2021MNRAS.502.3885S} {502, 3885}

\bibitem[\protect\citeauthoryear{{Shivashankar}, {Pranav}, {Natarajan}, {van de
  Weygaert}, {Bos}  \& {Rieder}}{{Shivashankar}
  et~al.}{2016}]{Shivashankar2016}
{Shivashankar} N.,  {Pranav} P.,  {Natarajan} V.,  {van de Weygaert} R.,  {Bos}
  E.~G.~P.,   {Rieder} S.,  2016, \mn@doi [IEEE Transactions on Visualizations
  and Computer Graphics. 2016. Vol. 22(6] {10.1109/TVCG.2015.2452919}, \href
  {https://ui.adsabs.harvard.edu/abs/2016ITVCG..22.1745S} {22, 1745}

\bibitem[\protect\citeauthoryear{{Sousbie}}{{Sousbie}}{2011}]{Sousbie2011a}
{Sousbie} T.,  2011, \mn@doi [\mnras] {10.1111/j.1365-2966.2011.18394.x}, \href
  {https://ui.adsabs.harvard.edu/abs/2011MNRAS.414..350S} {414, 350}

\bibitem[\protect\citeauthoryear{{Sousbie}, {Pichon}  \& {Kawahara}}{{Sousbie}
  et~al.}{2011}]{Sousbie2011b}
{Sousbie} T.,  {Pichon} C.,   {Kawahara} H.,  2011, \mn@doi [\mnras]
  {10.1111/j.1365-2966.2011.18395.x}, \href
  {https://ui.adsabs.harvard.edu/abs/2011MNRAS.414..384S} {414, 384}

\bibitem[\protect\citeauthoryear{{Springel}}{{Springel}}{2005}]{Springel2005}
{Springel} V.,  2005, \mn@doi [\mnras] {10.1111/j.1365-2966.2005.09655.x},
  \href {https://ui.adsabs.harvard.edu/abs/2005MNRAS.364.1105S} {364, 1105}

\bibitem[\protect\citeauthoryear{{Tegmark} et~al.,}{{Tegmark}
  et~al.}{2004}]{tegmark2004cosmological}
{Tegmark} M.,  et~al., 2004, \mn@doi [\prd] {10.1103/PhysRevD.69.103501}, \href
  {https://ui.adsabs.harvard.edu/abs/2004PhRvD..69j3501T} {69, 103501}

\bibitem[\protect\citeauthoryear{{Tempel}, {Stoica}  \& {Saar}}{{Tempel}
  et~al.}{2013}]{Tempel2013}
{Tempel} E.,  {Stoica} R.~S.,   {Saar} E.,  2013, \mn@doi [\mnras]
  {10.1093/mnras/sts162}, \href
  {https://ui.adsabs.harvard.edu/abs/2013MNRAS.428.1827T} {428, 1827}

\bibitem[\protect\citeauthoryear{{Tempel}, {Libeskind}, {Hoffman},
  {Liivam{\"a}gi}  \& {Tamm}}{{Tempel} et~al.}{2014a}]{Tempel2014}
{Tempel} E.,  {Libeskind} N.~I.,  {Hoffman} Y.,  {Liivam{\"a}gi} L.~J.,
  {Tamm} A.,  2014a, \mn@doi [\mnras] {10.1093/mnrasl/slt130}, \href
  {https://ui.adsabs.harvard.edu/abs/2014MNRAS.437L..11T} {437, L11}

\bibitem[\protect\citeauthoryear{{Tempel}, {Stoica}, {Mart{\'\i}nez},
  {Liivam{\"a}gi}, {Castellan}  \& {Saar}}{{Tempel}
  et~al.}{2014b}]{Tempel2014b}
{Tempel} E.,  {Stoica} R.~S.,  {Mart{\'\i}nez} V.~J.,  {Liivam{\"a}gi} L.~J.,
  {Castellan} G.,   {Saar} E.,  2014b, \mn@doi [\mnras]
  {10.1093/mnras/stt2454}, \href
  {https://ui.adsabs.harvard.edu/abs/2014MNRAS.438.3465T} {438, 3465}

\bibitem[\protect\citeauthoryear{{Tully}, {Shaya}, {Karachentsev}, {Courtois},
  {Kocevski}, {Rizzi}  \& {Peel}}{{Tully} et~al.}{2008}]{Tully2008localvoid}
{Tully} R.~B.,  {Shaya} E.~J.,  {Karachentsev} I.~D.,  {Courtois} H.~M.,
  {Kocevski} D.~D.,  {Rizzi} L.,   {Peel} A.,  2008, \mn@doi [\apj]
  {10.1086/527428}, \href
  {https://ui.adsabs.harvard.edu/abs/2008ApJ...676..184T} {676, 184}

\bibitem[\protect\citeauthoryear{{Tully}, {Courtois}  \& {Sorce}}{{Tully}
  et~al.}{2016}]{Cosmicflows3}
{Tully} R.~B.,  {Courtois} H.~M.,   {Sorce} J.~G.,  2016, \mn@doi [\aj]
  {10.3847/0004-6256/152/2/50}, \href
  {https://ui.adsabs.harvard.edu/abs/2016AJ....152...50T} {152, 50}

\bibitem[\protect\citeauthoryear{{Wang}, {Szalay}, {Arag{\'o}n-Calvo},
  {Neyrinck}  \& {Eyink}}{{Wang} et~al.}{2014}]{wang2014}
{Wang} X.,  {Szalay} A.,  {Arag{\'o}n-Calvo} M.~A.,  {Neyrinck} M.~C.,
  {Eyink} G.~L.,  2014, \mn@doi [\apj] {10.1088/0004-637X/793/1/58}, \href
  {https://ui.adsabs.harvard.edu/abs/2014ApJ...793...58W} {793, 58}

\bibitem[\protect\citeauthoryear{{Welker} et~al.,}{{Welker}
  et~al.}{2020}]{Welker2020}
{Welker} C.,  et~al., 2020, \mn@doi [\mnras] {10.1093/mnras/stz2860}, \href
  {https://ui.adsabs.harvard.edu/abs/2020MNRAS.491.2864W} {491, 2864}

\bibitem[\protect\citeauthoryear{{White}}{{White}}{1984}]{White1984ttt}
{White} S.~D.~M.,  1984, \mn@doi [\apj] {10.1086/162573}, \href
  {https://ui.adsabs.harvard.edu/abs/1984ApJ...286...38W} {286, 38}

\bibitem[\protect\citeauthoryear{{White}, {Frenk}, {Davis}  \&
  {Efstathiou}}{{White} et~al.}{1987}]{defw1987}
{White} S. D.~M.,  {Frenk} C.~S.,  {Davis} M.,   {Efstathiou} G.,  1987,
  \mn@doi [\apj] {10.1086/164990}, \href
  {https://ui.adsabs.harvard.edu/abs/1987ApJ...313..505W} {313, 505}

\bibitem[\protect\citeauthoryear{{Wilde}, {Elek}, {Burchett}, {Nagai},
  {Prochaska}, {Werk}, {Tuttle}  \& {Forbes}}{{Wilde} et~al.}{2023}]{Wilde2023}
{Wilde} M.~C.,  {Elek} O.,  {Burchett} J.~N.,  {Nagai} D.,  {Prochaska} J.~X.,
  {Werk} J.,  {Tuttle} S.,   {Forbes} A.~G.,  2023, \mn@doi [arXiv e-prints]
  {10.48550/arXiv.2301.02719}, \href
  {https://ui.adsabs.harvard.edu/abs/2023arXiv230102719W} {p. arXiv:2301.02719}

\bibitem[\protect\citeauthoryear{Wilding}{Wilding}{2022}]{Wilding2022}
Wilding G.,  2022, PhD thesis, University of Groningen,
  \mn@doi{10.33612/diss.250010290}

\bibitem[\protect\citeauthoryear{{Wilding}, {Nevenzeel}, {van de Weygaert},
  {Vegter}, {Pranav}, {Jones}, {Efstathiou}  \& {Feldbrugge}}{{Wilding}
  et~al.}{2021}]{wilding2021}
{Wilding} G.,  {Nevenzeel} K.,  {van de Weygaert} R.,  {Vegter} G.,  {Pranav}
  P.,  {Jones} B. J.~T.,  {Efstathiou} K.,   {Feldbrugge} J.,  2021, \mn@doi
  [\mnras] {10.1093/mnras/stab2326}, \href
  {https://ui.adsabs.harvard.edu/abs/2021MNRAS.507.2968W} {507, 2968}

\bibitem[\protect\citeauthoryear{{Zeldovich}}{{Zeldovich}}{1970}]{zeldovich1970}
{Zeldovich} Y.~B.,  1970, \aap, \href
  {https://ui.adsabs.harvard.edu/abs/1970A&A.....5...84Z} {500, 13}

\bibitem[\protect\citeauthoryear{{Zhang}, {Lee}, {Krolewski}, {Shi}, {Horowitz}
   \& {Kooistra}}{{Zhang} et~al.}{2023}]{zhang2023}
{Zhang} B.,  {Lee} K.-G.,  {Krolewski} A.,  {Shi} J.,  {Horowitz} B.,
  {Kooistra} R.,  2023, \mn@doi [\apj] {10.3847/1538-4357/ace695}, \href
  {https://ui.adsabs.harvard.edu/abs/2023ApJ...954...49Z} {954, 49}

\bibitem[\protect\citeauthoryear{{de Graaff}, {Cai}, {Heymans}  \&
  {Peacock}}{{de Graaff} et~al.}{2019}]{degraaff2019}
{de Graaff} A.,  {Cai} Y.-C.,  {Heymans} C.,   {Peacock} J.~A.,  2019, \mn@doi
  [\aap] {10.1051/0004-6361/201935159}, \href
  {https://ui.adsabs.harvard.edu/abs/2019A&A...624A..48D} {624, A48}

\bibitem[\protect\citeauthoryear{{de Lapparent}, {Geller}  \& {Huchra}}{{de
  Lapparent} et~al.}{1986}]{Lapperent1986}
{de Lapparent} V.,  {Geller} M.~J.,   {Huchra} J.~P.,  1986, \mn@doi [\apjl]
  {10.1086/184625}, \href
  {https://ui.adsabs.harvard.edu/abs/1986ApJ...302L...1D} {302, L1}

\bibitem[\protect\citeauthoryear{{de la Torre} et~al.,}{{de la Torre}
  et~al.}{2013}]{vipers2013}
{de la Torre} S.,  et~al., 2013, \mn@doi [\aap] {10.1051/0004-6361/201321463},
  \href {https://ui.adsabs.harvard.edu/abs/2013A&A...557A..54D} {557, A54}

\bibitem[\protect\citeauthoryear{{van de Weygaert}}{{van de
  Weygaert}}{2016}]{weygaert2016}
{van de Weygaert} R.,  2016, {Voids and the Cosmic Web: cosmic depression \&
  spatial complexity}.
pp 493--523 (\mn@eprint {arXiv} {1611.01222}),
  \mn@doi{10.1017/S1743921316010504}

\bibitem[\protect\citeauthoryear{{van de Weygaert} \& {Babul}}{{van de
  Weygaert} \& {Babul}}{1994}]{weybabul1994}
{van de Weygaert} R.,  {Babul} A.,  1994, \mn@doi [\apjl] {10.1086/187310},
  \href {https://ui.adsabs.harvard.edu/abs/1994ApJ...425L..59V} {425, L59}

\bibitem[\protect\citeauthoryear{{van de Weygaert} \& {Bernardeau}}{{van de
  Weygaert} \& {Bernardeau}}{1998}]{bernwey1998}
{van de Weygaert} R.,  {Bernardeau} F.,  1998, in Proceedings of the 12th
  Potsdam Cosmology Workshop. pp 207--216 (\mn@eprint {arXiv}
  {astro-ph/9803143}), \mn@doi{10.48550/arXiv.astro-ph/9803143}

\bibitem[\protect\citeauthoryear{{van de Weygaert} \& {Bond}}{{van de Weygaert}
  \& {Bond}}{2008a}]{weybond2008}
{van de Weygaert} R.,  {Bond} J.~R.,  2008a, {Clusters and the Theory of the
  Cosmic Web}.
p.~335, \mn@doi{10.1007/978-1-4020-6941-3_10}

\bibitem[\protect\citeauthoryear{{van de Weygaert} \& {Bond}}{{van de Weygaert}
  \& {Bond}}{2008b}]{WeygaertBond2005}
{van de Weygaert} R.,  {Bond} J.~R.,  2008b, {Clusters and the Theory of the
  Cosmic Web}.
p.~335, \mn@doi{10.1007/978-1-4020-6941-3_10}

\bibitem[\protect\citeauthoryear{{van de Weygaert} \& {Platen}}{{van de
  Weygaert} \& {Platen}}{2011}]{weygaert2011}
{van de Weygaert} R.,  {Platen} E.,  2011, in International Journal of Modern
  Physics Conference Series. pp 41--66 (\mn@eprint {arXiv} {0912.2997}),
  \mn@doi{10.1142/S2010194511000092}

\bibitem[\protect\citeauthoryear{{van de Weygaert} \& {Schaap}}{{van de
  Weygaert} \& {Schaap}}{2009}]{weyschaap2009}
{van de Weygaert} R.,  {Schaap} W.,  2009, in {Mart{\'\i}nez} V.~J.,  {Saar}
  E.,  {Mart{\'\i}nez-Gonz{\'a}lez} E.,   {Pons-Border{\'\i}a} M.~J.,  eds, ,
  Vol.~665, Data Analysis in Cosmology.
pp 291--413, \mn@doi{10.1007/978-3-540-44767-2_11}

\bibitem[\protect\citeauthoryear{{van de Weygaert} \& {van Kampen}}{{van de
  Weygaert} \& {van Kampen}}{1993}]{weykamp1993}
{van de Weygaert} R.,  {van Kampen} E.,  1993, \mn@doi [\mnras]
  {10.1093/mnras/263.2.481}, \href
  {https://ui.adsabs.harvard.edu/abs/1993MNRAS.263..481V} {263, 481}

\bibitem[\protect\citeauthoryear{van~de Weygaert, Shandarin, Saar  \&
  Einasto}{van~de Weygaert et~al.}{2014}]{iau308}
van~de Weygaert R.,  Shandarin S.,  Saar E.,   Einasto J.,  eds, 2014,
  {Proceedings, IAU Symposium 308: The Zeldovich Universe: Genesis and Growth
  of the Cosmic Web}: {Tallinn, Estonia, June 23-28, 2014}

\makeatother
\end{thebibliography}




\appendix

\section[MMF/Nexus]{\\ MMF/Nexus \& Cosmic Web Classification}
\label{app:mmfnexus}

The \Nexus{} suite of cosmic web identifiers represents an elaboration and extension of the original Multiscale Morphology Filter
\citep{MMFa,MMF3} algorithm and was developed with the goal of obtaining a more physically motivated and robust method.
\nexus{} is the principal representative of the full \Nexus{} suite of cosmic web identifiers \citep[see][]{NEXUS_MAIN}. These include the 
options for multiscale morphology identifiers on the basis of the linear density, the logarithmic density, the velocity divergence, the velocity shear and tidal force field. \Nexus{} has incorporated these options in a versatile code for the analysis of cosmic web structure and 
dynamics following the realisation that they are significant physical influences in shaping the cosmic mass distribution into the complexity
of the cosmic web.

\subsection{Hessian Geometry and Morphological Identity}
The basic setup of MMF/Nexus is that of defining a four-dimensional scale-space representation of the input field $f(\vec{x})$.
In nearly all implementations this achieved by means of a Gaussian filtering of $f(\vec{x})$ over a set of
scales $[R_0,R_1,...,R_N]$, 
\begin{equation}
    f_{R_n}(\vec{x}) = \int \frac{{\rm d}^3k}{(2\pi)^3} e^{-k^2R_n^2/2} \hat{f}(\vec{k})  e^{i\vec{k}\cdot\vec{x}} ,
    \label{eq:filtered_field}
\end{equation}
\noindent where $\hat{f}(\vec{k})$ is the Fourier transform of the input field $f(\vec{x})$. The
Hessian $H_{ij,R_n}(\vec{x})$ of the filtered field on the scale $R_n$ is computed in Fourier space
on the basis of the corresponding Fourier components $\hat{H}_{ij,R_n}(\vec{k})$,
\begin{eqnarray}
    H_{ij,R_n}(\vec{x})&\,=\,&R_n^2 \; \frac{\partial^2f_{R_n}(\vec{x})}{\partial x_i\partial x_j}\,.
    \ \\
    \hat{H}_{ij,R_n}(\vec{k})&\,=\,& -k_ik_j R_n^2 \hat{f}(\vec{k}) e^{-k^2R_n^2/2}\,.\nonumber
    \label{eq:hessian_general}
\end{eqnarray}
\noindent Note that the definition for the Hessian includes the normalisation term $R_n^2$. The key element of the MMF/Nexus formalism
is the morphological information contained in the eigenvalues of the Hessian matrix, $h_1 \le h_2 \le h_3$. By applying a set of morphology
filters on these scaled eigenvalues \citep[see][]{MMFb,NEXUS_MAIN} this is translated into a scale dependent environment signature
$\mathcal{S}_{R_n}(\mathbf{ x})$ that represents the geometry at the corresponding scale. 

\subsection{Scale Space and Multiscale Structure}
To analyse the multiscale nature of the cosmic web, the \textit{Scale-Space} representation of the cosmic mass distributions produces a
sequence of copies of the data having different resolutions \citep{florack1992,Lindeberg1994}. At each location $\vec{x}$ in the
probed volume, it involves an extra dimension, scale, that yields the eigenvalues of the Hessian filtered at the corresponding
(Gaussian) scale and the scale dependent environment signature $\mathcal{S}_{R_n}(\mathbf{ x})$.

A feature searching algorithm is applied to the combined set of scaled copies in order to identify the scale at which, locally, we find
the strongest morphological signature. It involves the combination of the complete set of scale-dependent environmental signatures to
find the maximum signature for all scales
\begin{equation}
    \mathcal{S}(\mathbf{ x}) = \max\limits_{\text{levels n}} \mathcal{S}_{R_n}(\mathbf{ x}).
    \label{eq:NEXUSsig}
\end{equation}

\subsection{Signature \& Versions} 
The final step in the MMF/Nexus procedure involves the use of criteria to find the threshold signature that discriminates 
between valid and invalid morphological detections. Signature values larger than the threshold correspond to real structures
while the rest are spurious detections. Different implementations and versions of the MMF/Nexus technique may differ in the 
definition of the threshold values. 

The final outcome of the MMF/Nexus procedure is a field which at each location $\vec{x}$ specifies what the local morphological
signature is, cluster node, filaments, wall or void. The resulting field $\delta^{\text{NEXUS}}_j(\mathbf{ x})$ is zero when the volume is not identified as cosmic web element j and is one when the volume elements is identified as element j. Here j is either filaments, nodes or walls. In this
identification we also intrinsically include the identification for voids which is defined as the volume elements that are neither a filament, node or wall.

Following the basic version of the MMF technique introduced by \cite{MMFb} it was applied to the analysis of the cosmic web
in simulations of cosmic structure formation \citep{MMF3} and for finding filaments and galaxy-filament alignments in the SDSS galaxy 
distribution \citep{Jones2010_allign}. The principal technique, and corresponding philosophy, has subsequently been branched in several 
further elaborations and developments \cite{NEXUS_CW_EVO,Aragon2014}. Nexus \citep{NEXUS_CW_EVO} has extended the MMF formalism to a
substantially wider range of physical agents involved in the formation of the cosmic web, along with a substantially firmer foundation
for the criteria used in identifying the various web-like structures. MMF-2 \citep{Aragon2014} focuses with even more attention than the
basic MMF formalism on the hierarchical nature of the cosmic web, by introducing and exploiting the concept of \textit{hierarchical 
  space}.

The \Nexus{} suite of cosmic web identifiers represents an elaboration and extension of the original Multiscale Morphology Filter
\citep{MMFa,MMF3} algorithm and was developed with the goal of obtaining a more physically motivated and robust method.
\nexus{} is the principal representative of the full \Nexus{} suite of cosmic web identifiers \citep[see][]{NEXUS_MAIN}. These include the 
options for multiscale morphology identifiers on the basis of the linear density, the logarithmic density, the velocity divergence, the velocity shear and tidal force field. \Nexus{} has incorporated these options in a versatile code for the analysis of cosmic web structure and 
dynamics following the realisation that they are significant physical influences in shaping the cosmic mass distribution into the complexity
of the cosmic web.

\subsection{NEXUS+}
\nexus{} is the density field \Nexus{} version with the highest dynamic range. As input it takes a regularly sampled density field. In a first step, the input field is Gaussian smoothed using a \logFilter{} filter that is applied over a set of scales $[R_0,R_1,...,R_N]$, with $R_n=2^{n/2}R_0$. It produces the logarithmic density field
\begin{equation}
  \delta_+\,=\,\log (1+\delta (\mathbf{ x}))\,,
\end{equation}
The logarithmic density field of \nexus{} is better equipped to take account of the wide dynamic range of the nonlinear hierarchically
evolved density field. The nonlinear field is highly non-Gaussian, with a large part of the volume having low-density values in
combination with long high-density tails in the high-density cluster and filament regions. It translates into a nonlinear density field
pdf that approaches a lognormal or skewed lognormal function. 

For each of the included scale-space scales, \nexus{} computes the eigenvalues of the Hessian matrix of the smoothed logarithmic density field. Using the Hessian eigenvalues of these, \nexus{} computes an environmental signature for each volume element that characterises how close this region is to an ideal knot, filament and wall. Then, for each point, the environmental signatures computed for each scale are combined to obtain a scale independent signature.

In the last step, physical criteria are used to determine a detection threshold. All points with signature values above the threshold are valid structures. For knots, the threshold is given by the requirement that most knot-regions should be virialized. For filaments and walls, the threshold is determined on the basis of the change in filament and wall mass as a function of signature. The peak of the mass variation with signature delineates the most prominent filamentary and wall features of the cosmic web.

The \nexus{} algorithm performs the environment detection by applying the above steps first to knots, then to filaments and finally to walls. Each volume element is assigned a single environment characteristic by requiring that filament regions cannot be knots and that walls regions cannot be either knots or filaments. The remaining regions are classified as voids.


\bsp	
\label{lastpage}
\end{document}